\documentclass[fleqn,10pt]{olplainarticle}
\usepackage[acronym]{glossaries}
\usepackage{booktabs}
\usepackage{graphicx} 
\usepackage{lipsum} 
\usepackage{afterpage}
\makeglossaries 
\usepackage{tikz}
\usepackage{array}
\usepackage{tabularx}
\usepackage{arydshln}
\usepackage{caption}
\title{The Biophysics of Cancer-Microbe Interfaces:
Current Understanding and Open Frontiers}

\author[1]{Juan Pablo Carmona-Almazán}
\author[1,2,3, *]{Anupam Sengupta}
\affil[1]{Physics of Living Matter, Department of Physics and Materials Science, University of Luxembourg, 162 A, Avenue de la Fa\"{i}encerie, L-1511, Luxembourg}
\affil[2]{Institute for Advanced Studies, University of Luxembourg, 2, Avenue de l’Université, L-4365, Esch-sur-Alzette, Luxembourg}
\affil[3]{Institute of Microbiology, University of Applied Sciences and Arts of Southern Switzerland (SUPSI), Via Flora Ruchat-Roncati 15, CH-6850, Mendrisio, Switzerland}
\affil[$\ast$]{Corresponding author: anupam.sengupta@uni.lu}


\begin{abstract}
The microbiome, the natural consortium of bacteria and other microorganisms, plays a fundamental role in sustaining ecosystems and human health. It is now recognized as a core component of the tumor microenvironment, culminating in its recent inclusion within the Hallmarks of Cancer. However, our current understanding of this active–active system remains dominated by biomolecular perspectives. Despite the well-established biophysical principles governing both microbial growth and dynamics as well as cancer progression, the biophysical interactions and reciprocal feedback mechanisms that emerge at the cancer–microbiome interface remain poorly understood. In particular, how these two active, interacting systems mutually shape each other's development, organization, and evolution has yet to be fully elucidated. In this review, we examine the tumor-associated microbiome through the lens of microbial active matter to decode its structural, geometric, and biomechanical attributes. Focusing on prevalent cancers, we comprehensively catalogue the cancer–microbe systems which are currently known, highlighting the biophysical traits and phenomena including microbial phenotypes, collective behavior and self-organization across multiple spatial and temporal scales, and their influence on the active dynamics of tumors and associated extracellular matrix. Building on this perspective, we evaluate a central hypothesis that microbes and cancer communicate through reciprocal biomechanical feedback loops that actively govern the emergence and evolution of oncological traits. We propose a set of testable hypotheses and future research directions, including several currently being pursued in the authors' laboratory. Finally, we discuss how structural traits and functional properties emerge from these complex interactions and highlight the growing importance of machine learning in interrogating this data-rich landscape. By integrating biomolecular insights with physics-based approaches, this review establishes a new conceptual framework for understanding the mechano-adaptive functions that emerge from this active–active system. The biophysical framework presented here complements traditional molecular approaches, enabling predictive mechanistic models of adaptive active matter while uncovering new translational opportunities for the diagnosis, prevention, and treatment of microbiome-associated cancers.
\end{abstract}

\usepackage{float}
\begin{document}
\flushbottom
\maketitle
\thispagestyle{empty}

\section{Introduction}

The understanding of the connection between microbes and cancer has a long and rich history.  More than 2,400 years ago, Hippocrates was the first to describe cancer clinically. He also coined terms like "carcinoma" and "cancer" itself.  Up until the Renaissance (XIV–XVII c.), cancer was described and explained using Galen's "humoral theory" (129–199 AD).  According to this theory, maintaining health requires that the four humors that compose the human body (blood, yellow bile, black bile, and phlegm) be in balance.  Overproduction of black bile has been associated with cancer \citep{Karpozilos2004TheAntiquity, Hajar2021MedicinePresent}. For centuries, this view dominated medical thinking, even after Antonie van Leeuwenhoek discovered bacteria and protists in 1675 using his self-designed microscopes \citep{Kutschera2023AntonieMicrobiology}. It was not until 1882 that Robert Koch described the tuberculosis bacillus, and show that a particular bacterium could cause an infectious disease.  Two years later, he published his four postulates, which formalized the concepts of germ theory and disease transmission \citep{Koch1882DieTuberkulose., Koch1884DieTuberkulose}. 

The notion that microbes cause diseases to spread quickly, and cancer was included in this theory.  Several hypotheses that suggested an infectious cause for cancer emerged in the 1890s.  Some of them suggested that parasitic protozoa might infiltrate the cytoplasm and nucleus of cells \citep{Steinhaus1891UeberCarcinom-Einschlusse, Ruffer1893OnTumours}. Others even suggested that cancer cells were parasites themselves, growing inside the body and eventually invading, damaging, and proliferating within host tissues \citep{Butlin1905CarcinomaEngland}. However, many of these early investigations emphasized the continued lack of evidence linking parasites and cancer cells directly. However, there have also been attempts to treat cancer using microorganisms, such as William B. Colley, who successfully reduced the size of the patient's tumors by injecting streptococcal organisms, being one of the pioneers in using immunotherapy \cite{McCarthy2006TheSarcomas}.

Almost a hundred years later, in 1983, Goodwin, Warren, and Marshall identified the \emph{Helicobacter pylori} bacterium (then called \emph{Campylobacter pyloridis}) and reported it as a cause of chronic gastritis, suggesting a possible association with gastric cancer \citep{JRobinWarren1983UNIDENTIFIEDGASTRITIS, Goodwin1986CampylobacterUlceration.}. During the following decade, the accumulation of evidence reinforced the role of \emph{H. pylori} as a major contributor to gastric carcinogenesis \citep{Correa1990HelicobacterRisks, Parsonnet1991HelicobacterCarcinoma, Tatsuta1993TheCancer, Asaka1994PossibleDevelopment}. In 1994, the International Agency for Research on Cancer formally recognized \emph{H. pylori} as the first bacterium classified as a carcinogen, along with the first parasites recognized as carcinogenic: \emph{Schistosoma haematobium}, whose role in bladder cancer had been suggested by Goebel and Ferguson nearly 90 years earlier due to the unusually high incidence of bladder carcinoma in endemic regions, and \emph{Opisthorchis viverrini} \citep{InternationalAgencyforResearchonCancer1994SchistosomesPylori, Berry2017SchistosomoseCancer}.

Based on these findings, the Hallmarks of Cancer conceptual framework was set in 2000 and identified six fundamental alterations in cell physiology that together contribute to the growth of cancer \citep{Hanahan2000TheCancer}. More than two decades later, in 2022, this framework increased to fourteen, with the addition of polymorphic microbes. Through the induction or inhibition of hallmark capabilities, these microbes have the ability to modulate the progression of cancer, making them potential tools and quasi-independent variables in the development, progression, and response to treatment of cancer \citep{Hanahan2022HallmarksDimensions}. Here, we will explore cancer as a complex ecosystem shaped not only by mechanical and metabolic factors, but also by its microbial component. From a biophysical perspective, a solid tumor functions as a living \textit{active solid}, a dynamic structure where self-driven units are embedded within a complex elastic matrix. While invasive cancer cells have long been recognized as the primary active units driving mechanical changes within this matrix, recent evidence points to a powerful, concurrent force: the cancer microbiome. Tumor-associated bacteria act as a secondary population of active mechanical agents that exert their own physical forces, alter fluid transport, and modulate the stiffness of the extracellular matrix \cite{Xu2022AutonomousSolids}. We will discuss how bacteria in tumors go far beyond simple associations, since they display distinct phenotypes, activate specific pathways, and generate physical consequences that can influence tumor behavior. The goal of this work is to highlight the biophysical perspective of cancer–microbiome interactions, with a specific focus on reciprocal biomechanical and topological interactions, whereby microbial communities influence tumor development while tumor progression, in turn, modulates the structural and organizational dynamics of associated biofilms.

\section{Microbes in cancer: evidence and associations}

\begin{figure}[ht]
    \centering
    \includegraphics[width=\linewidth]{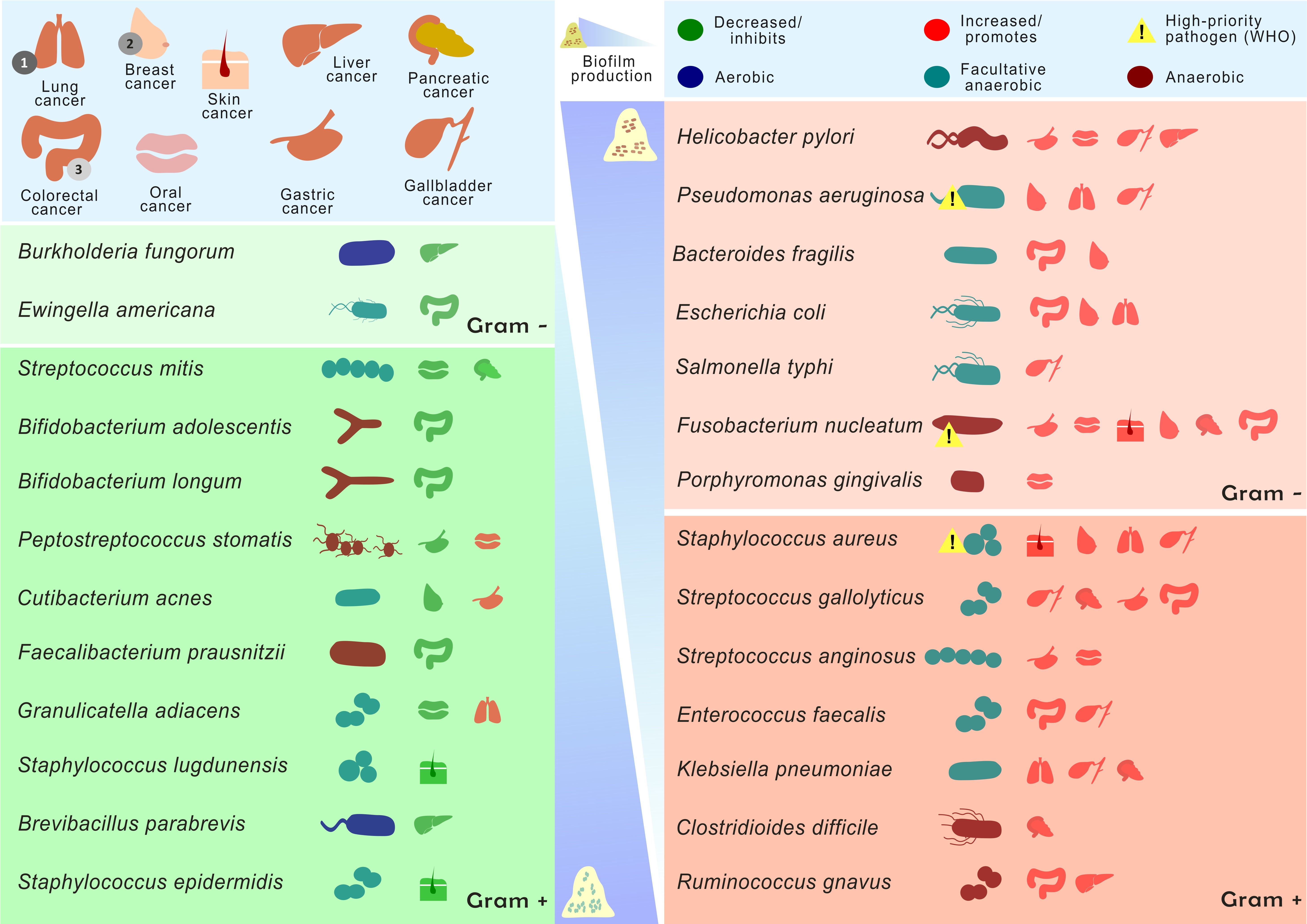}
    \caption{\textbf{Associations between bacterial species and cancer types in the human body.} This schematic summarizes the roles of selected bacteria in various cancers. The green panel lists bacteria associated with anti-tumor properties, while the red panel contains species linked to tumor progression or enriched in cancer tissues. Lighter background denote Gram-negative (-) bacteria, whereas darker background indicate Gram-positive (+) species. Organ icons map the specific cancers reported for each bacterium, color-coded by their functional effect. Bacterial shapes color indicate respiratory metabolism, and the central blue gradient reflects biofilm production capacity. Numbers adjacent to the lung, breast, and colorectal cancer icons denote their global incidence ranking \cite{Bray2024GlobalCountries}. Yellow triangles highlight WHO high-priority pathogens \cite{WHO2024}}
    \label{fig:bacteria-phenotypes}
\end{figure}

Having outlined the conceptual foundations of cancer as a mechanically and microbially influenced ecosystem, we now turn to the evidence linking specific bacteria to tumor biology. \textit{H. pylori} remains the only bacterium formally recognized as a human carcinogen; nevertheless, microbiome dysbiosis, an imbalance among resident microbial species, drives malignancy across multiple anatomical sites \cite{ElTekle2023BacteriaProgression}. In \textbf{gastric cancer (GC)}, \textit{H. pylori} damages the gastric mucosa by inducing chronic inflammation and atrophy, while reducing stomach acidity (achlorhydria), ultimately triggering widespread gastric dysbiosis. Interestingly, its abundance within tumor tissue is reported to be lower than in adjacent non-cancerous tissue, pointing to a "hit-and-run" mechanism \cite{Gunathilake2021AssociationCancer}. This pattern suggests that non-\textit{H. pylori} pathogens accelerate tumorigenesis; for example, \textit{Streptococcus anginosus} disrupts epithelial tight junctions to increase carcinogen permeability and promote cell proliferation \cite{Fu2024StreptococcusMice}, while \textit{Fusobacterium nucleatum}, a common oral resident bacteria, is enriched in GC patients and correlated with poor prognosis \cite{Lehr2023MicrobialCancer}. \textit{F. nucleatum} has been primarily linked to \textbf{colorectal cancer (CRC)}, where it secretes metabolites such as formate alongside virulence factors via outer membrane vesicles to drive oncogenic signaling, metabolic reprogramming, proinflammatory profiles, and immune evasion \cite{Weinberg2023ComprehensiveCRC., Ternes2022TheProgression}. Furthermore, \textit{F. nucleatum} induces cancer-associated fibroblast (CAF) remodeling within the tumor stroma, promoting subsequent CRC invasiveness \cite{Karta2025FusobacteriumCancer}. Together with \textit{pks}$^+$ \textit{Escherichia coli}, which induces chromosomal damage through colibactin production \cite{Mauri202582PCancer}, \textit{F. nucleatum} acts as a key proinflammatory driver in CRC, triggering intestinal dysbiosis by displacing beneficial commensals such as \textit{Bifidobacterium} and \textit{Anoxybacillus}. Additionally, studies using a multiorgan-on-a-chip model have demonstrated that \textit{E. coli} alters the metabolism of CRC therapeutics such as irinotecan, converting it into its toxic metabolite SN-38 \cite{Lucchetti2024An20/2024}. The oral cavity also sustains a dense microbiome. In \textbf{oral squamous cell carcinoma (OSCC)}, anoxic and acidic tumor microniches create favorable conditions for proliferating populations of \textit{F. nucleatum} and \textit{Porphyromonas gingivalis}. These species activate the E-cadherin/$\beta$-catenin pathway and upregulate epithelial–mesenchymal transition (EMT) genes, enhancing cancer cell invasiveness and survival \cite{Mivehchi2025ExploringReview, Shao2021ConversionCells}. Conversely, refluxed extragastric \textit{H. pylori} exacerbates oral tissue inflammation and reactive oxygen species (ROS) production, whereas commensal species such as \textit{Streptococcus mitis} exert protective effects by inducing G2/M phase cell-cycle arrest \cite{Inui2025InModulation}. In \textbf{lung cancer}, distinct microbial profiles are observed across different compartments of the respiratory tract. For example, the lower tract is predominantly colonized by Gram-positive cocci such as \textit{Streptococcus pneumoniae} and \textit{Staphylococcus aureus}. Additionally, chemotherapy has been reported to enrich opportunistic Gram-negative pathogens, including \textit{Klebsiella pneumoniae}, \textit{E. coli}, and \textit{Pseudomonas aeruginosa} \cite{Liu2020DistributionPatients, Ma2025DissectionCancer}. Furthermore, murine studies show that commensal bacteria from distal tissues can translocate to lung tumors and modulate metabolic pathways that inhibit tumorigenesis \cite{Zhu2023MicrobiomeMice}. \textbf{Skin cancers} broadly encompass actinic keratosis, melanoma, and squamous cell carcinoma. Across these malignancies, microbial dysbiosis is frequently characterized by the outgrowth of \textit{S. aureus}, against which commensal \textit{Staphylococcus epidermidis} provides antimicrobial defense \cite{Nakatsuji2018ANeoplasia, Byrd2018TheMicrobiome}. Notably, \textit{F. nucleatum} also contributes to melanoma progression by migrating from the gut to the tumor niche \cite{Mekadim2022DysbiosisProgression}. In \textbf{breast cancer}, dysbiosis is closely linked to aggressive triple-negative breast cancer (TNBC), marked by an enrichment of \textit{Lactobacillus} species and \textit{S. aureus}, the latter being implicated in metastatic dissemination \cite{Gao2025AMicrobiota, Fu2022Tumor-residentCancer}. Moreover, toxin-secreting bacteria further modulate breast cancer progression; for instance, intratumoral \textit{E. coli} alters TNBC metabolism, while \textit{P. aeruginosa} secretes biomolecules that induce resistance to trastuzumab therapy \cite{AlMalki2023E.Metabolism, Qin2025ACells}. An overview of these microbial associations is provided in Fig. \ref{fig:bacteria-phenotypes}.

\raggedbottom
\section{The cancer microbiome: characteristics and organization}

\subsection{Tumor microenvironment: An overview}


When referring to a tumor, it is often understood as a mass of cancer cells that proliferate and grow uncontrollably. However, these epithelial-derived cells at the tumor core represent only one component of the tumor microenvironment (TME). The TME also includes several non-cancerous cell types, among which stromal cells constitute a significant proportion, varying across tumor types (Fig. \ref{fig:2_TME}, bottom left). The most abundant are the cancer-associated fibroblasts (CAFs), which produce most of the extracellular components, such as growth factors, cytokines, and extracellular matrix (ECM), and are involved in tumor proliferation, ECM remodeling, immunosuppression, and angiogenesis \cite{Anderson2020TheMicroenvironment}. This last one, orchestrated by vascular endothelial cells, refers to the formation of new blood vessels that connect the tumor to the vascular system, thereby promoting cell migration, invasion, and metastasis \cite{Giubelan2023TumorCancer}. As in any living system, tumors require energetic support, and here cancer-associated adipocytes (CAAs) play a central role. These specialized stromal cells regulate energy balance and store excess energy as fat. Their role is particularly relevant in the breast cancer TME, where they promote the secretion of metabolites, enzymes, hormones, growth factors, and cytokines. Evidence suggests that adipocytes may protect breast cancer cells from ferroptosis through oleic acid \cite{Xie2022MammaryFerroptosis}.

Stromal cells are not the only non-malignant cellular components of the TME. Immune cells also play a crucial role in tumor progression, exerting both pro- and anti-tumorigenic effects. T cells can recognize and destroy cancer cells (such as CD8+ cytotoxic T cells) \cite{Mahadevan2023KRASG12DCells} or regulate the immune response (as regulatory T cells, Tregs), which suppress excessive immune activity but can also inhibit anti-tumor responses, thereby favoring tumor development \cite{Sun2024GALNT6Cells}. B cells, frequently located in lymph nodes near the TME, produce antibodies, antigens, and cytokines that can be recognized by other immune cells to suppress tumor progression \cite{Yang2024Pan-cancerSubtypes}. Macrophages, key components of the innate immune system, are responsible for pathogen phagocytosis and antigen presentation. Within tumors, hypoxia and cytokine secretion promote the infiltration of M2 macrophages, sometimes comprising up to 50$\%$ of tumor mass, which are generally associated with poor patient prognosis in several cancers, including breast, lung, and colorectal cancers \cite{Zhou2024TheTherapy, Cao2023TumorImmunotherapy, Feng2024M2-TypeMetastases}. Another innate immune cell type, neutrophils, are recruited at early stages of tumor growth, promoting inflammation through the release of cytokines and ROS to induce cell apoptosis\cite{Wu2023MUC16Cancer}. In later stages, however, neutrophils may support tumor progression by promoting glycolisis \cite{Ou2021C5aR1-positiveENO1} and secreting matrix metalloproteases (MMPs) that remodel the ECM, promoting angiogenesis, and facilitating local invasion  \cite{Rawat2022HyperactiveActivity}.

Beyond the cellular composition of the TME, it is important to establish the physico-chemical conditions within it. Once tumors reach a volume of about 1–2 mm$^3$, poor vessel perfusion combined with the rapid uptake of O$_2$ leads to a hypoxic condition \cite{Anderson2023TheDifficile, Bian2022AntihypertensionCancer}. Another affected parameter is pH, which becomes acidic due to the accumulation of acids (such as lactate) and CO$_2$, produced as a result of the high glucose consumption through both aerobic and anaerobic pathways. In the TME, the extracellular pH can decrease to around 6.6, impairing the infiltration and tumor-suppressive activity of immune cells. An acidic environment enhances ECM degradation and cell motility when it increases the expression of membrane-anchored membrane type 1-matrix metalloproteinase \cite{Liang2021AcidicAxis, Tang2023TumorImmunotherapy}. The accumulation of metabolites is another defining feature of the TME. There is an excess of lactate, produced by highly glycolytic cancer cells; adenosine, generated from the degradation of extracellular ATP; and ROS, which promote inflammation \cite{Fernandez2019HepaticFunction}.

Among the non-cellular components of the TME, one element to highlight in this review is the ECM, particularly given its physical properties, which is a complex, three-dimensional macromolecular network that is continuously remodeled. It provides a structural scaffold for cells and plays a fundamental role in tumor progression (Fig. \ref{fig:2_TME}, top left) \cite{Mittal2018TheTherapy}. As previously mentioned, the ECM is a dynamic structure in continuous remodeling processes such as deposition, degradation, and renewal to maintain tissue homeostasis. The group of matrix-associated proteins expressed by the genome of an organism is called the matrisome, and it contains approximately 4$\%$ of the entire human proteome \cite{Popova2022TheCancer}. The protein composition of the matrisome can be categorized into three groups. First, the structural proteins, which provide the scaffolding and tensile strength of the matrix, such as collagens and fibronectin, or confer pliability, as in the case of elastin. Secondly, the proteoglycans (PGs), which are specialized glycoproteins attached to long glycosaminoglycan chains. These proteins provide viscosity and allow tissues to resist compressive forces and buffer mechanical stress. Finally, the matricellular proteins consist of non-collagenous components that modulate cell–matrix interactions and growth factor activity \cite{Cox2021TheCancer}.

Collagen is the most abundant protein in the ECM. When collagen crosslinking increases, ECM stiffness also increases \cite{Wu2023MatrixPathway}. During cancer progression, both its density and architecture are altered. For example, collagen type I shows increased fiber density, alignment, and width in primary tumors and metastatic sites of patients with ovarian cancer. Higher fiber density enhances cancer cell adhesion, attributed to the fiber structures themselves, thereby supporting the formation of additional metastases \cite{Abbaspour2025CollagenAdhesion, Alkmin2022AssessingModels}. In late-stage breast cancer, there is an increase in anisotropic collagen fibers in the tissue microenvironment, suggesting that fiber organization is also a key determinant of metastasis and invasion. In healthy tissues, these fibrillar collagens are generally disordered or isotropically oriented \cite{Zhou2024EmulationMicroenvironments}.

\subsubsection{Biophysical properties of the ECM}
\paragraph{Mechanical properties.}

\begin{figure}[ht]
    \centering    \includegraphics[width=1\linewidth]{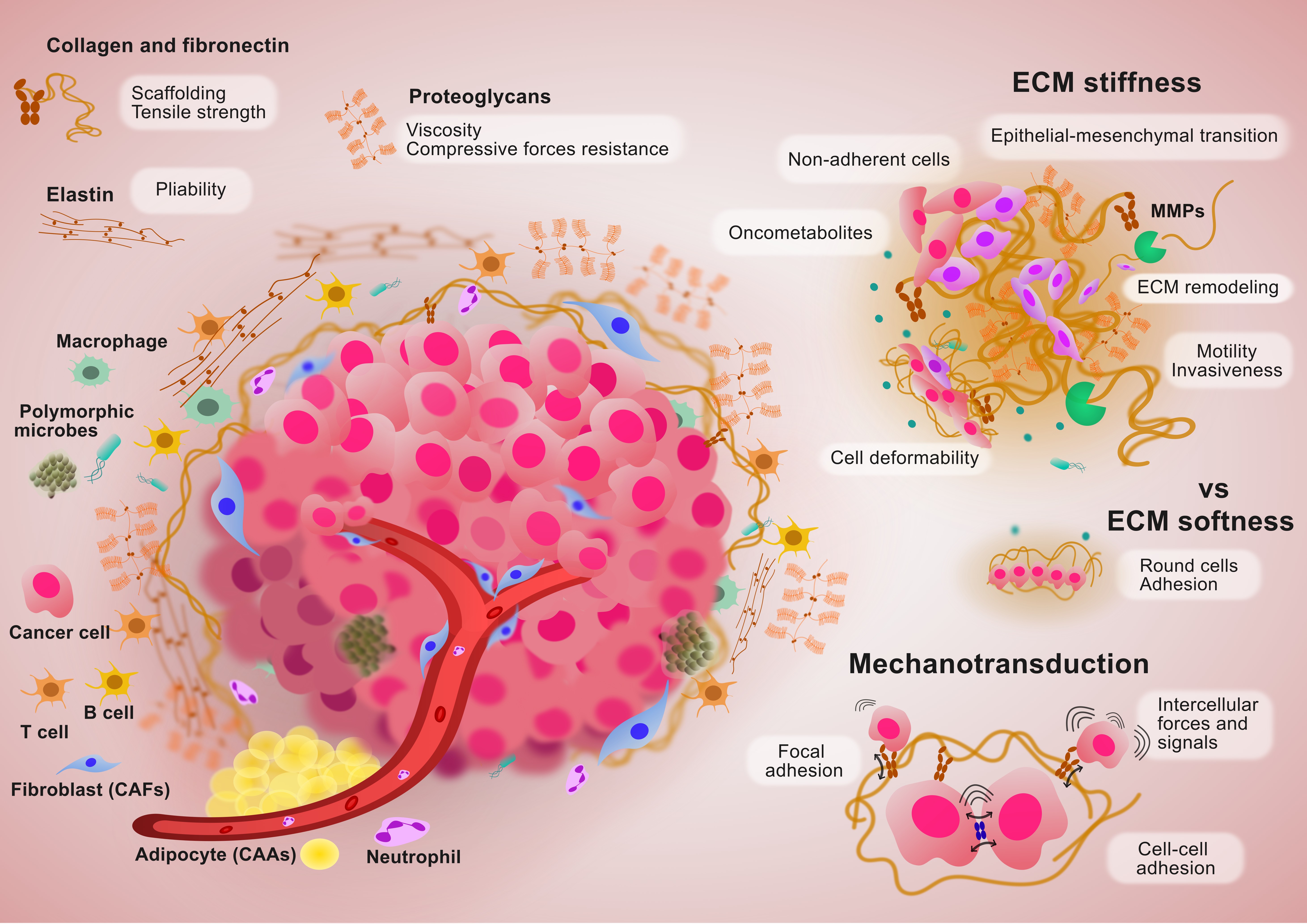}
    \caption{\textbf{Cancer micro-environments: Biophysical components and processes}. Conceptualizing the tumor as an active solid, the extracellular matrix (ECM) functions as the elastic network that holds the active units, namely cells (cancer, immune, epithelial, adipocytes) and bacteria. Structurally, the ECM relies on collagen and fibronectin for tensile strength, elastin for pliability, and proteoglycans for its viscous properties. Biomechanically, a stiffer ECM alters cell morphology (making them softer and more spread) and promotes epithelial-mesenchymal transition (EMT). To facilitate invasion through this stiffened network, matrix metalloproteinases (MMPs) degrade matrix proteins. Additionally, the ECM acts as the primary medium for mechanotransduction, enabling cells to communicate and adapt to physical forces across the microenvironment.}
    \label{fig:2_TME}
\end{figure}

The ECM provides essential mechanical support to cells. Mechanical properties have a physiological effect on cells through mechanotransduction, a process by which cells sense, transduce, and assimilate external forces, activating specific internal signals in response to this information (Fig. \ref{fig:2_TME}, bottom right). Cell–cell and ECM–cell interactions generate intercellular tension and extracellular traction on the matrix, transmitted through cell–cell adhesions or focal adhesions, respectively \cite{Zhang2019ExtracellularEpithelia}. One crucial mechanical property is stiffness, also referred to as rigidity or elasticity in this context, which is significantly altered in cancer (Fig. \ref{fig:2_TME}, top right). Typically, the stiffness of individual cancer cells is lower than healthy cells \cite{Kwon2020ComparisonType}, however human tumors are stiffer than healthy tissues, mainly due to increased collagen deposition and cross-linking, a condition known as desmoplasia \cite{Xie2019TheCells}. An increased ECM stiffness can affect multiple aspects of cancer cell behavior. It directly modifies cell morphology, leading to enhanced adhesion, proliferation, and invasive capabilities, thereby facilitating tumor progression. Several models have demonstrated that the spatiotemporal dynamics of cell growth and volume adaptation in 3D microenvironments are regulated by ECM stiffness. On stiffer matrices, cancer cells tend to be flatter and more spread, whereas on softer matrices they adopt a rounder and more compact morphology \cite{Mai2024ModulatingImmunotherapy, Wang2021MechanicalMatrices}. ECM stiffening and remodeling can also reprogram cancer cell metabolism. For instance, it has been associated with increased aerobic glycolysis through YAP activation, which contributes to cell migration. This condition also promotes the production of metabolites such as aspartate and glutamate, supporting cell proliferation and maintaining redox balance \cite{Mai2024ModulatingImmunotherapy, Liu2020StifferSignaling}. Another consequence of ECM stiffening is the  EMT, a process by which epithelial cells lose adhesiveness and acquire mesenchymal traits like mobility upon contact with collagen IV at the basal. This transition is triggered by signals produced during ECM stiffening and enhances the invasive and migratory behaviors of cancer cells \cite{Mai2024ModulatingImmunotherapy, Rice2017MatrixCells}. In cancer, ECM stiffening is positively correlated with tumor aggressiveness and invasiveness. For example, in breast cancer, metabolic enzymes that remodel the ECM are upregulated during EMT and are highly expressed in patients with aggressive disease \cite{Mai2024ModulatingImmunotherapy, Acerbi2015HumanInfiltration, Arnold2020UDP-glucoseProgression}. Similarly, in gallbladder cancer, a desmoplastic microenvironment rich in collagen fibers promotes EMT and supports a malignant phenotype \cite{Sun2025MatrixSignalling}. 

The ECM is governed by other mechanical properties in addition to the stiffness, such as viscoelasticity, a behavior that combines the elastic properties of a stiff material, represented with the elastic modulus $G'$ and the viscous characteristics of a fluid, represented by the loss modulus $G''$. Both, elastic and loss moduli, are related with the damping coefficient ($tan\delta=G''/G'$), where generally, stiffer viscoelastic materials show a lower damping coefficient than soft materials. Within the ECM, weak and strong crosslinks enable stress relaxation in response to deformation, and provide these viscoelastic properties. Also, elastic fibers confer pliability and non-linear elasticity to the tissues, allowing them to bend and twist \cite{Yuan2023ExtracellularTreatments}.  In metastatic breast cancer cell lines, the damping coefficient is the driving force of cellular migration, even more than the stiffness itself. A lower damping coefficient enhances the migration ability of cancer cells. The traction forces, or contractility, that the cells receive and extert with the ECM can change their morphologies and cytoskeletal structure, which influence cell response. Especiffcaly, a lower damping coefficient reduces the roundness and increases the spread, and in consequence the migration rate \cite{Chung2019Interface-derivedCeramics}. Another critical parameter that influences the early stages of metastasis is the shear stress originated from the interstitial fluid flow that moves through the ECM typically between the blood and lymphatic vessels. Progressive and active tumors are often characterized by increased intratumoral flow that exposes the cancer cells to a constant fluid shear stress, which is suggested to promote migration and invasion of cancer cells \cite{Ildiz2023TravellingJourney}.

\paragraph{ECM remodeling.} 

During development, wound healing, and tissue repair, the ECM undergoes continuous synthesis and degradation, known as turnover, and responds to perturbations through remodeling. These processes are driven by enzymes and involve several post-translational modifications \cite{Radzki2024MatrixPeriodontitis}. On the other hand, there are reciprocal interactions between cells and the ECM, known as dynamic reciprocity, where cells deposit, degrade, and remodel the matrix, while the ECM in turn influences cellular behavior. In cancer, ECM remodeling is significantly dysregulated, with contributions from both tumor cells and non-malignant stromal cells \cite{Cox2021TheCancer}. The main proteins involved in ECM remodeling are the previously mentioned MMPs, a group of zinc-dependent enzymes, which can degrade structural proteins, matricellular components, and proteoglycans. To maintain ECM homeostasis, their activity is regulated by inhibitors and activators, such as the tissue inhibitors of metalloproteinases (TIMPs). Disruption of this balance can promote tumor progression, since the ability of MMPs to degrade the ECM facilitates tumor cell infiltration into surrounding tissue, enhancing invasion and metastasis. For instance, in glioblastoma multiforme, known for its diffuse and infiltrative growth, MMP9 is upregulated, and its inhibition has been shown to reduce invasion in brain tumors \cite{Taheri2025EffectCells}. There is also evidence that primary tumors secrete factors that prepare the pre-metastatic niche before the arrival of metastatic cancer cells. For example, primary breast tumors can prime pre-metastatic lung tissue, increasing interstitial fibrillar collagen and ECM stiffness \cite{Cai2023PrimaryLungs}. A similar phenomenon occurs in uveal melanoma, an aggressive intraocular cancer that typically metastasizes to the liver. A recent study revealed a specific upregulated matrisome signature in high–metastatic risk liver tumors, related to ECM binding and stiffening, including MMP1 and collagen-1A1 \cite{Aughton2024RevealingDecellularisation}.

\subsection{Bacterial phenotypes within the microbiome}

The observable properties of an organism, like the morphology, development, or behavior, is what we typically consider as phenotype, result of the expression of the genotype, that contains its full hereditary information, even if not expressed. In terms of bacteria, the microbial phenotype can be the cell size and shape (or morphology), motility, biofilm formation, oxygen tolerance, Gram stain, related to sensitivity to antimicrobial agents, cellular composition, and biochemical activity, and frequently depends on the growth conditions, which include variables such as temperature, pH, redox potential and nutrient availability \cite{Sandle2014BIOCHEMICALMicroorganisms} (Fig. \ref{fig:bacteria-phenotypes}). 

\definecolor{shadecolor}{RGB}{246,243,226}
\colorbox{shadecolor}{%
\begin{minipage}{\textwidth}
\vspace{0.4em}

\begin{center}
\textbf{\large Box 1. Physical and chemical guidance taxis and biophysical modeling in the tumor microenvironment}
\end{center}

\vspace{0.6em}

The tumor microenvironment (TME) is characterized by strong chemical, mechanical, and electrical heterogeneities that act as physical guidance cues for both microbial and host cells \ref{fig:microbial-taxis}. To describe how bacteria and cells interact, migrate, and compete within these complex environments, biophysical and computational models have become essential. These models capture different physical guidance mechanisms, collectively referred to as \emph{taxis}, providing a quantitative framework to understand tumor colonization, infection dynamics, and therapeutic applications.

\vspace{0.5em}
\textbf{Chemotaxis and transport models.}
Chemical gradients generated by tumor metabolism, hypoxia, necrosis, and nutrient depletion guide bacterial migration. Chemotaxis enables bacteria such as \textit{Salmonella} and \textit{Escherichia coli} to preferentially colonize tumor tissue and establish protected microniches. Chemotaxis-driven models describe this navigation using analytical and stochastic frameworks to capture drift velocities, traveling-wave solutions, and population growth coupled to nutrient sensing \cite{Silva-Valenzuela2016SolidSalmonella, Song2018ATargeting, Kasinskas2006SalmonellaVitro, Narla2021AGrowth, Bains2025DriftEnvironments}.

\vspace{0.5em}
\textbf{Mechanotaxis (Haptotaxis and Durotaxis).}
Gradients of extracellular matrix (ECM) density and ligand availability (haptotaxis), as well as spatial variations in ECM stiffness (durotaxis), provide mechanical cues that direct migration. In cancer cells, these cues promote invasion toward fibrotic areas and support epithelial–mesenchymal transition. In bacteria, stiffness sensing and transient adhesion to ECM components affect their invasion modes, spatial distribution, and dissemination through dense polymeric environments \cite{Oudin2016MenaINVHaptotaxis, Fortunato2025Single-cellGradients, Harman2012TheTissue, Masters2021StaphylococcusPathogenesis, Ji2023DurotaxisGo, Song2017HowMotility, Liu2021ExtracellularInternalization}.

\vspace{0.5em}
\textbf{Rheotaxis and hydrodynamic navigation.}
Fluid flow and shear stress gradients guide bacterial movement in vascularized or interstitial environments. Rheotaxis models explain how bacteria respond to fluid velocity gradients through purely physical mechanisms involving flagellar geometry and shear-induced torques. These approaches show how flow conditions can strongly drive bacterial migration, promote upstream motion, and shape spatial organization near tumor vasculature \cite{Marcos2012BacterialRheotaxis, Siryaporn2015ColonizationNetworks, Shen2012FlowUpstream}.

\vspace{0.5em}
\textbf{Curvotaxis and geometry-guided migration.}
Substrate curvature at cellular and supracellular scales directs migration toward concave regions, guiding cancer and stromal cells along folded ECM structures and tumor boundaries. Geometry-based models describe this phenomenon by translating geometrical constraints into directed migration, highlighting the role of nuclear mechanics, intracellular pressure gradients, and adhesion dynamics in guiding long-range motion \cite{Pieuchot2018CurvotaxisLandscapes, Vassaux2019AMigration, Werner2019Cell-PerceivedMigration, Frey2025CurvatureTopographies, Manifacier2024InDistances, Schamberger2023CurvatureScales}.

\vspace{0.5em}
\textbf{Electrotaxis.}
Endogenous electric fields generated by tissue damage, inflammation, or infection guide the directional migration of immune cells, cancer cells, and bacteria. In the TME, electrical cues may cooperate with chemical and mechanical signals to regulate invasion, immune recruitment, and microbial localization \cite{Garg2019ElectromagneticCells, Huang2016CellularCells, Sun2025BacterialProcess}.

\vspace{0.5em}
\textbf{Infection mechanics and bacteria-enabled therapeutics.}
Integrating these physical cues, mechanistic models of infection focus on bacterial adhesion, intracellular dynamics, and the biomechanical response of host cell monolayers, capturing behaviors such as collective cell extrusion. Furthermore, these frameworks are used to evaluate bacteria-assisted drug delivery strategies, quantifying the trade-offs between deep tumor penetration, spatial release of metabolites, and local bacterial density (Fig. \ref{state-of-art}c) \cite{Galvez1997MathematicalLines, Galvez1997PenetrationInfection, Aparicio-Yuste2022ComputationalMonocytogenes, Aparicio-Yuste2025HybridMonolayers, Dominguez-Huttinger2017MathematicalTreatment, Toley2012MotilityTissue, Suh2018MathematicalSpheroids}.

\label{Box:taxis-models}
\end{minipage}}

In this subsection, we describe the main bacterial phenotypes and the characteristics that define them, with the aim of establishing a clear framework before discussing their relevance in more complex biological environments. By outlining these phenotypes, we create a reference point for understanding why certain groups behave as they do and how their structural and functional features influence their ecological roles. It is imperative to take into account the associated statistical noise in the distribution of the phenotypic traits that are captured experimentally. At the cellular scales, phenotypic traits shape properties at the population level, wherein systemic noise may self-regulate growth, collective structural organization, as well as the kinectics of emergent topological defects \cite{Dhar2022Self-regulationEnvironments}. This overview also helps contextualize the species listed in Fig. \ref{fig:bacteria-phenotypes}, which are commonly discussed in cancer-associated microbiomes.

\subsubsection{Morphology}

The shape of bacteria has been widely studied in recent decades, as it is known to play an important biological role and may have an adaptive origin. Although nature presents a broad diversity of bacterial shapes, each genus tends to favor one form. This preference is shaped by selective forces and passed from generation to generation through the genome. These adaptive forces include biological factors such as nutrient acquisition, cell division, and exposure to predators, as well as physical mechanisms that help bacteria cope with these conditions, including surface adhesion, passive dispersion, and active motility.

Molecular self-organization and the ability to respond to the cell’s innate geometry have allowed bacteria to develop shapes more complex than simple spheres. Rod-shaped bacteria, for example, can be seen as elongated spheres with an improved capacity to organize their internal contents \cite{Garner2021TowardRegulation}. From this basic form, many variations have evolved, each offering specific advantages. Long or filamentous bacteria tend to acquire nutrients more efficiently and adhere more easily to surfaces. Wider or more symmetrical bacteria show improved division and better motility. Helical or curved species move more effectively in viscous environments, while extremely large or very small bacteria can escape more easily from predators such as protists. Some species even develop branched forms or network-like structures to enhance buoyancy or remain anchored in soil (See Fig. \ref{fig:bacterial-motility}) \cite{Young2007BacterialShapes, Kysela2016DiversityMorphology, Jin2024MicrobesFeedback}.

Bacterial shape can also change over short timescales depending on the environment. Several species modify their morphology in response to environmental stress or during infection processes \cite{Yang2016StayingEnvironments}. For instance, nitric oxide, which is released during infections, can structurally damage bacteria, causing them to bend or fold and disrupting the cell wall of both Gram-positive and Gram-negative species \cite{Nguyen2023NitricBacteria}. Antibiotics can also affect morphology. In \textit{S. aureus}, for example, exposure to ceftriaxone results in cell elongation \cite{Azzam2023Sub-inhibitoryAureus}. Furthermore, recent studies have revealed that internal biomechanical forcing within expanding bacterial colonies plays a key role in the dynamical self-regulation of cell size in growing bacterial colonies. This, together with external biotic and abiotic factors, are thus expected to influence the biomechanical response of co-existing cancer cells as part of the cancer-microbiome ecosystems \cite{Wittmann2023CollectiveColonies}. 

\subsubsection{Motility}

Bacterial motility, together with chemotaxis (the attraction to regions enriched in specific metabolites or nutrients, see Box 1), provides bacteria with a strong ecological advantage. These traits allow cells to migrate, explore their surroundings, and locate favorable niches for colonization, as well as areas with higher resource availability. In contrast, non-motile or sessile strains generally show a reduced capacity to colonize and present attenuated virulence factors, which are often required for survival and pathogenicity in host-associated environments \cite{Tan2024StructuralSwitching, Palma2022MethodsInteractions}.

\begin{figure}[t]
    \centering
    \includegraphics[width=1\linewidth]{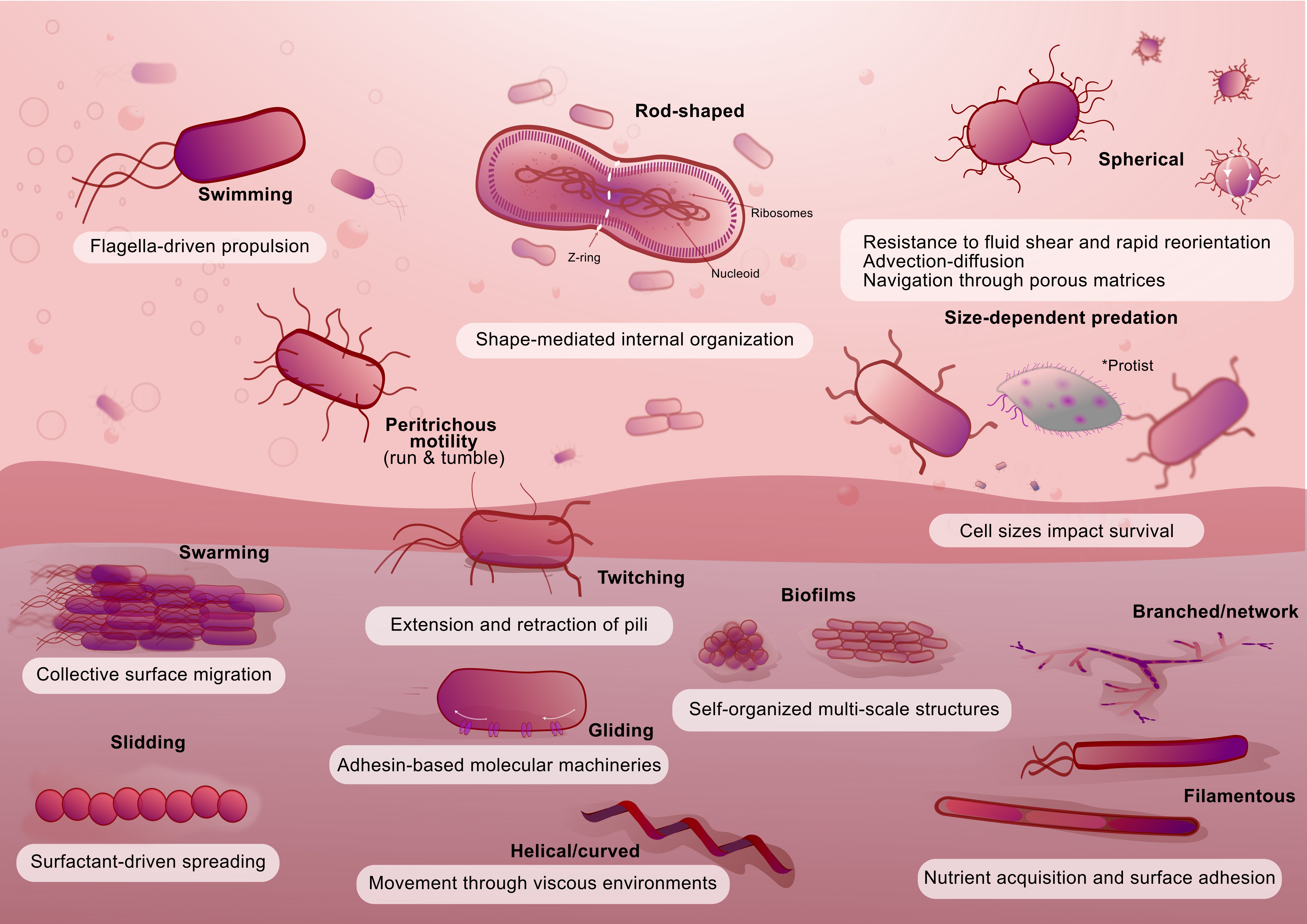}
    \caption{\textbf{Cancer-associated microbial phenotypes}. Specific bacterial traits enable adaptation to host tissues. Within the tumor microenvironment, microbes utilize various modes of motility to penetrate deeper into the tumor core or to co-migrate alongside metastatic cells. Furthermore, physical parameters such as bacterial shape and size directly influence their ability to navigate porous extracellular matrices and withstand fluid shear stress.}
    \label{fig:bacterial-motility}
\end{figure}

Bacterial motility can be classified into swimming, swarming, twitching, gliding, and sliding (Fig. \ref{fig:bacterial-motility}). Swimming occurs in liquid environments and is mainly mediated by flagella. Swarming is a coordinated movement over surfaces and often involves differentiation into a hyperflagellated phenotype. Active surface-associated motility includes twitching, driven by the extension and retraction of type IV pili, and gliding, which relies on different adhesin-based molecular machineries. In contrast, sliding is a passive spreading mechanism promoted by the secretion of surfactants or osmolytes that reduce surface tension and friction. Some bacteria present flagella distributed over the entire cell surface and are known as peritrichous bacteria, such as \textit{E. coli} and \textit{Salmonella}. In these species, flagellar-driven locomotion consists of directed “runs” interrupted by random “tumbles” that reorient the cell \cite{Chang2021StructuralSwitching}.

These motility behaviors are enabled by a highly efficient nanoscale machinery capable of rapidly driving the flagellar motor and switching its rotational direction to control cell migration. The flagellar motor is composed of a bidirectional rotating rotor surrounded by multiple torque-generating stator units. Each stator is formed by MotA/MotB protein complexes that function as ion-conducting channels. Ion flow through these channels converts electrochemical potential into mechanical torque, enabling flagellar rotation \cite{Tan2024StructuralSwitching, Chang2021StructuralSwitching} and defining motile bacteria as autonomous self-propelled units within active matter physics. This mechanism defines motile bacteria as autonomous self-propelled units in active matter physics \cite{Sengupta2020MicrobialFramework}. Directional switching is further regulated by the chemotaxis signaling network, which integrates environmental cues and guides motor rotation to guide bacterial movement \cite{Minamino2019DirectionalMotor}.

In terms of pathogenicity, motility plays an important role, as active movement within host tissues is often required to establish effective patterns of adhesion, invasion, and niche colonization. A representative example is \textit{Vibrio cholerae}, whose virulence is strongly reduced or abolished in non-motile strains. In contrast, strains with a polar flagellum show enhanced surface attachment and colonization capacity. Similar observations have been reported for the invasion and colonization of the gastric mucosa by \textit{Salmonella enterica} and \textit{H. pylori} \cite{Haiko2013TheVirulence}.

Despite its advantages, motility must be downregulated during the transition from motile planktonic cells to sessile biofilms, where bacteria lose mobility and assemble into multicellular aggregates. In the context of microbial biophysics, this shift marks a physical transition from dispersed self-propelled agents navigating interstitial spaces to locally immobile, force-generating aggregates that integrate directly into the biofilm's viscoelastic framework. Additionally, motility is beneficial during the initial approach to a surface, but its inhibition at later stages stabilizes cell aggregates and reduces the energetic cost associated with flagellar synthesis and operation. This transition is mainly driven by increased intracellular levels of cyclic-di-GMP (c-di-GMP), which simultaneously suppress motility and promote the production of extracellular polymeric substance (EPS) required for biofilm maturation. Motility repression occurs through a dual regulatory mechanism: a rapid, short-term inhibition of flagellar rotation (mot phenotype), followed by a slower, long-term transcriptional repression of flagellar genes. This ensures the progressive dilution and loss of flagella through cell division \cite{Guttenplan2013RegulationFormation}.

\subsubsection{Biofilm formation}

Many biofilms are structured, multicellular microbial communities that attach to biotic or abiotic surfaces. This growth mode is highly prevalent given the Earth hosts an estimated $1.2 \times 10^{30}$ bacterial and archaeal cells, 40–80$\%$ of which exist in biofilm form. Is not surprising that biofilms dominate many surface habitats, including soil and upper oceanic sediments, where they play major roles in biogeochemical cycles \cite{Jin2024MicrobesFeedback, Ramirez-Larrota2022AnEvasion}. Biofilms are embedded in a self-produced EPS matrix which provides mechanical stability and is composed of hydrated biopolymers, including polysaccharides, extracellular DNA (eDNA), and proteins. Its composition varies with species, strain, and environmental conditions \cite{Flemming2019BacteriaBiofilms}. Biofilms represent a collective microbial lifestyle, with emergent properties absent in single cells, such as enhanced communication, antimicrobial tolerance, and phenotypic differentiation \cite{Dumaru2019StudyDharan}. 

Mechanical cues and physical interactions, both externally applied as well as self-generated, strongly influence biofilm formation \cite{You2018GeometryColonies, Araujo2023SteeringConfinement, You2021Mono-to-multilayerColonies}. Initial attachment relies on non-specific forces, such as electrostatic interactions and hydrophobicity, as well as bacterial growth stage, specific adhesins and appendages like pili and flagella \cite{Riedel2025BacterialStage, El-Kirat-Chatel2020TheAdhesion,Ruhal2021BiofilmBacteria}. During maturation, growth within the confined EPS generates internal stress that shapes structural organization. Simulations show that hydrostatic stress accumulates in the interior, producing dense, disordered cell packing, while shear stress dominates at the periphery, promoting cell reorientation and a nematic surface layer reminiscent of liquid-crystalline ordering \cite{Sengupta2020MicrobialFramework, Li2024Agent-basedBiofilms}. These stress patterns provide a self-organizing framework for biofilm development. Drawing inspiration from the multi-field topological coupling demonstrated in nematic liquid-crystal systems \cite{Giomi2017Cross-talkMicrofluidics}, cancer–microbiome interactions can be conceptualized as a coupled active-matter system in which microbial organization, tumor architecture, and their associated biomechanical fields mutually influence one another. In this framework, topological defects and structural heterogeneities may act as mediators of bidirectional feedback, linking microbial activity to tumor development and, conversely, tumor progression to the reorganization of microbial communities and their surrounding biofilms. Such a perspective opens a new avenue for understanding cancer–microbiome interactions through the physics of coupled liquid-crystalline and active matter systems.

Biofilms are central to microbial pathogenicity, providing protection from immune defenses and antimicrobial agents. Approximately 80$\%$ of chronic and device-associated infections involve biofilms. The EPS matrix restricts drug penetration, making biofilm-embedded cells hundreds to thousands times less susceptible to antibiotics than planktonic bacteria. As a result, treatment frequently requires prolonged antimicrobial therapy and often the removal of contaminated implants \cite{Oliva2021Biofilm-RelatedDalbavancin}. Notably, exposure to several antibiotics,including $\beta$-lactams (ampicillin, ceftriaxone), gentamicin, and norfloxacin, can stimulate biofilm formation in major pathogens such as \textit{S. aureus}, complicating clinical management \cite{Azzam2023Sub-inhibitoryAureus}. Biofilm-associated microbes also secrete extracellular enzymes, such as proteases, which degrade the matrix or disrupt host defenses, enhancing virulence \cite{Ramirez-Larrota2022AnEvasion}.

 Biofilms can also be formed by two or more species, and their organization often depends on physical cell-to-cell recognition. When bacteria bind to each other while in suspension, the process is known as co-aggregation, and when this interaction occurs on a surface it is referred to as co-adhesion. In many communities there are primary colonizers that attach first to the substrate, such as those in the oral cavity, and create a favorable environment for secondary colonizers to join and expand the biofilm \cite{Soleimani2023NumericalCo-aggregation}. These cooperative interactions can strongly influence biofilm structure and behavior. For example, the co-culture of \textit{P. aeruginosa} and \textit{F. nucleatum} promotes bacterial proliferation and induces antibiotic tolerance through the formation of a dense biofilm enriched in polysaccharides, showing how interspecies interactions can modulate both biofilm architecture and resistance to treatment \cite{Li2020FusobacteriumA}.

\subsubsection{Oxygen tolerance}

Bacteria have persisted on Earth for millions of years, throughout constant fluctuations in environmental oxygen levels. In consequence, they have developed metabolic adaptations that allow them to take full advantage of oxygen when it is available, and even to live and thrive in its absence, as occurs with anaerobic bacteria \cite{Sperling2022BreathlessHistory}. However, in nature not everything is black or white. Besides strictly aerobic bacteria, which require oxygen to grow, and strictly anaerobic bacteria, which grow only in its absence, there are also aerotolerant bacteria, able to survive to oxygen exposure without using it in their metabolism. In addition, facultative anaerobic bacteria can be considered the most versatile, since they can switch their metabolism between aerobic and anaerobic modes depending on environmental conditions \cite{Andre2021TheInfection}.

This facultative ability contributes directly to the virulence of multiple pathogenic bacteria, some of which are listed by the WHO as high-risk pathogens \cite{WHO2024}, such as \textit{P. aeruginosa}, \textit{Salmonella typhi}, and \textit{S. aureus} (see Figure \ref{fig:bacteria-phenotypes}). Under hypoxic conditions, the antimicrobial activity of macrophages and neutrophils decreases because these cells depend on oxygen to produce ROS \cite{Hayek2020MechanismsConditions}. Even so, bacterial infections usually begin in normoxic environments, and the invading bacteria adapt their metabolism to the affected organ and its supply of oxygenated blood and nutrients (perfusion). Organs such as the liver and kidneys are highly perfused, whereas the colon is considered poorly perfused \cite{Andre2022TheInfections}.

Tissue invasion is an early and crucial step in the infectious process. Bacterial virulence depends on the ability to adhere to and persist on the host’s epithelial barrier. To do so, bacteria must resist the host immune response and subsequently colonize and spread through the tissues. Among the most common strategies is the formation of biofilms on organ surfaces or infection foci once the epithelial barrier has been crossed. At this stage, bacteria begin to reduce local levels of oxygen, pH, and glucose. One of the main causes of hypoxia in this context is bacterial aerobic respiration, which depletes oxygen locally. For instance, it has been shown that oxygen is practically absent in the deepest regions of biofilms, generating a gradient toward the periphery \cite{Ren2025TheEnvironments}. This explains why strictly aerobic bacteria rarely form biofilms, with the exception of those that generate infection foci surrounded by immune cells, which can lead to granuloma formation \cite{Qualls2015ImmunometabolismRespiration}. As mentioned previously, some biofilms are formed by mixed bacterial communities whose members can benefit one another by facilitating access to nutrients or providing more suitable surfaces for adhesion \cite{Soleimani2023NumericalCo-aggregation}. Commensal species such as \textit{Streptococcus gordonii} can also alter oxygen bioavailability in tissues, promoting colonization by pathogenic organisms. A clear example is \textit{Aggregatibacter actinomycetemcomitans}, which is able to shift from fermentative to respiratory metabolism, thus improving its growth and persistence \cite{Stacy2016ACross-respiration}.

Nevertheless, bacterial respiration is not the only cause of hypoxia. The immune system can induce this condition through the activation of hypoxia-driven signaling via hypoxia-inducible factors (HIF), in both infiltrating immunocytes and inflamed resident cells. In this sense, hypoxia can alter the efficiency of bacterial infection, reducing the disease progression. Nevertheless, it may impede the elimination of bacteria and result in enhanced replication of facultative anaerobic bacteria. Additionaly, could trigger bacterial dormanty, prolonging their lifes and protecting them against antibiotic treatment. That is why the net effect of immune cells caused hypoxia is debatably beneficial and depends on the nature of the pathogen and the characteristics of the infection \cite{Hayek2020MechanismsConditions, Schaffer2015TheInfection}. Finally, the facultative tolerance to oxygen observed in many bacteria has important implications in the TME. As previously noted, the TME is characterized by hypoxic conditions, which may facilitate bacterial invasion \cite{Anderson2023TheDifficile, Bian2022AntihypertensionCancer}.

\subsubsection{Cell wall composition (Gram stain)}

Among the multiple defense mechanisms, bacteria possess a sophisticated membrane that protects them against external agents and regulates certain metabolic functions. In general, bacterial membranes can be divided into two major groups depending on their molecular content and architecture (Fig. \ref{fig:bacteria-phenotypes}). These differences can be detected using the Gram staining assay. Gram-positive bacteria retain the violet dye due to their high peptidoglycan content, a polymer of glycans crosslinked by peptides that forms a thick outer membrane. Gram-negative bacteria, in contrast, remain colorless in this assay and possess a more complex, asymmetric outer membrane composed of a thin peptidoglycan layer surrounded by a selective lipid layer rich in lipopolysaccharides (LPS) \cite{Tavares2020ActivityBacteria}.

The thick peptidoglycan membrane of Gram-positive bacteria, only a few tens of nanometers thick, provides resistance to osmotic pressure and turgor and regulates cell shape \cite{Mueller2020BacterialStress}. Atomic force microscopy (AFM) studies in mature \textit{S. aureus} and \textit{Bacillus subtilis} cells revealed large and deep porosities of approximately 60 nm in diameter at the outermost part, which gradually close toward the inner membrane at a depth of about 23 nm, making the interior much denser \cite{Pasquina-Lemonche2020TheWall}. The peptidoglycan layer also allows adhesion of lipoteichoic acids, which give amphiphilicity and an anionic character to the membrane. Many antimicrobial agents are lipophilic and cationic, which facilitates their adhesion to and destabilization of the positively charged Gram-positive membrane. For this reason, antibiotics and other antimicrobial compounds target this membrane, leading to bacterial cell death \cite{Tavares2020ActivityBacteria}.

In Gram-negative bacteria, the outermost membrane is an asymmetric bilayer, with phospholipids on the inner side and LPS on the outer side. LPS molecules contain a polysaccharide core linked to a hydrophobic domain oriented inward and a chain of specific oligosaccharides facing outward. This hydrophilic domain is called the O-antigen. LPS provides structural integrity and rigidity to the membrane and plays a role in determining bacterial shape. For example, rod-shaped bacteria tend to become more spherical when LPS lipids are degraded \cite{Fivenson2023ADetermination}. The outer membrane also allows interaction with surfaces and protects the bacteria from toxic compounds. Its complexity and selectivity reduce or inhibit the effect of antibiotics \cite{Kaur2024Envelope, Farhana2023BiochemistryLipopolysaccharide}. The composition, size, and abundance of LPS vary between species and can serve as a biomarker for bacterial identification in host-pathogen interactions, including in cancer environments such as pancreatic ductal adenocarcinoma \cite{Guenther2020BacterialStudy}.

\clearpage
\section{Biophysics of the cancer microbiome: state of the art}
\raggedbottom

Cancer microbiome is one of the key regulators of oncogenesis. The molercular and pathogenic mechanisms across microbial nches have been described extensively \cite{Cullin2021MicrobiomeCancer}, but the biophysical description of this relationship remains underexplored. Matrix stiffness, viscoelasticity, and shear stress fundamentally alter tumor phenotypes and therapeutic efficacy \cite{Singh2018BeyondResistance, Cheng2025AdvancementsInteraction}, while simultaneously governing bacterial adhesion, invasion, and biofilm architecture \cite{Bastounis2022MechanicalPathogens}. The local microbiota can remodel the ECM through mechanisms like inflammation, and altering cell-matrix adhesion, promoting, in consequence, alterations in tumor mechanosignaling, circulation survival and distant colonization \cite{Herrera-Quintana2024BreastInteractions} (Fig. \ref{fig:mechanical-forces-bacteria}). While experimental insights into these biomechanical feedbacks are still emerging, integrating physical cues with microbial dynamics provides an essential framework to understand the mechanobiology of tumor progression \cite{Li2026TheTherapy, Zhou2025MechanicalProgression}. Key experimental advances bridging bacterial mechanics and host-tumor interactions are summarized in Fig. \ref{state-of-art}.

\subsubsection{Interactions and feedback between bacteria and TME}

The TME is a complex ecosystem with microbiomes that show tissue-specific profiles. The relationship between the TME and the bacteria that live in it is not one-way. There is evidence that both influence each other, creating bidirectional feedback loops. These interactions include changes in the physical structure of the TME and in its metabolic and immunological components. In turn, these factors affect the survival, distribution, and responses of the associated microbes, either inside cancer and immune cells \cite{Fu2022Tumor-residentCancer} or outside them \cite{GaleanoNino2022EffectCancer}. 

\begin{figure}[t]
    \centering    \includegraphics[width=1\linewidth]{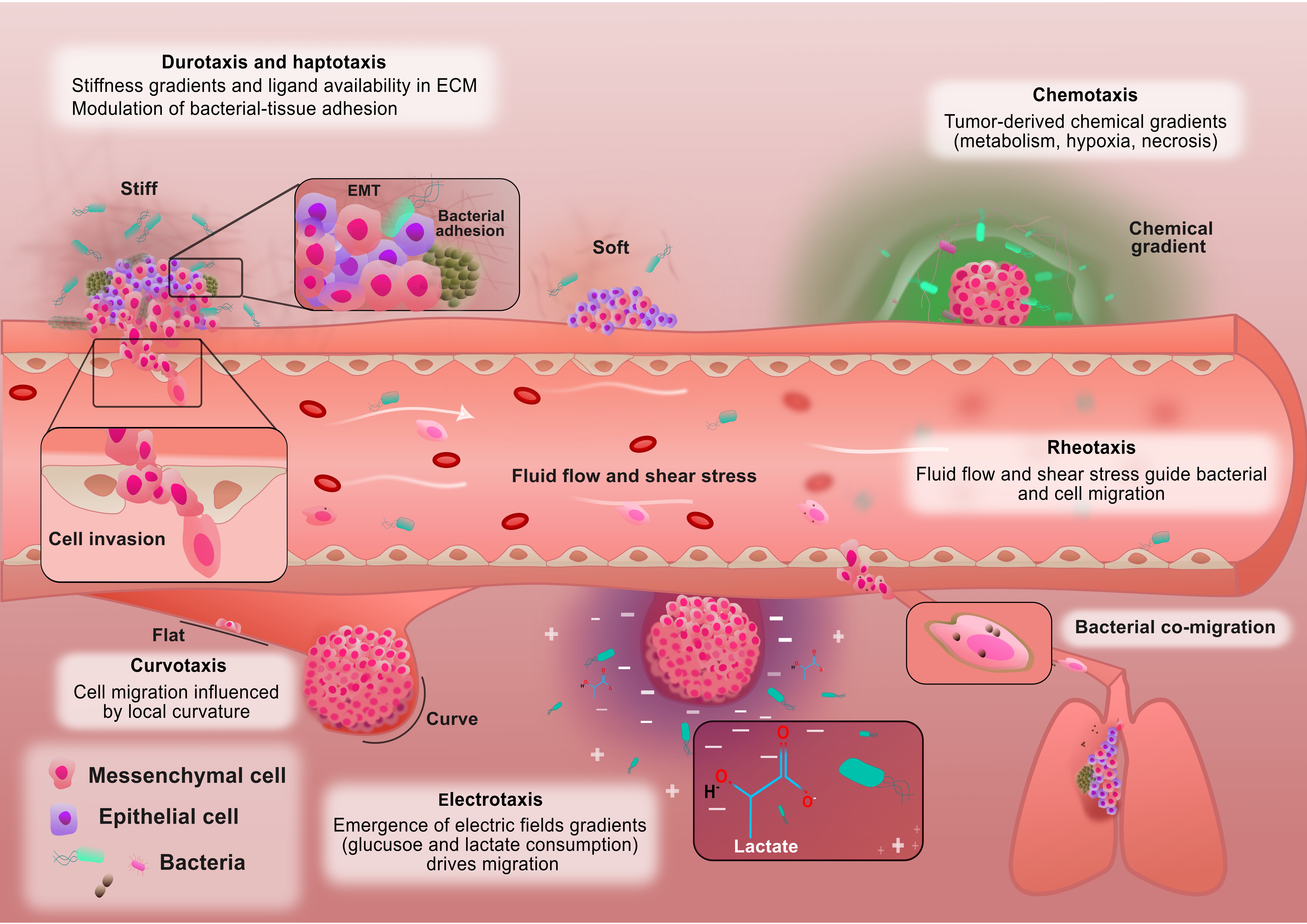}
    \caption{\textbf{Microbial taxis in cancer environments}. Microbial and cellular migration within the tumor microenvironment (TME) is directed by a combination of physical and chemical gradients. Bacteria and cancer cells navigate host tissues through physical cues such as matrix stiffness (durotaxis), ligand availability (haptotaxis), fluid shear stress (rheotaxis), and local tissue curvature (curvotaxis). Simultaneously, chemical gradients derived from tumor metabolism, hypoxia, and electric fields (chemotaxis and electrotaxis) guide bacterial invasion, proliferation, and co-migration with metastatic cells. These directional guidance mechanisms allow microbes to exploit the TME, driving local dysbiosis and favoring pathogenic colonization.}
    \label{fig:microbial-taxis}
\end{figure}

Importantly, taxis-driven behaviors such as chemotaxis, rheotaxis, and mechanically guided migration shape bacterial transport, localization, and competition within the TME. A visual representation of these directional cues is provided in Fig. \ref{fig:microbial-taxis}, while a more detailed description of these physical phenomena and their corresponding biophysical models can be found in Box 1. These physical processes contribute to interconnected physical, metabolic, and immunological feedback loops that continuously remodel both the TME and its associated microbiota. Many of the reciprocal bacteria-TME interactions described below have been captured by biophysical models that explicitly couple transport, tissue mechanics, and infection dynamics.

When dysbiosis happens, it promotes early and sustained inflammation, fibrosis, and collagen deposition in the TME \cite{Rosean2019PreexistingCancer}. The physical structure of the TME, especially the abundance and organization of collagen, may variate between cancer types and affects the growth of bacteria such as \textit{E. coli} and \textit{Salmonella typhimurium} (Fig. \ref{state-of-art}g) \cite{Takahashi2023ImpactTumours}. A clear example comes from pancreatic ductal adenocarcinoma, where cancer cells do not produce the normal heterotrimeric type I collagen ($\alpha1/\alpha2/\alpha1$) typical of fibroblasts. Instead, they produce a unique homotrimeric form of Col1 ($\alpha1/\alpha1/\alpha1$). This homotrimer supports oncogenic signaling, cancer cell proliferation, and tumor growth. It also correlates with a tumor microbiome enriched in anaerobic \textit{Bacteroidales} under hypoxic and immunosuppressive conditions. When Col1 homotrimers are deleted in PDAC models, hypoxia is reduced, T cell infiltration increases, and the tumor microbiome shifts toward microaerophilic \textit{Campylobacterales}. This shift is associated with improved survival and restored gut microbiome homeostasis, and it increases the response to immunotherapy \cite{Chen2022OncogenicCancer}.

The immune response in the TME is also strongly influenced by microorganisms. A clear example is the production of cytokines, which are small immune proteins that act as messengers and regulate inflammation and growth processes. The control and balance of cytokine production is very important, because it determines if the immune system will act in a pro-tumorigenic or anti-tumorigenic way. When there is dysbiosis, immune cell functions are affected, and this changes the response to therapy and the production of cytokines \cite{Iida2013CommensalMicroenvironment}. For instance, \textit{Fusobacterium periodonticum} can increase cytokine production, while \textit{B. fragilis} and \textit{P. asaccharolytica} can have inhibitory effects \cite{Sulit2023BacterialMicroenvironment}.

More than the effect of a single bacterial species, studies in CRC have shown that a high collective activity of the bacterial biomass in the TME has a stronger immunosuppressive effect than individual pathogens \cite{Kvich2024BiofilmsCancer}. Some studies suggest that bacterial communities are located in specific microniches of the TME, where they can attract immune cells and increase immunosuppressive activity, which promotes tumor progression \cite{GaleanoNino2022EffectCancer}. On the other hand, this same ability to attract immune cells has been used to design attenuated bacteria with immune-cell receptors that reduce tumor progression instead of supporting it \cite{Lin2024InImmunity}.

\subsubsection{Mechanical interactions}

The mechanical properties of the substrate, such as the ECM, influence both the mode of adhesion and the dynamics of growth and invasion of bacteria and their biofilms. During the invasion process, bacteria use force-sensitive mechanisms to adjust their adhesion in response to the mechanical stress around them. Some pathogens such as \textit{S. aureus} use surface proteins to attach to host cells. The strength of this adhesion is influenced by the tension generated by physical stress, for example fluid flow. The binding force between this pathogen and the host can increase from about 0.1 nN to 1.5 nN when exposed to mechanical tension, such as an increase in flow. This ability to modulate adhesion allows the bacterium to remain firmly attached under high shear stress and to detach more easily when shear stress decreases. Techniques such as single-cell force spectroscopy (SCFS) and single-molecule force spectroscopy (SMFS) have been widely used to characterize these forces (Fig. \ref{state-of-art}h) \cite{Vitry2017Force-inducedLoricrin, Herman-Bausier2018StaphylococcusAdhesion, daCosta2022FibronectinAureus}.

An increase in ECM stiffness correlates positively with faster bacterial expansion and movement once bacteria are attached to a surface. Asp \textit{et al}. reported that biofilms of \textit{Serratia marcescens} develop more rapidly on harder and more porous substrates. Under these conditions, the poroelastic properties of the network enhance fluid flow, which facilitates nutrient distribution throughout the biofilm and supports faster spreading. At the same time, higher substrate stiffness increases the internal mechanical stress within the biofilm, further contributing to its growth dynamics. Interestingly, Asp \textit{et al}. also identified a coupling between stiffness and porosity, as materials such as agar lose porosity when stiffened, thereby limiting nutrient availability and slowing bacterial growth \cite{Asp2022SpreadingPermeability}.  

Beyond bacterial collectives, ECM stiffness also modulates host cell mechanics and cell–cell competition in infected tissues. Increasing matrix stiffness promotes collective squeezing and extrusion of infected cells, driven by differential cell displacements and force generation between infected and uninfected populations. This effect is further enhanced when the stiffness contrast between uninfected and infected cells increases, or when the infection focus is small, and it is consistently more pronounced on stiff compared to soft matrices \cite{Aparicio-Yuste2022ACells}. ECM stiffness is a key regulator of both bacterial expansion and host cell mechanical responses, shaping the balance between microbial persistence and tissue-level defense mechanisms. Consistently, bacterial surface motility is also sensitive to substrate mechanics. For instance, the twitching velocity of \textit{P. aeruginosa} increases markedly on stiffer surfaces, rising from 0.02 to 0.4~$\mu$m/min when substrate stiffness increases from 3 to 80~kPa \cite{Gomez2023SubstrateMotility}.

However, the physical and mechanical environment of the ECM influences not only the growth rate but also the colony shape, spatial distribution, and invasion patterns of bacteria, as well as their sensitivity to antibiotics. Liu, \textit{et al}, showed that bacteria tend to adhere and invade host cells in areas with relatively high traction forces. They also found a lower percentage of infected cells when cultured on soft substrates, although these cells contained a larger number of internalized bacteria. In other words, as ECM stiffness increases, there is a shift in the invasion mode, moving from a more clustered pattern on soft surfaces to a more dispersed pattern on stiff surfaces. This change in invasion behavior has a direct effect on therapeutic efficacy, which is considerably higher on soft substrates \cite{Liu2021ExtracellularInternalization}.

Importantly, the TME shares these same mechanical characteristics. Metastatic cell lines from breast, lung, and prostate cancers show higher traction forces compared with their non metastatic counterparts and healthy cells. This correlates with the characteristic increase in ECM stiffness and collagen density in the TME, where cells tend to generate stronger traction forces \cite{Kraning-Rush2012CellularPotential}. This suggests that bacteria may invade the TME more easily, because the mechanical conditions of the tumor, with higher stiffness and higher traction forces, are similar to those that favor bacterial invasion.

\subsubsection{Biophysical consequences of bacterial phenotypes}

\begin{figure}[t!]
    \centering    \includegraphics[width=1\linewidth]{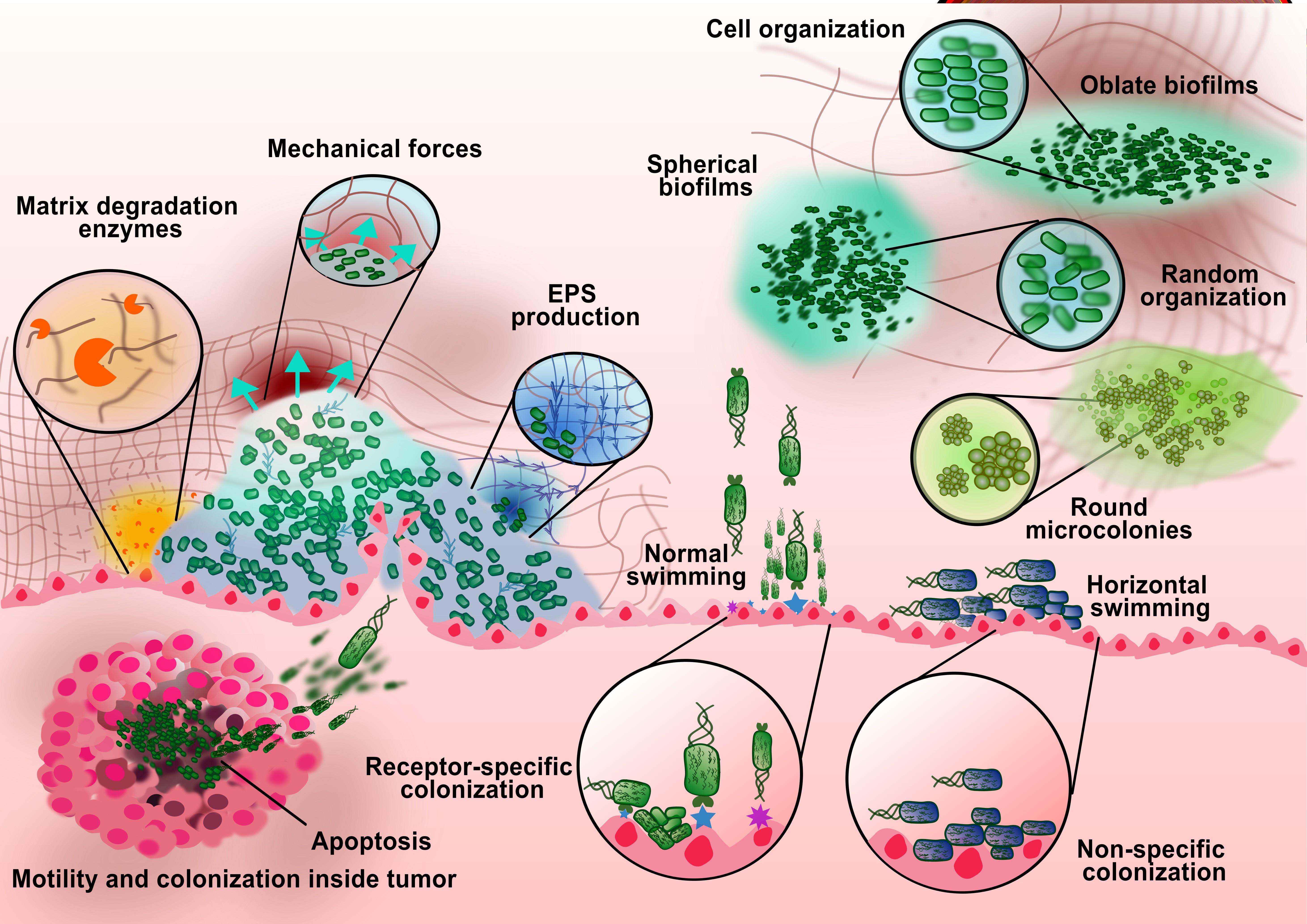}
    \caption{\textbf{Biophysical consequences of bacterial phenotypes in cancer environments}. Bacteria utilize distinct physical strategies to colonize and invade the TME. Biofilm formation allows bacteria to exert mechanical forces that disrupt tissue architecture and secrete matrix degradation enzymes for deeper invasion. Interestingly, macroscopic and microscopic biofilm morphology is responsive to both mechanical cues and individual cell shape. Stiffer matrices induce oblate biofilms characterized by organized bacterial growth, while softer tissues promote spherical biofilms with random cell organization and localized EPS secretion. Furthermore, individual bacterial geometry influences internal structure; cocci typically form local spherical microcolonies, although the overall bulk biofilm may remain oblate or amorphous. Beyond biofilms, bacterial motility drives deep tumor penetration, establishing cytotoxic niches that trigger apoptosis in cancer cells. Finally, during the landing stage, swimming kinematics determine adherence: motility directed perpendicular to the local cell membrane (normal swimming) enables receptor-specific colonization, whereas motion parallel to the tissue (horizontal swimming) results in faster, non-specific colonization.}
    \label{fig:mechanical-forces-bacteria}
\end{figure}

\paragraph{Mechanical deformation of the ECM by biofilms.} Biofilms exert strong mechanical effects on their surroundings, reshaping soft tissues and modifying the properties of the ECM. These interactions become particularly relevant when considering how bacterial communities may behave inside the TME, which is also a mechanically abnormal and often stiff environment. Biofilms can generate large internal mechanical stresses as cells proliferate within an elastic matrix, and these forces are transmitted to the substrate through adhesion (Fig. \ref{fig:mechanical-forces-bacteria}). Studies have shown that biofilms from pathogens such as \textit{Vibrio cholerae} and \textit{P. aeruginosa} can build up surface stresses close to 100 kPa, enough to deform soft hydrogels and induce buckling of the colony. These stresses can disrupt epithelial integrity by causing epithelial monolayers to delaminate from the ECM, bend, or lose cell–cell contacts, ultimately leading to cell death (Fig. \ref{state-of-art}e) \cite{Cont2020BiofilmsEpithelia}. Together, these observations indicate that biofilms may contribute to tissue damage through a mechanical mode of infection, not only through biochemical factors such as toxins.

Biofilms also display strong genotypic and phenotypic heterogeneity, and the spatial arrangement of cells plays a key role in nutrient access, antimicrobial exposure, and interspecies interactions. Mechanical forces such as flow or direct contact can deform the EPS matrix and redistribute cells within the biofilm. The EPS primarily functions as a mechanical scaffold that maintains cohesion between bacterial cells. By analogy to mammalian mechanosensing, it has been proposed that bacteria may regulate EPS production in response to environmental stiffness: cells sensing low elasticity may reinforce the matrix by producing more components, whereas those experiencing high stiffness might secrete matrix-degrading enzymes to permit further expansion (Fig. \ref{fig:mechanical-forces-bacteria}) \cite{Dufrene2020Mechanomicrobiology:Forces}.

The mechanical properties of the environment strongly influence biofilm morphology. In stiff matrices, embedded biofilms adopt oblate shapes and show bipolar cellular ordering, while in soft matrices they remain spherical with random organization \cite{Zhang2021MorphogenesisBiofilms} (Fig. \ref{fig:mechanical-forces-bacteria}). Importantly, this anisotropic shaping arises from biofilm–gel interactions rather than the intrinsic shape of individual bacterial cells. Although these effects have been mainly studied in synthetic hydrogels and epithelial monolayers, the same principles may apply inside tumors, where increased stiffness, high collagen density, and altered mechanical stress fields are common. Intratumoral bacteria can form localized microniches, and dysbiosis contributes to fibrosis and ECM remodeling, leading to metastasis and epithelial-mesenchymal transition (Fig. \ref{state-of-art}b) \cite{GaleanoNino2022EffectCancer}.By locally degrading structural fibers or depositing dense biopolymers, bacterial aggregates directly alter the elastic network of the ECM. This local remodeling feeds back into the macroscopic properties of the tumor as an active solid, shifting its effective stiffness, internal stress distribution, and resistance to mechanical failure. Even if direct evidence linking biofilm-induced buckling or delamination to tumor progression is not yet available, these mechanical similarities suggest that biofilms could interact with the TME, amplifying local matrix disruption or influencing therapeutic response.

\paragraph{Influence of bacterial morphology.}
Bacterial cell shape also affects colony expansion under confinement. Low aspect ratio (spherical) bacteria tend to form compact and rounded colonies, whereas high aspect ratio (rod-shaped) bacteria produce elongated colonies with larger surface area, improving nutrient access (\cite{Sreepadmanabh2024CellConfinement}. Rod-shaped bacteria also show more robust population growth when exposed to mechanical confinement. However, while morphology influences how cells expand locally, the global anisotropic shape of biofilms, oblate or spherical, is governed by the mechanical properties of the surrounding gel rather than by the anisotropy of the bacterial cells themselves \cite{Zhang2021MorphogenesisBiofilms}.

\paragraph{Bacterial motility and diffusion through the TME.} How microorganisms interact with their physical environment, explore surfaces, and eventually colonize the TME, depends on bacterial motility and transport through the TME. Viewed through the lens of an active solid, motile bacteria function as self-propelled units that actively navigate and exert forces against the elastic matrix of the tumor, much like bacterial biofilms have previously been described as active solids \cite{Xu2022AutonomousSolids}. Motility allows bacteria to navigate complex landscapes, sense mechanical and chemical cues, and initiate the first stages of attachment to surfaces resembling the extracellular matrix. During this early interaction, bacteria employ distinct kinematic strategies to approach and explore the tissue interface. Some species utilize a surface-normal (perpendicular) swimming strategy, characterized by rapid transitions from the bulk fluid to the near-surface region. This orthogonal movement allows them to quickly probe the physical properties of the surface and increases the probability of engaging with specific receptors. While previous literature frequently describes this behavior as "vertical" swimming, we use the term surface-normal to explicitly define the motion relative to the local tissue boundary and to clarify that this kinematic pattern is independent of gravitaxis. Conversely, other species rely on a surface-parallel (horizontal) strategy, executing coordinated movements along the tissue boundary to promote efficient but non-specific colonization. These distinct motility patterns act as critical biophysical virulence factors, facilitating the initial mechanical steps required for bacteria to adhere to, invade, and penetrate host tissues \cite{Abdulkadieva2022StrainColi}.

Inside tumors, motility becomes a key determinant of how efficiently bacteria colonize and distribute within the TME (see Fig. \ref{fig:mechanical-forces-bacteria}) . More motile strains show higher colonization capacity, even when they are introduced at very low densities. This behavior has been described for species like \textit{Salmonella}, which display high intrinsic tumor tropism and can reach tumor tissues efficiently by actively swimming through the microenvironment \cite{Silva-Valenzuela2016SolidSalmonella}. Motility also shapes the spatial distribution of bacteria inside tumors. Highly motile cells can move away from entry points, penetrate deeper layers, and disperse across regions that would otherwise remain inaccessible to non motile variants (Fig. \ref{fig:mechanical-forces-bacteria}). Mathematical models support this idea by showing that dispersion, which depends directly on motility, governs how deeply bacteria can infiltrate dense tissues, while growth only amplifies the local density after motile cells reach specific regions. This deeper penetration has important consequences for therapeutic applications. A more uniform distribution of bacteria throughout the tumor correlates with stronger tumor regression, since the accumulation of bacteria inside the tissue promotes apoptosis of cancer cells (Fig. \ref{state-of-art}b) \cite{Toley2012MotilityTissue}.
In a recent study, the antitumor efficacy of \textit{Ewingella americana}, a bacterium isolated from the gut of frogs, was evaluated in colorectal cancer (CRC) models, including organoids and mice. The bacterium eliminated tumors through a multifaceted mechanism, combining direct cytotoxicity driven by intratumoral proliferation with a robust activation of the host immune response. Compared to other bacterial candidates and conventional treatments such as doxorubicin, \textit{E. americana} demonstrated greater tumor clearance and higher selectivity, highlighting its potential as a promising therapeutic strategy (Fig. \ref{state-of-art}f) \cite{Iwata2025DiscoveryProperties}.
Motile bacteria also display specific targeting behavior. They often migrate toward spatially heterogeneous regions, especially necrotic zones, guided by chemotaxis toward chemical gradients produced by dying or stressed cells (Fig. \ref{fig:microbial-taxis}). At the same time, they tend to move away from areas with functional vasculature, where conditions are less favorable for stable colonization \cite{Toley2012MotilityTissue}.

Bacteria use a variety of strategies to infiltrate host cells, combining motility and surface interactions to overcome both physical and biological barriers. Motile species, such as \textit{V. cholerae}, \textit{Campylobacter jejuni}, and \textit{Salmonella enterica}, exploit active swimming, chemotaxis, or enzymatic degradation of barriers to reach and penetrate host tissues. At the same time, many bacteria rely on adhesion to host receptors or the cell surface to trigger internalization. For example, \textit{Listeria monocytogenes} binds specific epithelial receptors to enter cells, while \textit{Shigella} and pathogenic \textit{E. coli} use contact-based systems to manipulate the host membrane and promote uptake. Other pathogens, like \textit{Clostridium difficile}, act indirectly by disrupting cell–cell contacts or the cytoskeleton to facilitate entry \cite{McCoy2024InInteractions}. Motility further work togehter with intracellular retention. By increasing the expression of master regulators of motility not only enhances bacterial movement but also promotes accumulation inside cancer cells, helping them remain within the tumor and reducing the likelihood of escaping back into circulation. Consequently, motile strains achieve higher retention and more efficient colonization of the TME \cite{Raman2019TheSalmonella}.

\subsubsection{Role in metastasis}

Bacteria are now recognized as active participants in metastatic progression, influencing how tumor cells spread to distant organs and how they adapt to new environments. Several works show that specific microbial communities inside tumors can modify signaling, immunity, and even the physical properties of cancer cells, creating conditions that favor dissemination \cite{Meyers2025MicrobiomeMethods}. A key process enhancing metastasis is the recruitment of M2 macrophages, which is heavily driven by bacterial metabolic byproducts. For example, butyrate produced by bacteria like \textit{Roseburia} in lung cancer, or lactate produced by \textit{E. coli} in CRC, promotes an immunosuppressive environment by stimulating M2 macrophages, weakening regulatory T cell activity, and reducing the antitumor response of CD8 T cells \cite{Ma2024IntratumorMetastasis, Gu2024Tumor-residentModulation}. Similarly, lactate production by tumor-resident \textit{Staphylococcus} directly promotes metastatic colonization in lung adenocarcinoma \cite{Yu2025LactateAdenocarcinoma}. Other studies show that bacteria can directly promote the early steps of dissemination. In cervical cancer, metastatic lymph nodes contain a higher abundance of Gram negative species such as \textit{E. coli}, \textit{Prevotella bivia}, and \textit{Fusobacterium nucleatum}. These bacteria enhance the formation of capillary like structures in lymphatic endothelial cells and facilitate the migration of uninfected tumor cells through paracrine signaling \cite{Liu2025Gram-negativeCancer}. In the same cancer type, \textit{Pseudomonas} has been found enriched in tumor tissue. \textit{In vitro}, \textit{Pseudomonas} increases the proliferation and migration of cervical cancer cells, suggesting that microbial composition in these tumors influences how aggressive they become \cite{Guo2025TheCancer}.

A growing number of studies point to a more direct role of bacteria in the physical process of metastasis. Some species of the genus \textit{Bacillus} are found specifically in metastatic breast tumors. When breast cancer cells are cultured with \textit{Bacillus thermoamylovorans} isolated from metastatic tissue, they produce almost three times more metastatic burden than controls. This observation indicates a direct interaction between bacteria and cancer cells that enhances metastatic potential \cite{Gerbec2025IdentificationMetastasis}. 

\begin{figure} [H]
  \centering 
\includegraphics[width=0.75\linewidth]{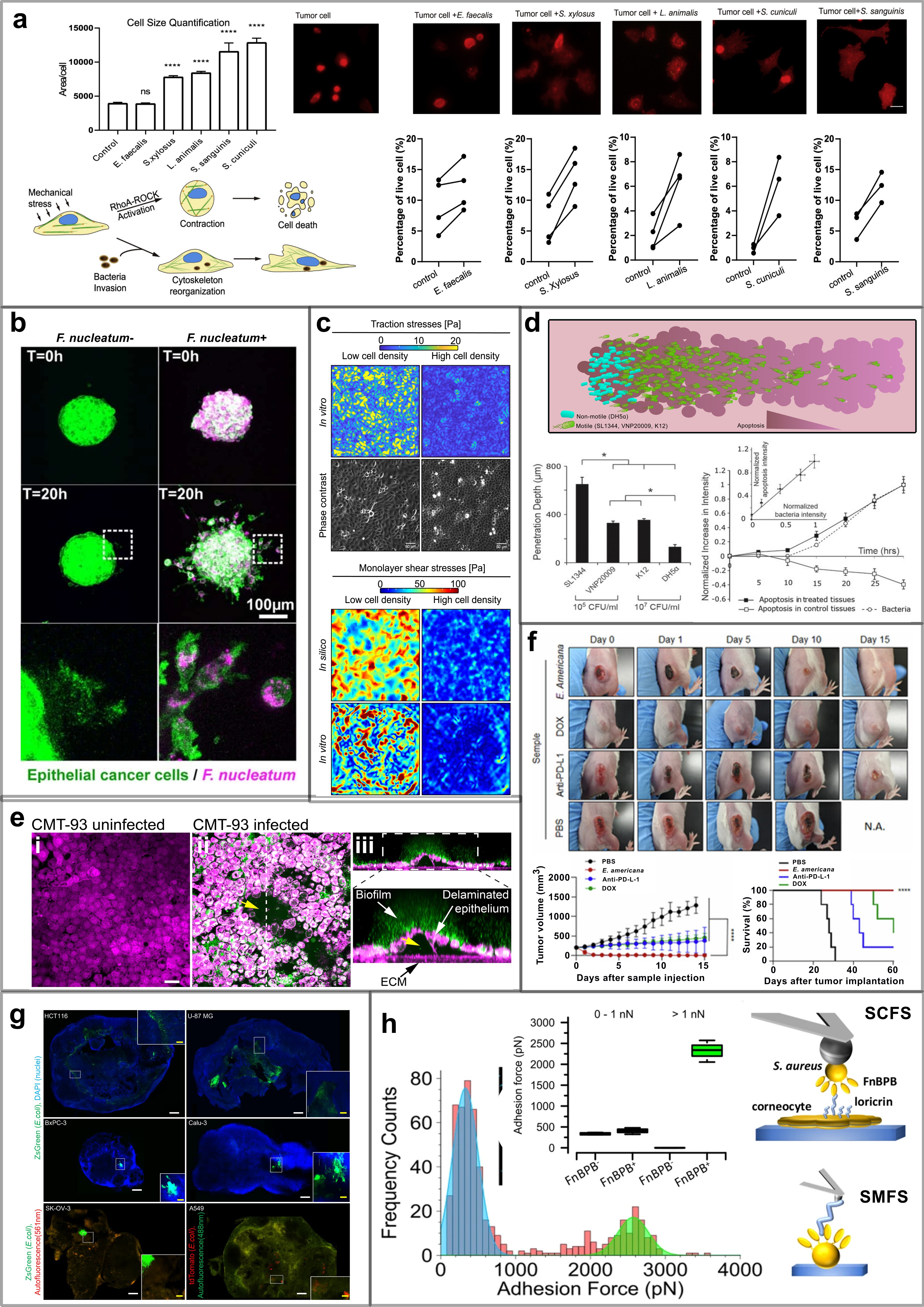}
  \caption{(Caption next page.)}
  \label{state-of-art}
\end{figure}

A clear biophysical mechanism has also been described in breast cancer. Tumor resident bacteria are carried by circulating tumor cells during metastasis. These intracellular bacteria help cancer cells resist fluid shear stress in the bloodstream by reorganizing the actin cytoskeleton, which increases cell survival during circulation (Fig. \ref{fig:microbial-taxis} and \ref{state-of-art}a). Depleting intratumor bacteria reduces lung metastasis, even if the primary tumor remains unchanged. 

Moreover, intratumor administration of strains such as \textit{Staphylococcus}, \textit{Lactobacillus}, \textit{Enterococcus}, and \textit{Streptococcus} promotes metastasis in animal models, even though the overall bacterial biomass is low \cite{Fu2022Tumor-residentCancer}, and extracellular bacteria like \textit{F. nucleatum} induce a quiescent state that protects the cells from chemotherapeutic agents \cite{Luis2025Tumor-infiltratingArrest}.

Bacteria can also influence metastatic progression through their toxins. Exposure to \textit{Bacteroides fragilis} toxin (BFT), induces epithelial hyperplasia in the mammary gland and increases metastatic behavior in cancer cells. Tumors implanted in mice colonized with entorotoxigenic \textit{B. fragilis} show faster growth and more metastasis compared with tumors exposed to non toxigenic strains \cite{Parida2021AAxes}. Finally, \textit{F. nucleatum}, which binds to tumors through the Fap2 to Gal GalNAc interaction, is found at higher levels in breast cancer. Its presence correlates with metastatic progression, and metronidazole treatment that targets this bacterium reduces metastasis, confirming a functional role \cite{Parhi2020BreastProgression}. Together, these works show that bacteria can promote metastasis in multiple ways, from epigenetic reprogramming and immune modulation to direct biophysical support during circulation and adaptation to distant organs like the liver or lungs. The metastatic process is therefore not only a property of cancer cells but also a consequence of their interaction with specific members of the tumor microbiome.

\addtocounter{figure}{-1}
\begin{figure} [t!]
\caption{\textbf{Experimental biophysics of the cancer microbiome interface.} Representative landmark findings elucidating mechanical, hydrodynamic, and structural interactions between intratumoral bacteria and cancer cells.
\textbf{a.} Bacterial internalization promotes host cell spreading (top and bottom left) and enhances resistance to fluid shear stress during circulation in the bloodstream (bottom right) \cite{Fu2022Tumor-residentCancer}.
\textbf{b.} Live-cell confocal imaging of a 3D collagen-embedded colorectal cancer (CRC) spheroid model reveals that \textit{Fusobacterium nucleatum} remodels the extracellular matrix (ECM), accelerating epithelial–mesenchymal transition (EMT) and metastasis \cite{GaleanoNino2022EffectCancer}.
\textbf{c.} Elevated epithelial cell densities suppress cellular traction forces \textit{in vitro} and decrease monolayer tension \textit{in vitro} and \textit{in silico}, demonstrating the density-dependent mechanics of confined tumor tissue \cite{Aparicio-Yuste2025HybridMonolayers}.
\textbf{d.} In microfluidic CRC spheroids, motile strains of \textit{Pseudomonas aeruginosa} and \textit{Escherichia coli} penetrate deep into hypoxic tissue to trigger localized cancer cell apoptosis (top and bottom left), with cytotoxicity scaling directly with bacterial infiltration depth (bottom right) \cite{Toley2012MotilityTissue}.
\textbf{e.} Confocal reconstruction of \textit{Vibrio cholerae} biofilms colonizing an epithelial monolayer demonstrates mechanical barrier disruption and local ECM deformation \cite{Cont2020BiofilmsEpithelia}.
\textbf{f.} Antitumor efficacy of gut-derived \textit{Ewingella americana} in a murine CRC model (top); bacterial colonization drives complete tumor regression within days (bottom left) and substantially prolongs survival relative to conventional chemotherapies (bottom right) \cite{Iwata2025DiscoveryProperties}.
\textbf{g.} Fluorescence microscopy demonstrating the cancer-type selectivity of \textit{E. coli}, illustrated by differential tissue accumulation across distinct tumor models \cite{Takahashi2023ImpactTumours}.
\textbf{h.} Nanoscale adhesion mechanics of \textit{Staphylococcus aureus} probed via atomic force microscopy (AFM). Single-cell force spectroscopy (SCFS) and single-molecule force spectroscopy (SMFS; top and bottom right, respectively) were used to characterize the binding forces between bacterial fibronectin-binding protein B (FnBPB) and the skin corneocyte protein loricrin, a key interaction in bacterial skin infections. The graphs display the adhesion force distribution (histogram) alongside a box plot comparing $\text{FnBPB}^+$ and $\text{FnBPB}^-$ strains, demonstrating significantly higher adhesion forces when FnBPB is expressed \cite{daCosta2022FibronectinAureus}.
Panels \textbf{a}, \textbf{b}, \textbf{e}, and \textbf{h} adapted from refs.~\cite{Fu2022Tumor-residentCancer,GaleanoNino2022EffectCancer,Cont2020BiofilmsEpithelia,daCosta2022FibronectinAureus} under a CC BY license; \textbf{f} adapted from ref.~\cite{Iwata2025DiscoveryProperties} under a CC BY-NC license; \textbf{g} reproduced without modification from ref.~\cite{Takahashi2023ImpactTumours} under a CC BY-NC-ND 4.0 license; \textbf{c} and \textbf{d} adapted with permission from ref.~\cite{Aparicio-Yuste2025HybridMonolayers} (CC BY-NC-ND 4.0) and ref.~\cite{Toley2012MotilityTissue} (Oxford University Press), respectively.}
\end{figure}

\section{Conclusions and future perspectives}

In this review, we establish cancer as an active solid system, composed of mechanical, metabolic, and microbial constituents, wherein reciprocal biomechanical and topological interactions: microbes shape tumor development, while tumor growth also reshapes microbial biofilm organization, remains the central theme. The emergence of cross-disciplinary approaches has opened new avenues for the simultaneous characterization of microbial and cancer landscapes, which, until now, have been investigated predominantly through biomolecular studies. Future research using multi-scale models is needed to understand the physical effects produced by these microbial components, effectively combining microbial features with tumor mechanics. To achieve this, quantitative methods exist to measure the bacterial mechanical impact, such as atomic force microscopy (AFM), bulk and microrheology, and traction force microscopy (TFM). Once standardized physical measurements are obtained, it will be possible to map the biophysical signatures of specific tumor-bacteria associations. State-of-the-art imaging techniques, including light-sheet microscopy for multiscale tracking, offer unprecedented opportunities to elucidate the behavior, physiology, and dynamics of cancer–microbe ensembles across multiple length and timescales. These approaches enable us to investigate living materials undergoing profound structural and functional transformations, drawing on the physics of liquid crystals to advance the growing exploration of topology-mediated phenomena in both fundamental research and applications aimed at tailoring material properties down to the molecular scale.

Our ability to resolve microscale dynamics will help uncover how microbes, through their structural and topological organization, shape the nonequilibrium developmental dynamics of tumors and their metastatic transitions over extended timescales. This investigation can span individual microbial species as well as oncologically relevant microbial consortia. Conversely, we are poised to determine whether and how tumor growth influences the microscale structural, transport, and topological properties of associated biofilms, and how these reciprocal interactions depend on microbial phenotypes, physiological states, and population-level fitness. By examining microbial communities together with their surrounding matrices, we can uncover active topological feedback mechanisms with potentially far-reaching implications for cancer diagnostics, therapeutics, and disease management. As this interdisciplinary frontier expands, several open questions emerge. First, from a predictive point of view, would it be possible to predict bacterial phenotypes and colonization patterns just by knowing the mechanical properties of a tumor? Second, focusing on tumor remodeling, are biomechanical alterations, like matrix stiffening and mechanotransduction, reversible when the bacterial influence is removed? Finally, analyzing these physical interactions opens promising therapeutic avenues: could engineered bacteria be used to provoke therapeutic mechanical disruptions in tumors, based on their phenotypes and their influence on tumor progression? Ultimately, by elucidating how biomechanical coupling governs the emergence of structure and function in cancer–microbiome systems, upcoming studies will establish a new, data-rich mechanistic paradigm in biophysics, complementing conventional molecular approaches. Such an interdisciplinary framework would not only advance fundamental mechanistic models of adaptive active matter but also create strategic translational opportunities for understanding and managing bacteria-associated cancers with cancer-microbiome as a core pillar of next-generation oncology. 

\clearpage
\section*{Acknowledgments}

This work was carried out with the support of the FNR-Luxembourg’s PRIDE Doctoral Training Unit CANBIO2 (CANBIO2, https://www.lih.lu/en/research-support/doctoral-training/canbio2/), with additional funding from the University of Luxembourg and the Luxembourg National Research Fund’s ATTRACT Investigator Grant (Grant no. A17/MS/ 11572821/MBRACE to AS) as well as the CORE Grant (Grant no. C19/MS/13719464/ TOPOFLUME/Sengupta to AS). The authors thank the members of the CANBIO2 consortium for valuable discussions during the course of this work. AS gratefully acknowledges the AUDACITY Grant (AUDACITY Grant no.: IAS-20/CAMEOS) from the Institute for Advanced Studies, University of Luxembourg for supporting this work.

\paragraph{AI Usage Disclosure:} Google Gemini was used solely for language editing, phrasing, and improving the clarity of the text. All scientific concepts, analyses, and conclusions were developed, verified, and approved by the authors.

\paragraph{Figure Permissions:} All third-party images included in this manuscript were either obtained from Open Access articles published under Creative Commons licenses (CC BY, CC BY-NC 4.0, and CC BY-NC-ND 4.0) or adapted with formal permission from the respective publishers (including Oxford University Press and RightsLink-participating journals).

\bibliography{references}

@article{Kaur2024Envelope,
    title = {{ Insight into the outer membrane asymmetry of P. aeruginosa and the role of MlaA in modulating the lipidic composition, mechanical, biophysical, and functional membrane properties of the cell envelope }},
    year = {2024},
    journal = {Microbiology Spectrum},
    author = {Kaur, M. and Mozaheb, N. and Paiva, T. O. and Herent, M.-F. and Goormaghtigh, F. and Paquot, A. and Terrasi, R. and Mignolet, E. and D{\'{e}}cout, J.-L. and Lorent, J. H. and Larondelle, Y. and Muccioli, G. G. and Quetin-Leclercq, J. and Dufr{\^{e}}ne, Y. F. and Mingeot-Leclercq, M.-P.},
    number = {11},
    month = {11},
    pages = {11},
    volume = {12},
    publisher = {American Society for Microbiology},
    url = {/doi/pdf/10.1128/spectrum.01484-24?download=true},
    doi = {10.1128/SPECTRUM.01484-24;JOURNAL:JOURNAL:SPECTRUM;WGROUP:STRING:PUBLICATION},
    issn = {2165-0497},
    pmid = {39373473}
}

@article{Mauri202582PCancer,
    title = {{82P Intratumoral presence of the genotoxic gut bacteria pks+ E. coli in very early onset colorectal cancer}},
    year = {2025},
    journal = {Annals of Oncology},
    author = {Mauri, G. and Sogari, A. and Crisafulli, G. and Monti, L. and Cavestro, G.M. and Puzzono, M. and Mariano, S. and Tosi, F.M. and Bonazzina, E.F. and Bencardino, K. and Amatu, A. and Ghezzi, S. and Aquilano, M.C. and Bonoldi, E. and Arena, S. and Siena, S. and Bardelli, A. and Bianchi, A. Sartore},
    month = {7},
    pages = {S42},
    volume = {36},
    publisher = {Elsevier BV},
    url = {https://www.annalsofoncology.org/action/showFullText?pii=S0923753425002972 https://www.annalsofoncology.org/action/showAbstract?pii=S0923753425002972 https://www.annalsofoncology.org/article/S0923-7534(25)00297-2/abstract},
    doi = {10.1016/j.annonc.2025.05.094},
    issn = {09237534}
}

@article{Vassaux2019AMigration,
    title = {{A Biophysical Model for Curvature-Guided Cell Migration}},
    year = {2019},
    journal = {Biophysical Journal},
    author = {Vassaux, Maxime and Pieuchot, Laurent and Anselme, Karine and Bigerelle, Maxence and Milan, Jean Louis},
    number = {6},
    month = {9},
    pages = {1136--1144},
    volume = {117},
    publisher = {Cell Press},
    url = {https://www.sciencedirect.com/science/article/pii/S0006349519305909?via%3Dihub},
    doi = {10.1016/J.BPJ.2019.07.022},
    issn = {0006-3495},
    pmid = {31400917}
}

@article{Stacy2016ACross-respiration,
    title = {{A commensal bacterium promotes virulence of an opportunistic pathogen via cross-respiration}},
    year = {2016},
    journal = {mBio},
    author = {Stacy, Apollo and Fleming, Derek and Lamont, Richard J. and Rumbaugh, Kendra P. and Whiteley, Marvin},
    number = {3},
    volume = {7},
    publisher = {American Society for Microbiology},
    url = {/doi/pdf/10.1128/mbio.00782-16?download=true},
    doi = {10.1128/MBIO.00782-16;WEBSITE:WEBSITE:ASMJ;ISSUE:ISSUE:DOI},
    issn = {21507511},
    pmid = {27353758}
}

@article{Nakatsuji2018ANeoplasia,
    title = {{A commensal strain of Staphylococcus epidermidis protects against skin neoplasia}},
    year = {2018},
    journal = {Science Advances},
    author = {Nakatsuji, Teruaki and Chen, Tiffany H. and Butcher, Anna M. and Trzoss, Lynnie L. and Nam, Sang Jip and Shirakawa, Karina T. and Zhou, Wei and Oh, Julia and Otto, Michael and Fenical, William and Gallo, Richard L.},
    number = {2},
    month = {2},
    volume = {4},
    publisher = {American Association for the Advancement of Science},
    url = {/doi/pdf/10.1126/sciadv.aao4502?download=true},
    doi = {10.1126/SCIADV.AAO4502/SUPPL{\_}FILE/AAO4502{\_}TABLES3.XLSX},
    issn = {23752548},
    pmid = {29507878}
}

@article{Song2018ATargeting,
    title = {{A microfluidic device for studying chemotaxis mechanism of bacterial cancer targeting}},
    year = {2018},
    journal = {Scientific Reports 2018 8:1},
    author = {Song, Jing and Zhang, Yu and Zhang, Chengqian and Du, Xiaohui and Guo, Zhe and Kuang, Yanbin and Wang, Yingyan and Wu, Peng and Zou, Kun and Zou, Lijuan and Lv, Jianxin and Wang, Qi},
    number = {1},
    month = {4},
    pages = {6394-},
    volume = {8},
    publisher = {Nature Publishing Group},
    url = {https://www.nature.com/articles/s41598-018-24748-7},
    doi = {10.1038/s41598-018-24748-7},
    issn = {2045-2322},
    pmid = {29686328}
}

@article{Gao2025AMicrobiota,
    title = {{A Nano-Strategy for Advanced Triple-Negative Breast Cancer Therapy by Regulating Intratumoral Microbiota}},
    year = {2025},
    journal = {Nano Letters},
    author = {Gao, Jifan and Tang, Lu and Fu, Cong and Cao, Yuqi and Liu, Hening and Yin, Yue and Li, Zixuan and Zhu, Yuanbo and Shu, Weijie and Zhang, Yi and Ru, Xinrong and Wang, Wei},
    number = {15},
    month = {4},
    pages = {6134--6144},
    volume = {25},
    publisher = {American Chemical Society},
    url = {/doi/pdf/10.1021/acs.nanolett.5c00298?ref=article_openPDF},
    doi = {10.1021/ACS.NANOLETT.5C00298/ASSET/IMAGES/LARGE/NL5C00298{\_}0005.JPEG},
    issn = {15306992},
    pmid = {40177896}
}

@article{Parida2021AAxes,
    title = {{A procarcinogenic colon microbe promotes breast tumorigenesis and metastatic progression and concomitantly activates notch and b-catenin axes}},
    year = {2021},
    journal = {Cancer Discovery},
    author = {Parida, Sheetal and Wu, Shaoguang and Siddharth, Sumit and Wang, Guannan and Muniraj, Nethaji and Nagalingam, Arumugam and Hum, Christina and Mistriotis, Panagiotis and Hao, Haiping and Conover Talbot, C. and Konstantopoulos, Konstantinos and Gabrielson, Kathleen L. and Sears, Cynthia L. and Sharma, Dipali},
    number = {5},
    month = {5},
    pages = {1138--1157},
    volume = {11},
    publisher = {American Association for Cancer Research Inc.},
    url = {/cancerdiscovery/article/11/5/1138/666276/A-Procarcinogenic-Colon-Microbe-Promotes-Breast https://dx.doi.org/10.1158/2159-8290.CD-20-0537},
    doi = {10.1158/2159-8290.CD-20-0537/333485/AM/A-PRO-CARCINOGENIC-COLON-MICROBE-PROMOTES-BREAST},
    issn = {21598290},
    pmid = {33408241}
}

@article{Fivenson2023ADetermination,
    title = {{A role for the Gram-negative outer membrane in bacterial shape determination}},
    year = {2023},
    journal = {Proceedings of the National Academy of Sciences of the United States of America},
    author = {Fivenson, Elayne M. and Rohs, Patricia D.A. and Vettiger, Andrea and Sardis, Marios F. and Torres, Grasiela and Forchoh, Alison and Bernhardt, Thomas G.},
    number = {35},
    month = {8},
    pages = {e2301987120},
    volume = {120},
    publisher = {National Academy of Sciences},
    url = {/doi/pdf/10.1073/pnas.2301987120?download=true},
    doi = {10.1073/PNAS.2301987120;WGROUP:STRING:PUBLICATION},
    issn = {10916490},
    pmid = {37607228}
}

@article{Qin2025ACells,
    title = {{A signaling molecule from intratumor bacteria promotes trastuzumab resistance in breast cancer cells}},
    year = {2025},
    journal = {Proceedings of the National Academy of Sciences of the United States of America},
    author = {Qin, Gege and Shao, Xiying and Liu, Xiaolong and Xu, Jiachao and Wang, Xiaojia and Wang, Wenxi and Gao, Lu and Liang, Yuxin and Xie, Lina and Su, Dan and Yang, Hongwei and Zhou, Wei and Fang, Xiaohong},
    number = {2},
    month = {1},
    pages = {e2421710122},
    volume = {122},
    publisher = {National Academy of Sciences},
    url = {/doi/pdf/10.1073/pnas.2421710122?download=true},
    doi = {10.1073/PNAS.2421710122/SUPPL{\_}FILE/PNAS.2421710122.SM01.AVI},
    issn = {10916490},
    pmid = {39786928}
}

@article{Aparicio-Yuste2022ACells,
    title = {{A Stiff Extracellular Matrix Favors the Mechanical Cell Competition that Leads to Extrusion of Bacterially-Infected Epithelial Cells}},
    year = {2022},
    journal = {Frontiers in Cell and Developmental Biology},
    author = {Aparicio-Yuste, Raúl and Muenkel, Marie and Clark, Andrew G. and G{\'{o}}mez-Benito, María J. and Bastounis, Effie E.},
    month = {6},
    pages = {912318},
    volume = {10},
    publisher = {Frontiers Media S.A.},
    url = {https://pmc.ncbi.nlm.nih.gov/articles/PMC9257086/},
    doi = {10.3389/FCELL.2022.912318/FULL},
    issn = {2296634X}
}

@article{Narla2021AGrowth,
    title = {{A traveling-wave solution for bacterial chemotaxis with growth}},
    year = {2021},
    journal = {Proceedings of the National Academy of Sciences of the United States of America},
    author = {Narla, Avaneesh V. and Cremer, Jonas and Hwa, Terence},
    number = {48},
    month = {11},
    pages = {e2105138118},
    volume = {118},
    publisher = {National Academy of Sciences},
    url = {/doi/pdf/10.1073/pnas.2105138118?download=true},
    doi = {10.1073/PNAS.2105138118;ISSUE:ISSUE:DOI},
    issn = {10916490},
    pmid = {34819366},
    arxivId = {2103.08100}
}

@article{Liang2021AcidicAxis,
    title = {{Acidic microenvironment enhances MT1-MMP-mediated cancer cell motility through integrin {$\beta$}1/cofilin/F-actin axis}},
    year = {2021},
    journal = {Acta Biochimica et Biophysica Sinica},
    author = {Liang, Lubiao and Sui, Ran and Song, Yongxiang and Zhao, Yajin},
    number = {11},
    month = {11},
    pages = {1558--1566},
    volume = {53},
    publisher = {Oxford Academic},
    url = {https://dx.doi.org/10.1093/abbs/gmab130},
    doi = {10.1093/ABBS/GMAB130},
    issn = {17457270},
    pmid = {34568889}
}

@article{Tavares2020ActivityBacteria,
    title = {{Activity of Specialized Biomolecules against Gram-Positive and Gram-Negative Bacteria}},
    year = {2020},
    journal = {Antibiotics 2020, Vol. 9, Page 314},
    author = {Tavares, Tânia D. and Antunes, Joana C. and Padr{\~{a}}o, Jorge and Ribeiro, Ana I. and Zille, Andrea and Amorim, M. Teresa P. and Ferreira, Fernando and Felgueiras, Helena P.},
    number = {6},
    month = {6},
    pages = {314},
    volume = {9},
    publisher = {Multidisciplinary Digital Publishing Institute},
    url = {https://www.mdpi.com/2079-6382/9/6/314/htm https://www.mdpi.com/2079-6382/9/6/314},
    doi = {10.3390/ANTIBIOTICS9060314},
    issn = {2079-6382}
}

@article{Cheng2025AdvancementsInteraction,
    title = {{Advancements and Challenges in Modeling Mechanobiology in Intestinal Host-Microbiota Interaction}},
    year = {2025},
    journal = {ACS Applied Materials {\&} Interfaces},
    author = {Cheng, Hong Sheng and Tey, Yee Han and Hu, Si Yuan and Yeo, Alethea Yen Ning and Ngo, Zong Heng and Kim, Joseph Han Sol and Tan, Nguan Soon},
    number = {22},
    month = {6},
    pages = {31698},
    volume = {17},
    publisher = {American Chemical Society},
    url = {https://pmc.ncbi.nlm.nih.gov/articles/PMC12147776/},
    doi = {10.1021/ACSAMI.4C20961},
    issn = {19448252},
    pmid = {40382722}
}

@article{Li2024Agent-basedBiofilms,
    title = {{Agent-based modeling of stress anisotropy driven nematic ordering in growing biofilms}},
    year = {2024},
    journal = {Soft Matter},
    author = {Li, Changhao and Nijjer, Japinder and Feng, Luyi and Zhang, Qiuting and Yan, Jing and Zhang, Sulin},
    number = {16},
    month = {4},
    pages = {3401--3410},
    volume = {20},
    publisher = {The Royal Society of Chemistry},
    url = {https://pubs.rsc.org/en/content/articlehtml/2024/sm/d3sm01535a https://pubs.rsc.org/en/content/articlelanding/2024/sm/d3sm01535a},
    doi = {10.1039/D3SM01535A},
    issn = {1744-6848},
    pmid = {38563244}
}

@article{Ramirez-Larrota2022AnEvasion,
    title = {{An Introduction to Bacterial Biofilms and Their Proteases, and Their Roles in Host Infection and Immune Evasion}},
    year = {2022},
    journal = {Biomolecules 2022, Vol. 12, Page 306},
    author = {Ram{\'{i}}rez-Larrota, Juan Sebastián and Eckhard, Ulrich},
    number = {2},
    month = {2},
    pages = {306},
    volume = {12},
    publisher = {Multidisciplinary Digital Publishing Institute},
    url = {https://www.mdpi.com/2218-273X/12/2/306/htm https://www.mdpi.com/2218-273X/12/2/306},
    doi = {10.3390/BIOM12020306},
    issn = {2218-273X},
    pmid = {35204806}
}

@article{Lucchetti2024An20/2024,
    title = {{An Organ-on-Chip Platform for Simulating Drug Metabolism Along the Gut–Liver Axis (Adv. Healthcare Mater. 20/2024)}},
    year = {2024},
    journal = {Advanced Healthcare Materials},
    author = {Lucchetti, Mara and Aina, Kehinde Oluwasegun and Grandmougin, Léa and J{\"{a}}ger, Christian and Escriva, Pau Pérez and Letellier, Elisabeth and Mosig, Alexander S. and Wilmes, Paul},
    number = {20},
    month = {8},
    pages = {2470129},
    volume = {13},
    publisher = {John Wiley {\&} Sons, Ltd},
    url = {/doi/pdf/10.1002/adhm.202470129 https://onlinelibrary.wiley.com/doi/abs/10.1002/adhm.202470129 https://advanced.onlinelibrary.wiley.com/doi/10.1002/adhm.202470129},
    doi = {10.1002/ADHM.202470129},
    issn = {2192-2659}
}

@article{Bian2022AntihypertensionCancer,
    title = {{Antihypertension Nanoblockers Increase Intratumoral Perfusion of Sequential Cytotoxic Nanoparticles to Enhance Chemotherapy Efficacy against Pancreatic Cancer}},
    year = {2022},
    journal = {Advanced Science},
    author = {Bian, Suchen and Dong, Haijiang and Zhao, Long and Li, Zequn and Chen, Jian and Zhu, Xingxin and Qiu, Nasha and Jia, Xing and Song, Wenfeng and Li, Zekuan and Zheng, Shusen and Wang, Hangxiang and Song, Penghong},
    number = {29},
    month = {10},
    pages = {2201931},
    volume = {9},
    publisher = {John Wiley {\&} Sons, Ltd},
    url = {https://advanced.onlinelibrary.wiley.com/doi/pdf/10.1002/advs.202201931 https://onlinelibrary.wiley.com/doi/abs/10.1002/advs.202201931 https://advanced.onlinelibrary.wiley.com/doi/10.1002/advs.202201931},
    doi = {10.1002/ADVS.202201931},
    issn = {2198-3844},
    pmid = {36026578}
}

@article{Kutschera2023AntonieMicrobiology,
    title = {{Antonie van Leeuwenhoek (1632–1723): Master of Fleas and Father of Microbiology}},
    year = {2023},
    journal = {Microorganisms},
    author = {Kutschera, Ulrich},
    number = {8},
    month = {8},
    pages = {1994},
    volume = {11},
    publisher = {Multidisciplinary Digital Publishing Institute (MDPI)},
    url = {https://pmc.ncbi.nlm.nih.gov/articles/PMC10458164/},
    doi = {10.3390/MICROORGANISMS11081994},
    issn = {20762607},
    pmid = {37630554}
}

@article{Alkmin2022AssessingModels,
    title = {{Assessing the roles of collagen fiber morphology and matrix stiffness on ovarian cancer cell migration dynamics using multiphoton fabricated orthogonal image-based models}},
    year = {2022},
    journal = {Acta Biomaterialia},
    author = {Alkmin, Samuel and Patankar, Manish S. and Campagnola, Paul J.},
    month = {11},
    pages = {342--354},
    volume = {153},
    publisher = {Elsevier},
    url = {https://www.sciencedirect.com/science/article/pii/S1742706122005992#sec0018},
    doi = {10.1016/J.ACTBIO.2022.09.037},
    issn = {1742-7061},
    pmid = {36152908}
}

@article{Gunathilake2021AssociationCancer,
    title = {{Association between bacteria other than Helicobacter pylori and the risk of gastric cancer}},
    year = {2021},
    journal = {Helicobacter},
    author = {Gunathilake, Madhawa and Lee, Jeonghee and Choi, Il Ju and Kim, Young Il and Kim, Jeongseon},
    number = {5},
    month = {10},
    pages = {e12836},
    volume = {26},
    publisher = {John Wiley and Sons Inc},
    url = {/doi/pdf/10.1111/hel.12836 https://onlinelibrary.wiley.com/doi/abs/10.1111/hel.12836 https://onlinelibrary.wiley.com/doi/10.1111/hel.12836},
    doi = {10.1111/HEL.12836;REQUESTEDJOURNAL:JOURNAL:15235378;WGROUP:STRING:PUBLICATION},
    issn = {15235378},
    pmid = {34268831}
}

@article{Xu2022AutonomousSolids,
    title = {{Autonomous waves and global motion modes in living active solids}},
    year = {2022},
    journal = {Nature Physics 2022 19:1},
    author = {Xu, Haoran and Huang, Yulu and Zhang, Rui and Wu, Yilin},
    number = {1},
    month = {11},
    pages = {46--51},
    volume = {19},
    publisher = {Nature Publishing Group},
    url = {https://www.nature.com/articles/s41567-022-01836-0},
    doi = {10.1038/s41567-022-01836-0},
    issn = {1745-2481},
    arxivId = {2208.09664}
}

@article{Flemming2019BacteriaBiofilms,
    title = {{Bacteria and archaea on Earth and their abundance in biofilms}},
    year = {2019},
    journal = {Nature Reviews Microbiology 2019 17:4},
    author = {Flemming, Hans Curt and Wuertz, Stefan},
    number = {4},
    month = {2},
    pages = {247--260},
    volume = {17},
    publisher = {Nature Publishing Group},
    url = {https://www.nature.com/articles/s41579-019-0158-9},
    doi = {10.1038/s41579-019-0158-9},
    issn = {1740-1534},
    pmid = {30760902}
}

@misc{ElTekle2023BacteriaProgression,
    title = {{Bacteria in cancer initiation, promotion and progression}},
    year = {2023},
    booktitle = {Nature Reviews Cancer},
    author = {El Tekle, Geniver and Garrett, Wendy S.},
    number = {9},
    month = {9},
    pages = {600--618},
    volume = {23},
    publisher = {Nature Research},
    doi = {10.1038/s41568-023-00594-2},
    issn = {14741768},
    pmid = {37400581}
}

@article{Riedel2025BacterialStage,
    title = {{Bacterial Adhesion on Soft Surfaces: The Dual Role of Substrate Stiffness and Bacterial Growth Stage}},
    year = {2025},
    journal = {Microorganisms 2025, Vol. 13, Page 637},
    author = {Riedel, René and Rani, Garima and Sengupta, Anupam},
    number = {3},
    month = {3},
    pages = {637},
    volume = {13},
    publisher = {Multidisciplinary Digital Publishing Institute},
    url = {https://www.mdpi.com/2076-2607/13/3/637},
    doi = {10.3390/MICROORGANISMS13030637},
    issn = {2076-2607}
}

@article{Mueller2020BacterialStress,
    title = {{Bacterial cell wall quality control during environmental stress}},
    year = {2020},
    journal = {mBio},
    author = {Mueller, Elizabeth A. and Levin, Petra Anne},
    number = {5},
    month = {9},
    pages = {1--15},
    volume = {11},
    publisher = {American Society for Microbiology},
    url = {/doi/pdf/10.1128/mbio.02456-20?download=true},
    doi = {10.1128/MBIO.02456-20;WEBSITE:WEBSITE:ASMJ;ISSUE:ISSUE:DOI},
    issn = {21507511},
    pmid = {33051371}
}

@article{Sun2025BacterialProcess,
    title = {{Bacterial galvanotaxis: a neglected biological process}},
    year = {2025},
    journal = {Trends in Microbiology},
    author = {Sun, Yaohui and Zhao, Min},
    number = {7},
    month = {7},
    pages = {713--717},
    volume = {33},
    publisher = {Elsevier Ltd},
    url = {https://www.cell.com/action/showFullText?pii=S0966842X25000708 https://www.cell.com/action/showAbstract?pii=S0966842X25000708 https://www.cell.com/trends/microbiology/abstract/S0966-842X(25)00070-8},
    doi = {10.1016/j.tim.2025.02.013},
    issn = {18784380},
    pmid = {40102094}
}

@article{Guenther2020BacterialStudy,
    title = {{Bacterial lipopolysaccharide as negative predictor of gemcitabine efficacy in advanced pancreatic cancer - translational results from the AIO-PK0104 Phase 3 study}},
    year = {2020},
    journal = {British journal of cancer},
    author = {Guenther, Michael and Haas, Michael and Heinemann, Volker and Kruger, Stephan and Westphalen, Christoph Benedikt and von Bergwelt-Baildon, Michael and Mayerle, Julia and Werner, Jens and Kirchner, Thomas and Boeck, Stefan and Ormanns, Steffen},
    number = {9},
    month = {10},
    pages = {1370--1376},
    volume = {123},
    publisher = {Br J Cancer},
    url = {https://pubmed.ncbi.nlm.nih.gov/32830200/},
    doi = {10.1038/S41416-020-01029-7},
    issn = {1532-1827},
    pmid = {32830200}
}

@article{Sulit2023BacterialMicroenvironment,
    title = {{Bacterial lipopolysaccharide modulates immune response in the colorectal tumor microenvironment}},
    year = {2023},
    journal = {npj Biofilms and Microbiomes 2023 9:1},
    author = {Sulit, A. K. and Daigneault, M. and Allen-Vercoe, E. and Silander, O. K. and Hock, B. and McKenzie, J. and Pearson, J. and Frizelle, F. A. and Schmeier, S. and Purcell, R.},
    number = {1},
    month = {8},
    pages = {59-},
    volume = {9},
    publisher = {Nature Publishing Group},
    url = {https://www.nature.com/articles/s41522-023-00429-w},
    isbn = {4152202300429},
    doi = {10.1038/s41522-023-00429-w},
    issn = {2055-5008},
    pmid = {37612266}
}

@article{Young2007BacterialShapes,
    title = {{Bacterial morphology: why have different shapes?}},
    year = {2007},
    journal = {Current Opinion in Microbiology},
    author = {Young, Kevin D.},
    number = {6},
    month = {12},
    pages = {596--600},
    volume = {10},
    publisher = {Elsevier Current Trends},
    url = {https://www.sciencedirect.com/science/article/pii/S1369527407001348},
    doi = {10.1016/J.MIB.2007.09.009},
    issn = {1369-5274},
    pmid = {17981076}
}

@article{Marcos2012BacterialRheotaxis,
    title = {{Bacterial rheotaxis}},
    year = {2012},
    journal = {Proceedings of the National Academy of Sciences of the United States of America},
    author = {{Marcos} and Fu, Henry C. and Powers, Thomas R. and Stocker, Roman},
    number = {13},
    month = {3},
    pages = {4780--4785},
    volume = {109},
    publisher = {National Academy of Sciences},
    url = {/doi/pdf/10.1073/pnas.1120955109?download=true},
    doi = {10.1073/PNAS.1120955109;WEBSITE:WEBSITE:PNAS-SITE;WGROUP:STRING:PUBLICATION},
    issn = {00278424},
    pmid = {22411815}
}

@article{Singh2018BeyondResistance,
    title = {{Beyond Tissue Stiffness and Bioadhesivity: Advanced Biomaterials to Model Tumor Microenvironments and Drug Resistance}},
    year = {2018},
    journal = {Trends in Cancer},
    author = {Singh, Ankur and Brito, Ilana and Lammerding, Jan},
    number = {4},
    month = {4},
    pages = {281--291},
    volume = {4},
    publisher = {Cell Press},
    url = {https://www.cell.com/action/showFullText?pii=S2405803318300232 https://www.cell.com/action/showAbstract?pii=S2405803318300232 https://www.cell.com/trends/cancer/abstract/S2405-8033(18)30023-2},
    doi = {10.1016/j.trecan.2018.01.008},
    issn = {24058033},
    pmid = {29606313}
}

@article{Sandle2014BIOCHEMICALMicroorganisms,
    title = {{BIOCHEMICAL AND MODERN IDENTIFICATION TECHNIQUES | Food-Poisoning Microorganisms}},
    year = {2014},
    journal = {Encyclopedia of Food Microbiology: Second Edition},
    author = {Sandle, T.},
    month = {1},
    pages = {238--243},
    publisher = {Academic Press},
    url = {https://www.sciencedirect.com/science/chapter/referencework/abs/pii/B9780123847300000367},
    isbn = {9780123847331},
    doi = {10.1016/B978-0-12-384730-0.00036-7}
}

@article{Farhana2023BiochemistryLipopolysaccharide,
    title = {{Biochemistry, Lipopolysaccharide}},
    year = {2023},
    journal = {StatPearls},
    author = {Farhana, Aisha and Khan, Yusuf S.},
    month = {4},
    publisher = {StatPearls Publishing},
    url = {https://www.ncbi.nlm.nih.gov/sites/books/NBK554414/},
    pmid = {32119301}
}

@article{Ruhal2021BiofilmBacteria,
    title = {{Biofilm patterns in gram-positive and gram-negative bacteria}},
    year = {2021},
    journal = {Microbiological Research},
    author = {Ruhal, Rohit and Kataria, Rashmi},
    month = {10},
    pages = {126829},
    volume = {251},
    publisher = {Urban {\&} Fischer},
    url = {https://www.sciencedirect.com/science/article/pii/S094450132100135X?via%3Dihub},
    doi = {10.1016/J.MICRES.2021.126829},
    issn = {0944-5013},
    pmid = {34332222}
}

@article{Oliva2021Biofilm-RelatedDalbavancin,
    title = {{Biofilm-Related Infections in Gram-Positive Bacteria and the Potential Role of the Long-Acting Agent Dalbavancin}},
    year = {2021},
    journal = {Frontiers in Microbiology},
    author = {Oliva, Alessandra and Stefani, Stefania and Venditti, Mario and Di Domenico, Enea Gino},
    month = {10},
    pages = {749685},
    volume = {12},
    publisher = {Frontiers Media S.A.},
    url = {https://pmc.ncbi.nlm.nih.gov/articles/PMC8569946/},
    doi = {10.3389/FMICB.2021.749685},
    issn = {1664302X},
    pmid = {34745053}
}

@article{Kvich2024BiofilmsCancer,
    title = {{Biofilms and core pathogens shape the tumor microenvironment and immune phenotype in colorectal cancer}},
    year = {2024},
    journal = {Gut Microbes},
    author = {Kvich, Lasse and Fritz, Blaine Gabriel and Zschach, Henrike and Terkelsen, Thilde and Raskov, Hans and H{\o}st-Rasmussen, Kathrine and Jakobsen, Morten Ragn and Gheorghe, Alexandra Gabriella and G{\"{o}}genur, Ismail and Bjarnsholt, Thomas},
    number = {1},
    month = {12},
    volume = {16},
    publisher = {Taylor and Francis Ltd.},
    url = {https://www.tandfonline.com/doi/pdf/10.1080/19490976.2024.2350156},
    doi = {10.1080/19490976.2024.2350156;PAGE:STRING:ARTICLE/CHAPTER},
    issn = {19490984},
    pmid = {38726597}
}

@article{Cont2020BiofilmsEpithelia,
    title = {{Biofilms deform soft surfaces and disrupt epithelia}},
    year = {2020},
    journal = {eLife},
    author = {Cont, Alice and Rossy, Tamara and Al-Mayyah, Zainebe and Persat, Alexandre},
    month = {10},
    pages = {1--22},
    volume = {9},
    publisher = {eLife Sciences Publications Ltd},
    doi = {10.7554/ELIFE.56533},
    issn = {2050084X},
    pmid = {33025904}
}

@article{Parhi2020BreastProgression,
    title = {{Breast cancer colonization by Fusobacterium nucleatum accelerates tumor growth and metastatic progression}},
    year = {2020},
    journal = {Nature Communications 2020 11:1},
    author = {Parhi, Lishay and Alon-Maimon, Tamar and Sol, Asaf and Nejman, Deborah and Shhadeh, Amjad and Fainsod-Levi, Tanya and Yajuk, Olga and Isaacson, Batya and Abed, Jawad and Maalouf, Naseem and Nissan, Aviram and Sandbank, Judith and Yehuda-Shnaidman, Einav and Ponath, Falk and Vogel, Jörg and Mandelboim, Ofer and Granot, Zvi and Straussman, Ravid and Bachrach, Gilad},
    number = {1},
    month = {6},
    pages = {1--12},
    volume = {11},
    publisher = {Nature Publishing Group},
    url = {https://www.nature.com/articles/s41467-020-16967-2},
    doi = {10.1038/s41467-020-16967-2},
    issn = {2041-1723},
    pmid = {32591509}
}

@article{Herrera-Quintana2024BreastInteractions,
    title = {{Breast Cancer: Extracellular Matrix and Microbiome Interactions}},
    year = {2024},
    journal = {International Journal of Molecular Sciences 2024, Vol. 25, Page 7226},
    author = {Herrera-Quintana, Lourdes and V{\'{a}}zquez-Lorente, Héctor and Plaza-Diaz, Julio},
    number = {13},
    month = {6},
    pages = {7226},
    volume = {25},
    publisher = {Multidisciplinary Digital Publishing Institute},
    url = {https://www.mdpi.com/1422-0067/25/13/7226/htm https://www.mdpi.com/1422-0067/25/13/7226},
    doi = {10.3390/IJMS25137226},
    issn = {1422-0067},
    pmid = {39000333}
}

@article{Sperling2022BreathlessHistory,
    title = {{Breathless through Time: Oxygen and Animals across Earth’s History}},
    year = {2022},
    journal = {https://doi.org/10.1086/721754},
    author = {Sperling, Erik A. and Boag, Thomas H. and Duncan, Murray I. and Endriga, Cecilia R. and Marquez, J. Andres and Mills, Daniel B. and Monarrez, Pedro M. and Sclafani, Judith A. and Stockey, Richard G. and Payne, Jonathan L.},
    number = {2},
    month = {10},
    pages = {184--206},
    volume = {243},
    publisher = {The University of Chicago PressChicago, IL},
    url = {/doi/pdf/10.1086/721754},
    doi = {10.1086/721754},
    issn = {19398697},
    pmid = {36548971}
}

@article{Ou2021C5aR1-positiveENO1,
    title = {{C5aR1-positive neutrophils promote breast cancer glycolysis through WTAP-dependent m6A methylation of ENO1}},
    year = {2021},
    journal = {Cell Death {\&} Disease 2021 12:8},
    author = {Ou, Baochi and Liu, Yuan and Yang, Xiaowei and Xu, Xiaojun and Yan, Yunwen and Zhang, Jingjie},
    number = {8},
    month = {7},
    pages = {737-},
    volume = {12},
    publisher = {Nature Publishing Group},
    url = {https://www.nature.com/articles/s41419-021-04028-5},
    doi = {10.1038/s41419-021-04028-5},
    issn = {2041-4889},
    pmid = {34312368}
}

@article{Goodwin1986CampylobacterUlceration.,
    title = {{Campylobacter pyloridis, gastritis, and peptic ulceration.}},
    year = {1986},
    journal = {Journal of Clinical Pathology},
    author = {Goodwin, C S and Armstrong, J A and Marshall, B J},
    number = {4},
    month = {4},
    pages = {353--365},
    volume = {39},
    url = {https://jcp.bmj.com/lookup/doi/10.1136/jcp.39.4.353},
    doi = {10.1136/jcp.39.4.353},
    issn = {0021-9746}
}

@article{Butlin1905CarcinomaEngland,
    title = {{Carcinoma is a Parasitic Disease: Being the Bradshaw Lecture delivered before the Royal College of Surgeons of England}},
    year = {1905},
    journal = {BMJ},
    author = {Butlin, H. T.},
    number = {2346},
    month = {12},
    pages = {1565--1571},
    volume = {2},
    url = {https://www.bmj.com/lookup/doi/10.1136/bmj.2.2346.1565},
    doi = {10.1136/bmj.2.2346.1565},
    issn = {0959-8138}
}

@article{Sreepadmanabh2024CellConfinement,
    title = {{Cell shape affects bacterial colony growth under physical confinement}},
    year = {2024},
    journal = {Nature communications},
    author = {Sreepadmanabh, M. and Ganesh, Meenakshi and Sanjenbam, Pratibha and Kurzthaler, Christina and Agashe, Deepa and Bhattacharjee, Tapomoy},
    number = {1},
    month = {12},
    pages = {9561},
    volume = {15},
    doi = {10.1038/s41467-024-53989-6},
    issn = {20411723},
    pmid = {39516204}
}

@article{Werner2019Cell-PerceivedMigration,
    title = {{Cell-Perceived Substrate Curvature Dynamically Coordinates the Direction, Speed, and Persistence of Stromal Cell Migration}},
    year = {2019},
    journal = {Advanced Biosystems},
    author = {Werner, Maike and Petersen, Ansgar and Kurniawan, Nicholas A. and Bouten, Carlijn V.C.},
    number = {10},
    month = {10},
    pages = {1900080},
    volume = {3},
    publisher = {Wiley-VCH Verlag},
    url = {/doi/pdf/10.1002/adbi.201900080 https://onlinelibrary.wiley.com/doi/abs/10.1002/adbi.201900080 https://onlinelibrary.wiley.com/doi/10.1002/adbi.201900080},
    doi = {10.1002/ADBI.201900080;PAGEGROUP:STRING:PUBLICATION},
    issn = {23667478}
}

@article{Huang2016CellularCells,
    title = {{Cellular microenvironment modulates the galvanotaxis of brain tumor initiating cells}},
    year = {2016},
    journal = {Scientific Reports 2016 6:1},
    author = {Huang, Yu Ja and Hoffmann, Gwendolyn and Wheeler, Benjamin and Schiapparelli, Paula and Quinones-Hinojosa, Alfredo and Searson, Peter},
    number = {1},
    month = {2},
    pages = {21583-},
    volume = {6},
    publisher = {Nature Publishing Group},
    url = {https://www.nature.com/articles/srep21583},
    doi = {10.1038/srep21583},
    issn = {2045-2322},
    pmid = {26898606}
}

@article{Kraning-Rush2012CellularPotential,
    title = {{Cellular Traction Stresses Increase with Increasing Metastatic Potential}},
    year = {2012},
    journal = {PLoS ONE},
    author = {Kraning-Rush, Casey M. and Califano, Joseph P. and Reinhart-King, Cynthia A.},
    number = {2},
    month = {2},
    pages = {e32572},
    volume = {7},
    url = {https://pmc.ncbi.nlm.nih.gov/articles/PMC3289668/},
    doi = {10.1371/JOURNAL.PONE.0032572},
    issn = {19326203},
    pmid = {22389710}
}

@article{Abbaspour2025CollagenAdhesion,
    title = {{Collagen fiber density observed in metastatic ovarian cancer promotes tumor cell adhesion}},
    year = {2025},
    journal = {Acta Biomaterialia},
    author = {Abbaspour, Ali and Martinez Cavazos, Ana L. and Patel, Roshan and Yang, Ning and McGregor, Stephanie M. and Brooks, Erin G. and Masters, Kristyn S. and Kreeger, Pamela K.},
    month = {6},
    pages = {299--312},
    volume = {200},
    publisher = {Elsevier},
    url = {https://www.sciencedirect.com/science/article/pii/S1742706125003605},
    doi = {10.1016/J.ACTBIO.2025.05.035},
    issn = {1742-7061},
    pmid = {40374134}
}

@article{Wittmann2023CollectiveColonies,
    title = {{Collective mechano-response dynamically tunes cell-size distributions in growing bacterial colonies}},
    year = {2023},
    journal = {Communications Physics 2023 6:1},
    author = {Wittmann, René and Nguyen, G. H.Philipp and L{\"{o}}wen, Hartmut and Schwarzendahl, Fabian J. and Sengupta, Anupam},
    number = {1},
    month = {11},
    pages = {331-},
    volume = {6},
    publisher = {Nature Publishing Group},
    url = {https://www.nature.com/articles/s42005-023-01449-w},
    doi = {10.1038/s42005-023-01449-w},
    issn = {2399-3650}
}

@article{Siryaporn2015ColonizationNetworks,
    title = {{Colonization, competition, and dispersal of pathogens in fluid flow networks}},
    year = {2015},
    journal = {Current Biology},
    author = {Siryaporn, Albert and Kim, Minyoung Kevin and Shen, Yi and Stone, Howard A. and Gitai, Zemer},
    number = {9},
    month = {5},
    pages = {1201--1207},
    volume = {25},
    publisher = {Cell Press},
    url = {https://www.cell.com/action/showFullText?pii=S0960982215002766 https://www.cell.com/action/showAbstract?pii=S0960982215002766 https://www.cell.com/current-biology/abstract/S0960-9822(15)00276-6},
    doi = {10.1016/j.cub.2015.02.074},
    issn = {18790445},
    pmid = {25843031}
}

@article{Iida2013CommensalMicroenvironment,
    title = {{Commensal bacteria control cancer response to therapy by modulating the tumor microenvironment}},
    year = {2013},
    journal = {Science},
    author = {Iida, Noriho and Dzutsev, Amiran and Stewart, C. Andrew and Smith, Loretta and Bouladoux, Nicolas and Weingarten, Rebecca A. and Molina, Daniel A. and Salcedo, Rosalba and Back, Timothy and Cramer, Sarah and Dai, Ren Ming and Kiu, Hiu and Cardone, Marco and Naik, Shruti and Patri, Anil K. and Wang, Ena and Marincola, Francesco M. and Frank, Karen M. and Belkaid, Yasmine and Trinchieri, Giorgio and Goldszmid, Romina S.},
    number = {6161},
    pages = {967--970},
    volume = {342},
    doi = {10.1126/SCIENCE.1240527},
    issn = {10959203},
    pmid = {24264989}
}

@article{Kwon2020ComparisonType,
    title = {{Comparison of Cancer Cell Elasticity by Cell Type}},
    year = {2020},
    journal = {Journal of Cancer},
    author = {Kwon, Sangwoo and Yang, Woochul and Moon, Donggerami and Kim, Kyung Sook},
    number = {18},
    pages = {5403},
    volume = {11},
    publisher = {Ivyspring International Publisher},
    url = {https://pmc.ncbi.nlm.nih.gov/articles/PMC7391204/},
    doi = {10.7150/JCA.45897},
    issn = {18379664},
    pmid = {32742487}
}

@article{Weinberg2023ComprehensiveCRC.,
    title = {{Comprehensive study of the intratumoral microbiome in early- vs. late-onset colorectal cancer: Final analysis of COSMO CRC.}},
    year = {2023},
    journal = {Journal of Clinical Oncology},
    author = {Weinberg, Benjamin Adam and Wang, Hongkun and Geng, Xue and Suter, Robert K. and Sortur, Shrayus and Green, Myra E. and Shokralla, Shadi and Bakhshi, Emily and Chaldekas, Krysta and Harris, Brent T. and Clagett, Dionyssia and Marshall, John},
    number = {16{\_}suppl},
    month = {6},
    pages = {3530--3530},
    volume = {41},
    publisher = {American Society of Clinical Oncology},
    url = {https://signin.asco.org/oauth2/aus1goap5nQ33vo7A297/v1/authorize/?client_id=0oa1lqh0mhJversiO297&response_type=code&scope=openid profile email&state=%2Fdoi%2Fpdf%2F10.1200%2FJCO.2023.41.16_suppl.3530&redirect_uri=https%3A%2F%2Fascopubs.org%2Faction%2FssoReturn},
    doi = {10.1200/JCO.2023.41.16{\_}SUPPL.3530},
    issn = {0732-183X}
}

@article{Aparicio-Yuste2022ComputationalMonocytogenes,
    title = {{Computational modelling of epithelial cell monolayers during infection with Listeria monocytogenes}},
    year = {2022},
    journal = {Computer Methods in Applied Mechanics and Engineering},
    author = {Aparicio-Yuste, Raul and Serrano-Alcalde, Francisco and Muenkel, Marie and Garcia-Aznar, Jose Manuel and Bastounis, Effie E. and Gomez-Benito, Maria Jose},
    month = {11},
    pages = {115477},
    volume = {401},
    publisher = {North-Holland},
    url = {https://www.sciencedirect.com/science/article/pii/S0045782522005011},
    doi = {10.1016/J.CMA.2022.115477},
    issn = {0045-7825}
}

@article{Shao2021ConversionCells,
    title = {{Conversion from epithelial to partial-EMT phenotype by Fusobacterium nucleatum infection promotes invasion of oral cancer cells}},
    year = {2021},
    journal = {Scientific Reports 2021 11:1},
    author = {Shao, Wenhua and Fujiwara, Natsumi and Mouri, Yasuhiro and Kisoda, Satoru and Yoshida, Kayo and Yoshida, Kaya and Yumoto, Hiromichi and Ozaki, Kazumi and Ishimaru, Naozumi and Kudo, Yasusei},
    number = {1},
    month = {7},
    pages = {1--14},
    volume = {11},
    publisher = {Nature Publishing Group},
    url = {https://www.nature.com/articles/s41598-021-94384-1},
    isbn = {0123456789},
    doi = {10.1038/s41598-021-94384-1},
    issn = {2045-2322},
    pmid = {34294795}
}

@article{Giomi2017Cross-talkMicrofluidics,
    title = {{Cross-talk between topological defects in different fields revealed by nematic microfluidics}},
    year = {2017},
    journal = {Proceedings of the National Academy of Sciences of the United States of America},
    author = {Giomi, Luca and Kos, Žiga and Ravnik, Miha and Sengupta, Anupam},
    number = {29},
    month = {7},
    pages = {E5771-E5777},
    volume = {114},
    publisher = {National Academy of Sciences},
    url = {/doi/pdf/10.1073/pnas.1702777114?download=true},
    doi = {10.1073/PNAS.1702777114;PAGE:STRING:ARTICLE/CHAPTER},
    issn = {10916490},
    pmid = {28674012},
    arxivId = {1612.00095}
}

@article{Schamberger2023CurvatureScales,
    title = {{Curvature in Biological Systems: Its Quantification, Emergence, and Implications across the Scales}},
    year = {2023},
    journal = {Advanced Materials},
    author = {Schamberger, Barbara and Ziege, Ricardo and Anselme, Karine and Ben Amar, Martine and Bykowski, Michał and Castro, André P.G. and Cipitria, Amaia and Coles, Rhoslyn A. and Dimova, Rumiana and Eder, Michaela and Ehrig, Sebastian and Escudero, Luis M. and Evans, Myfanwy E. and Fernandes, Paulo R. and Fratzl, Peter and Geris, Liesbet and Gierlinger, Notburga and Hannezo, Edouard and Igli{\v{c}}, Aleš and Kirkensgaard, Jacob J.K. and Kollmannsberger, Philip and Kowalewska, Łucja and Kurniawan, Nicholas A. and Papantoniou, Ioannis and Pieuchot, Laurent and Pires, Tiago H.V. and Renner, Lars D. and Sageman-Furnas, Andrew O. and Schr{\"{o}}der-Turk, Gerd E. and Sengupta, Anupam and Sharma, Vikas R. and Tagua, Antonio and Tomba, Caterina and Trepat, Xavier and Waters, Sarah L. and Yeo, Edwina F. and Roschger, Andreas and Bidan, Cécile M. and Dunlop, John W.C.},
    number = {13},
    month = {3},
    pages = {2206110},
    volume = {35},
    publisher = {John Wiley and Sons Inc},
    url = {/doi/pdf/10.1002/adma.202206110 https://onlinelibrary.wiley.com/doi/abs/10.1002/adma.202206110 https://advanced.onlinelibrary.wiley.com/doi/10.1002/adma.202206110},
    doi = {10.1002/ADMA.202206110;REQUESTEDJOURNAL:JOURNAL:15214095;WGROUP:STRING:PUBLICATION},
    issn = {15214095}
}

@article{Frey2025CurvatureTopographies,
    title = {{Curvature Perception of Mesenchymal Cells on Mesoscale Topographies}},
    year = {2025},
    journal = {Advanced healthcare materials},
    author = {Frey, Kerstin and Brunner, Michael and Curio, Christóbal and Kemkemer, Ralf},
    number = {3},
    month = {1},
    volume = {14},
    publisher = {Adv Healthc Mater},
    url = {https://pubmed.ncbi.nlm.nih.gov/39659136/},
    doi = {10.1002/ADHM.202402865},
    issn = {2192-2659},
    pmid = {39659136}
}

@article{Pieuchot2018CurvotaxisLandscapes,
    title = {{Curvotaxis directs cell migration through cell-scale curvature landscapes}},
    year = {2018},
    journal = {Nature communications},
    author = {Pieuchot, Laurent and Marteau, Julie and Guignandon, Alain and Dos Santos, Thomas and Brigaud, Isabelle and Chauvy, Pierre François and Cloatre, Thomas and Ponche, Arnaud and Petithory, Tatiana and Rougerie, Pablo and Vassaux, Maxime and Milan, Jean Louis and Tusamda Wakhloo, Nayana and Spangenberg, Arnaud and Bigerelle, Maxence and Anselme, Karine},
    number = {1},
    month = {12},
    volume = {9},
    publisher = {Nat Commun},
    url = {https://pubmed.ncbi.nlm.nih.gov/30266986/},
    doi = {10.1038/S41467-018-06494-6},
    issn = {2041-1723},
    pmid = {30266986}
}

@article{Koch1884DieTuberkulose,
    title = {{Die Aetiologie der Tuberkulose}},
    year = {1884},
    journal = {Mittheilungen aus dem Kaiserlichen Gesundheitsamte},
    author = {Koch, Robert},
    pages = {1--88}
}

@article{Koch1882DieTuberkulose.,
    title = {{Die {\"{A}}tiologie der Tuberkulose.}},
    year = {1882},
    journal = {Berliner klinische Wochenschrift},
    author = {Koch, Robert},
    pages = {429--445},
    volume = {15},
    language = {German}
}

@article{Minamino2019DirectionalMotor,
    title = {{Directional switching mechanism of the bacterial flagellar motor}},
    year = {2019},
    journal = {Comput. Struct. Biotechnol. J.},
    author = {Minamino, Tohru and Kinoshita, Miki and Namba, Keiichi},
    month = {1},
    pages = {1075--1081},
    volume = {17},
    publisher = {Elsevier B.V.},
    doi = {10.1016/j.csbj.2019.07.020},
    issn = {20010370}
}

@article{Iwata2025DiscoveryProperties,
    title = {{Discovery and characterization of antitumor gut microbiota from amphibians and reptiles: Ewingella americana as a novel therapeutic agent with dual cytotoxic and immunomodulatory properties}},
    year = {2025},
    journal = {Gut Microbes},
    author = {Iwata, Seigo and Yamasita, Nagi and Asukabe, Kensuke and Sakari, Matomo and Miyako, Eijiro},
    number = {1},
    month = {12},
    volume = {17},
    publisher = {Taylor {\&} Francis},
    url = {https://www.tandfonline.com/doi/pdf/10.1080/19490976.2025.2599562},
    doi = {10.1080/19490976.2025.2599562},
    issn = {19490984},
    pmid = {41376334}
}

@article{Ma2025DissectionCancer,
    title = {{Dissection of intratumor microbiome–host interactions at the single-cell level in lung cancer}},
    year = {2025},
    journal = {hLife},
    author = {Ma, Yong Jing and Sun, Yuan Chen and Wang, Lu and Xu, Wan Xing and Fan, Xiao Dan and Ding, Jun and Heeschen, Christopher and Wu, Wen Juan and Zheng, Xiao Qi and Liu, Ning Ning},
    number = {8},
    month = {8},
    pages = {391--406},
    volume = {3},
    publisher = {Elsevier},
    url = {https://www.sciencedirect.com/science/article/pii/S2949928324000750},
    doi = {10.1016/J.HLIFE.2024.09.001},
    issn = {2949-9283}
}

@article{Liu2020DistributionPatients,
    title = {{Distribution of pathogenic bacteria in lower respiratory tract infection in lung cancer patients after chemotherapy and analysis of integron resistance genes in respiratory tract isolates of uninfected patients}},
    year = {2020},
    journal = {Journal of Thoracic Disease},
    author = {Liu, Haiyan and Liu, Bin and Zheng, Fade and Chen, Xiaoxia and Ye, Lingyun and He, Yayi},
    number = {8},
    month = {8},
    pages = {4216},
    volume = {12},
    publisher = {AME Publishing Company},
    url = {https://pmc.ncbi.nlm.nih.gov/articles/PMC7475539/},
    doi = {10.21037/JTD-20-928},
    issn = {20776624},
    pmid = {32944333}
}

@article{Kysela2016DiversityMorphology,
    title = {{Diversity Takes Shape: Understanding the Mechanistic and Adaptive Basis of Bacterial Morphology}},
    year = {2016},
    journal = {PLOS Biology},
    author = {Kysela, David T. and Randich, Amelia M. and Caccamo, Paul D. and Brun, Yves V.},
    number = {10},
    month = {10},
    pages = {e1002565},
    volume = {14},
    publisher = {Public Library of Science},
    url = {https://journals.plos.org/plosbiology/article?id=10.1371/journal.pbio.1002565},
    doi = {10.1371/JOURNAL.PBIO.1002565},
    issn = {1545-7885},
    pmid = {27695035}
}

@article{Bains2025DriftEnvironments,
    title = {{Drift velocity of bacterial chemotaxis in dynamic chemical environments}},
    year = {2025},
    journal = {Philosophical Transactions of the Royal Society A: Mathematical, Physical and Engineering Sciences},
    author = {Bains, Jason Singh and Baggaley, Andrew William and Croze, Ottavio Alfred},
    number = {2304},
    month = {9},
    volume = {383},
    publisher = {Royal Society Publishing},
    url = {https://dx.doi.org/10.1098/rsta.2024.0261},
    doi = {10.1098/RSTA.2024.0261/234912},
    issn = {1364503X},
    pmid = {40931662}
}

@article{Ji2023DurotaxisGo,
    title = {{Durotaxis and negative durotaxis: where should cells go?}},
    year = {2023},
    journal = {Communications Biology 2023 6:1},
    author = {Ji, Congcong and Huang, Yuxing},
    number = {1},
    month = {11},
    pages = {1169-},
    volume = {6},
    publisher = {Nature Publishing Group},
    url = {https://www.nature.com/articles/s42003-023-05554-y},
    doi = {10.1038/s42003-023-05554-y},
    issn = {2399-3642},
    pmid = {37973823}
}

@article{Mekadim2022DysbiosisProgression,
    title = {{Dysbiosis of skin microbiome and gut microbiome in melanoma progression}},
    year = {2022},
    journal = {BMC Microbiology 2022 22:1},
    author = {Mekadim, Chahrazed and Skalnikova, Helena Kupcova and Cizkova, Jana and Cizkova, Veronika and Palanova, Anna and Horak, Vratislav and Mrazek, Jakub},
    number = {1},
    month = {2},
    pages = {1--19},
    volume = {22},
    publisher = {BioMed Central},
    url = {https://link.springer.com/articles/10.1186/s12866-022-02458-5 https://link.springer.com/article/10.1186/s12866-022-02458-5},
    doi = {10.1186/S12866-022-02458-5},
    issn = {1471-2180},
    pmid = {35216552}
}

@article{AlMalki2023E.Metabolism,
    title = {{E. coli Secretome Metabolically Modulates MDA-MB-231 Breast Cancer Cells’ Energy Metabolism}},
    year = {2023},
    journal = {International Journal of Molecular Sciences},
    author = {AlMalki, Reem H. and Sebaa, Rajaa and Al-Ansari, Mysoon M. and Al-Alwan, Monther and Alwehaibi, Moudi A. and Rahman, Anas M.Abdel},
    number = {4},
    month = {2},
    volume = {24},
    publisher = {Multidisciplinary Digital Publishing Institute (MDPI)},
    doi = {10.3390/ijms24044219},
    issn = {14220067},
    pmid = {36835626}
}

@article{GaleanoNino2022EffectCancer,
    title = {{Effect of the intratumoral microbiota on spatial and cellular heterogeneity in cancer}},
    year = {2022},
    journal = {Nature},
    author = {Galeano Ni{\~{n}}o, Jorge Luis and Wu, Hanrui and LaCourse, Kaitlyn D. and Kempchinsky, Andrew G. and Baryiames, Alexander and Barber, Brittany and Futran, Neal and Houlton, Jeffrey and Sather, Cassie and Sicinska, Ewa and Taylor, Alison and Minot, Samuel S. and Johnston, Christopher D. and Bullman, Susan},
    number = {7937},
    month = {11},
    pages = {810--817},
    volume = {611},
    publisher = {Nature Research},
    url = {https://www.nature.com/articles/s41586-022-05435-0},
    doi = {10.1038/S41586-022-05435-0;TECHMETA},
    issn = {14764687},
    pmid = {36385528}
}

@article{Taheri2025EffectCells,
    title = {{Effect of TIMPs and their minimally engineered variants in blocking invasion and migration of brain cancer cells}},
    year = {2025},
    journal = {Oncotarget},
    author = {Taheri, Elham and Raeeszadeh-Sarmazdeh, Maryam},
    number = {1},
    pages = {118--130},
    volume = {16},
    publisher = {Oncotarget},
    url = {https://pubmed.ncbi.nlm.nih.gov/40019229/},
    doi = {10.18632/ONCOTARGET.28691},
    issn = {1949-2553},
    pmid = {40019229}
}

@article{Garg2019ElectromagneticCells,
    title = {{Electromagnetic fields alter the motility of metastatic breast cancer cells}},
    year = {2019},
    journal = {Communications Biology 2019 2:1},
    author = {Garg, Ayush Arpit and Jones, Travis H. and Moss, Sarah M. and Mishra, Sanjay and Kaul, Kirti and Ahirwar, Dinesh K. and Ferree, Jessica and Kumar, Prabhat and Subramaniam, Deepa and Ganju, Ramesh K. and Subramaniam, Vish V. and Song, Jonathan W.},
    number = {1},
    month = {8},
    pages = {303-},
    volume = {2},
    publisher = {Nature Publishing Group},
    url = {https://www.nature.com/articles/s42003-019-0550-z},
    doi = {10.1038/s42003-019-0550-z},
    issn = {2399-3642},
    pmid = {31428691}
}

@article{Zhou2024EmulationMicroenvironments,
    title = {{Emulation and evaluation of tumor cell combined chemotherapy in isotropic/anisotropic collagen fiber microenvironments}},
    year = {2024},
    journal = {Lab on a Chip},
    author = {Zhou, Lianjie and Li, Guoqiang and Yao, Jingru and Wang, Jing and Yao, Xiyao and Ye, Zhikai and Zheng, Dongtian and Song, Kena and Zhang, Hongfei and Zhang, Xianquan and Shuai, Jianwei and Ye, Fangfu and Li, Ming and Li, Yufeng and Chen, Guo and Cheng, Yuyan and Liu, He and Shaw, Peter and Liu, Liyu},
    number = {11},
    month = {5},
    pages = {2999--3014},
    volume = {24},
    publisher = {The Royal Society of Chemistry},
    url = {https://pubs.rsc.org/en/content/articlehtml/2024/lc/d4lc00051j https://pubs.rsc.org/en/content/articlelanding/2024/lc/d4lc00051j},
    doi = {10.1039/D4LC00051J},
    issn = {1473-0189},
    pmid = {38742451}
}

@article{Mivehchi2025ExploringReview,
    title = {{Exploring the role of oral bacteria in oral cancer: a narrative review}},
    year = {2025},
    journal = {Discover Oncology 2025 16:1},
    author = {Mivehchi, Hassan and Eskandari-Yaghbastlo, Aisan and Pour Bahrami, Parnian and Elhami, Anis and Faghihinia, Farbod and Nejati, Seyedeh Tabasom and Kazemi, Kimia Sadat and Nabi Afjadi, Mohsen},
    number = {1},
    month = {2},
    pages = {1--24},
    volume = {16},
    publisher = {Springer},
    url = {https://link.springer.com/article/10.1007/s12672-025-01998-2},
    isbn = {0123456789},
    doi = {10.1007/S12672-025-01998-2},
    issn = {2730-6011}
}

@article{Zhang2019ExtracellularEpithelia,
    title = {{Extracellular and intercellular force distribution in circularly shaped epithelia}},
    year = {2019},
    journal = {Extreme Mechanics Letters},
    author = {Zhang, Yao and Wei, Qiong and Zhao, Tiankai and Zhao, Peng and Zhang, Sulin},
    month = {9},
    pages = {100526},
    volume = {31},
    publisher = {Elsevier},
    url = {https://www.sciencedirect.com/science/article/pii/S2352431619300471#fig3},
    doi = {10.1016/J.EML.2019.100526},
    issn = {2352-4316}
}

@article{Yuan2023ExtracellularTreatments,
    title = {{Extracellular matrix remodeling in tumor progression and immune escape: from mechanisms to treatments}},
    year = {2023},
    journal = {Molecular Cancer 2023 22:1},
    author = {Yuan, Zhennan and Li, Yingpu and Zhang, Sifan and Wang, Xueying and Dou, He and Yu, Xi and Zhang, Zhiren and Yang, Shanshan and Xiao, Min},
    number = {1},
    month = {3},
    pages = {1--42},
    volume = {22},
    publisher = {BioMed Central},
    url = {https://link.springer.com/articles/10.1186/s12943-023-01744-8 https://link.springer.com/article/10.1186/s12943-023-01744-8},
    doi = {10.1186/S12943-023-01744-8},
    issn = {1476-4598},
    pmid = {36906534}
}

@article{Liu2021ExtracellularInternalization,
    title = {{Extracellular matrix stiffness modulates host-bacteria interactions and antibiotic therapy of bacterial internalization}},
    year = {2021},
    journal = {Biomaterials},
    author = {Liu, Xiaoye and Zhu, Kui and Duan, Xiaocen and Wang, Pudi and Han, Yiming and Peng, Wenjing and Huang, Jianyong},
    month = {10},
    pages = {121098},
    volume = {277},
    publisher = {Elsevier},
    url = {https://www.sciencedirect.com/science/article/pii/S0142961221004543?via%3Dihub},
    doi = {10.1016/J.BIOMATERIALS.2021.121098},
    issn = {0142-9612},
    pmid = {34478931}
}

@article{daCosta2022FibronectinAureus,
    title = {{Fibronectin binding protein B binds to loricrin and promotes corneocyte adhesion by Staphylococcus aureus}},
    year = {2022},
    journal = {Nature Communications 2022 13:1},
    author = {da Costa, Thaina M. and Viljoen, Albertus and Towell, Aisling M. and Dufr{\^{e}}ne, Yves F. and Geoghegan, Joan A.},
    number = {1},
    month = {5},
    pages = {2517-},
    volume = {13},
    publisher = {Nature Publishing Group},
    url = {https://www.nature.com/articles/s41467-022-30271-1},
    doi = {10.1038/s41467-022-30271-1},
    issn = {2041-1723},
    pmid = {35523796}
}

@article{Shen2012FlowUpstream,
    title = {{Flow directs surface-attached bacteria to twitch upstream}},
    year = {2012},
    journal = {Biophysical Journal},
    author = {Shen, Yi and Siryaporn, Albert and Lecuyer, Sigolene and Gitai, Zemer and Stone, Howard A.},
    number = {1},
    month = {7},
    pages = {146--151},
    volume = {103},
    publisher = {Elsevier},
    url = {https://www.cell.com/action/showFullText?pii=S000634951200625X https://www.cell.com/action/showAbstract?pii=S000634951200625X https://www.cell.com/biophysj/abstract/S0006-3495(12)00625-X},
    doi = {10.1016/j.bpj.2012.05.045},
    issn = {00063495},
    pmid = {22828341}
}

@article{Vitry2017Force-inducedLoricrin,
    title = {{Force-induced strengthening of the interaction between Staphylococcus aureus clumping factor B and loricrin}},
    year = {2017},
    journal = {mBio},
    author = {Vitry, Pauline and Valotteau, Claire and Feuillie, Cécile and Bernard, Simon and Alsteens, David and Geoghegan, Joan A. and Dufr{\^{e}}ne, Yves F.},
    number = {6},
    month = {11},
    volume = {8},
    publisher = {American Society for Microbiology},
    url = {/doi/pdf/10.1128/mbio.01748-17?download=true},
    doi = {10.1128/MBIO.01748-17;SUBPAGE:STRING:FULL},
    issn = {21507511},
    pmid = {29208742}
}

@article{Li2020FusobacteriumA,
    title = {{Fusobacterium nucleatum Interaction with Pseudomonas aeruginosa Induces Biofilm-Associated Antibiotic Tolerance via Fusobacterium Adhesin A}},
    year = {2020},
    journal = {ACS infectious diseases},
    author = {Li, Qian and Tan, Lisi and Wang, Hongyan and Kou, Yurong and Shi, Xiaoting and Zhang, Shuwei and Pan, Yaping},
    number = {7},
    month = {7},
    pages = {1686--1696},
    volume = {6},
    publisher = {ACS Infect Dis},
    url = {https://pubmed.ncbi.nlm.nih.gov/32320601/},
    doi = {10.1021/ACSINFECDIS.9B00402},
    issn = {2373-8227},
    pmid = {32320601}
}

@article{Karta2025FusobacteriumCancer,
    title = {{Fusobacterium nucleatum interacts with cancer-associated fibroblasts to promote colorectal cancer}},
    year = {2025},
    journal = {The EMBO Journal 2025 44:19},
    author = {Karta, Jessica and Meyers, Marianne and Rodriguez, Fabien and Koncina, Eric and Gilson, Cedric and Klein, Eliane and Gabola, Monica and Benzarti, Mohaned and P{\'{e}}rez Escriva, Pau and Molina Tijeras, Jose Alberto and Correia Tavares Bernardino, Catarina and Ponath, Falk and Carpentier, Anais and Pujabet, Mònica Aguilera and Schmoetten, Maryse and Tsenkova, Mina and Saoud, Perla and Gaigneaux, Anthoula and Ternes, Dominik and Alonso, Lidia and Z{\"{u}}gel, Nikolaus and Willemssen, Eric and Koppes, Philippe and L{\'{e}}onard, Daniel and Casanova, Luis Perez and Haan, Serge and Mittelbronn, Michel and Meiser, Johannes and Pozdeev, Vitaly I. and Vogel, Jörg and Nuciforo, Paolo G. and Wilmes, Paul and Letellier, Elisabeth},
    number = {19},
    month = {8},
    pages = {5375--5393},
    volume = {44},
    publisher = {Springer},
    url = {https://link.springer.com/article/10.1038/s44318-025-00542-w},
    doi = {10.1038/S44318-025-00542-W},
    issn = {1460-2075},
    pmid = {40846900}
}

@article{Sun2024GALNT6Cells,
    title = {{GALNT6 promotes bladder cancer malignancy and immune escape by epithelial-mesenchymal transition and CD8+ T cells}},
    year = {2024},
    journal = {Cancer Cell International 2024 24:1},
    author = {Sun, Xiaoxin and Wu, Haotian and Tang, Ling and Al-Danakh, Abdullah and Jian, Yuli and Gong, Li and Li, Congchen and Yu, Xiao and Zeng, Guang and Chen, Qiwei and Yang, Deyong and Wang, Shujing},
    number = {1},
    month = {9},
    pages = {308-},
    volume = {24},
    publisher = {BioMed Central},
    url = {https://link.springer.com/articles/10.1186/s12935-024-03492-1 https://link.springer.com/article/10.1186/s12935-024-03492-1},
    doi = {10.1186/S12935-024-03492-1},
    issn = {1475-2867}
}

@article{You2018GeometryColonies,
    title = {{Geometry and Mechanics of Microdomains in Growing Bacterial Colonies}},
    year = {2018},
    journal = {Physical Review X},
    author = {You, Zhihong and Pearce, Daniel J.G. and Sengupta, Anupam and Giomi, Luca},
    number = {3},
    month = {9},
    pages = {031065},
    volume = {8},
    publisher = {American Physical Society},
    url = {https://journals.aps.org/prx/abstract/10.1103/PhysRevX.8.031065},
    doi = {10.1103/PHYSREVX.8.031065/LD-5.AVI},
    issn = {21603308},
    arxivId = {1703.04504}
}

@article{Bray2024GlobalCountries,
    title = {{Global cancer statistics 2022: GLOBOCAN estimates of incidence and mortality worldwide for 36 cancers in 185 countries}},
    year = {2024},
    journal = {CA: A Cancer Journal for Clinicians},
    author = {Bray, Freddie and Laversanne, Mathieu and Sung, Hyuna and Ferlay, Jacques and Siegel, Rebecca L. and Soerjomataram, Isabelle and Jemal, Ahmedin},
    number = {3},
    month = {5},
    pages = {229--263},
    volume = {74},
    publisher = {Wiley},
    doi = {10.3322/caac.21834},
    issn = {0007-9235},
    pmid = {38572751}
}

@article{Liu2025Gram-negativeCancer,
    title = {{Gram-negative intratumoral bacteria mediate lymph node metastasis through LPS-TLR4/MAPK signaling pathway in cervical cancer}},
    year = {2025},
    journal = {Journal of Infection},
    author = {Liu, Yaqin and Ma, Peng and Liu, Dongdong and Liu, Yongzhu and Ran, Ziwei and Yang, Lunhao and Xu, Lingqing and Yin, Weiguo and Chen, Fu and Li, Linhai and Lu, Yang},
    number = {2},
    month = {8},
    pages = {106532},
    volume = {91},
    publisher = {W.B. Saunders},
    url = {https://www.sciencedirect.com/science/article/pii/S0163445325001264},
    doi = {10.1016/J.JINF.2025.106532},
    issn = {0163-4453},
    pmid = {40513620}
}

@article{Hanahan2022HallmarksDimensions,
    title = {{Hallmarks of Cancer: New Dimensions}},
    year = {2022},
    journal = {Cancer Discovery},
    author = {Hanahan, Douglas},
    number = {1},
    month = {1},
    pages = {31--46},
    volume = {12},
    publisher = {American Association for Cancer Research},
    url = {/cancerdiscovery/article/12/1/31/675608/Hallmarks-of-Cancer-New-DimensionsHallmarks-of https://dx.doi.org/10.1158/2159-8290.CD-21-1059},
    doi = {10.1158/2159-8290.CD-21-1059},
    issn = {2159-8274},
    pmid = {35022204}
}

@article{Correa1990HelicobacterRisks,
    title = {{Helicobacter pylori and gastric carcinoma: Serum antibody prevalence in populations with contrasting cancer risks}},
    year = {1990},
    journal = {Cancer},
    author = {Correa, Pelayo and Fox, James and Fontham, Elizabeth and Ruiz, Bernardo and Lin, Youping and Zavala, Diego and Taylor, Nancy and Mackinley, Dawn and De Lima, Eduardo and Portilla, Humberto and Zarama, Guillermo},
    number = {12},
    pages = {2569--2574},
    volume = {66},
    doi = {10.1002/1097-0142(19901215)66:12<2569::AID-CNCR2820661220>3.0.CO;2-I},
    issn = {10970142},
    pmid = {2249197}
}

@article{Parsonnet1991HelicobacterCarcinoma,
    title = {{Helicobacter pylori Infection and the Risk of Gastric Carcinoma}},
    year = {1991},
    journal = {New England Journal of Medicine},
    author = {Parsonnet, Julie and Friedman, Gary D. and Vandersteen, Daniel P. and Chang, Yuan and Vogelman, Joseph H. and Orentreich, Norman and Sibley, Richard K.},
    number = {16},
    month = {10},
    pages = {1127--1131},
    volume = {325},
    url = {http://www.nejm.org/doi/abs/10.1056/NEJM199110173251603},
    doi = {10.1056/NEJM199110173251603},
    issn = {0028-4793}
}

@article{Fernandez2019HepaticFunction,
    title = {{Hepatic Tumor Microenvironments and Effects on NK Cell Phenotype and Function}},
    year = {2019},
    journal = {International Journal of Molecular Sciences 2019, Vol. 20, Page 4131},
    author = {Fern{\'{a}}ndez, Julián Piñeiro and Luddy, Kimberly A. and Harmon, Cathal and O’Farrelly, Cliona},
    number = {17},
    month = {8},
    pages = {4131},
    volume = {20},
    publisher = {Multidisciplinary Digital Publishing Institute},
    url = {https://www.mdpi.com/1422-0067/20/17/4131/htm https://www.mdpi.com/1422-0067/20/17/4131},
    doi = {10.3390/IJMS20174131},
    issn = {1422-0067},
    pmid = {31450598}
}

@article{Song2017HowMotility,
    title = {{How Bacteria Respond to Material Stiffness during Attachment: A Role of Escherichia coli Flagellar Motility}},
    year = {2017},
    journal = {ACS Applied Materials and Interfaces},
    author = {Song, Fangchao and Brasch, Megan E. and Wang, Hao and Henderson, James H. and Sauer, Karin and Ren, Dacheng},
    number = {27},
    month = {7},
    pages = {22176--22184},
    volume = {9},
    publisher = {American Chemical Society},
    url = {https://pubs.acs.org/doi/abs/10.1021/acsami.7b04757},
    doi = {10.1021/ACSAMI.7B04757/SUPPL{\_}FILE/AM7B04757{\_}SI{\_}007.AVI},
    issn = {19448252},
    pmid = {28636823}
}

@article{Acerbi2015HumanInfiltration,
    title = {{Human breast cancer invasion and aggression correlates with ECM stiffening and immune cell infiltration}},
    year = {2015},
    journal = {Integrative Biology},
    author = {Acerbi, I. and Cassereau, L. and Dean, I. and Shi, Q. and Au, A. and Park, C. and Chen, Y. Y. and Liphardt, J. and Hwang, E. S. and Weaver, V. M.},
    number = {10},
    month = {10},
    pages = {1120--1134},
    volume = {7},
    publisher = {Oxford Academic},
    url = {https://dx.doi.org/10.1039/c5ib00040h},
    doi = {10.1039/C5IB00040H},
    issn = {17579708},
    pmid = {25959051}
}

@article{Aparicio-Yuste2025HybridMonolayers,
    title = {{Hybrid model to simulate host cell biomechanics and infection spread during intracellular infection of epithelial monolayers}},
    year = {2025},
    journal = {Computers in Biology and Medicine},
    author = {Aparicio-Yuste, Raul and Hundsdorfer, Lara and Bastounis, Effie E. and Gomez-Benito, Maria Jose},
    month = {2},
    pages = {109506},
    volume = {185},
    publisher = {Pergamon},
    url = {https://www.sciencedirect.com/science/article/pii/S0010482524015919#sec2},
    doi = {10.1016/J.COMPBIOMED.2024.109506},
    issn = {0010-4825},
    pmid = {39662314}
}

@article{Rawat2022HyperactiveActivity,
    title = {{Hyperactive neutrophils infiltrate vital organs of tumor bearing host and contribute to gradual systemic deterioration via upregulated NE, MPO and MMP-9 activity}},
    year = {2022},
    journal = {Immunology Letters},
    author = {Rawat, Kavita and Syeda, Saima and Shrivastava, Anju},
    month = {1},
    pages = {35--48},
    volume = {241},
    publisher = {Elsevier},
    url = {https://www.sciencedirect.com/science/article/pii/S0165247821001814},
    doi = {10.1016/J.IMLET.2021.12.001},
    issn = {0165-2478},
    pmid = {34890699}
}

@article{Gerbec2025IdentificationMetastasis,
    title = {{Identification of intratumoral bacteria that enhance breast tumor metastasis}},
    year = {2025},
    journal = {mBio},
    author = {Gerbec, Zachary J. and Serapio-Palacios, Antonio and Metcalfe-Roach, Avril and Krekhno, Zakhar and Bar-Yoseph, Haggai and Woodward, Sarah E. and Pena-D{\'{i}}az, Jorge and Nemirovsky, Oksana and Awrey, Shannon and Moreno, Sebastian H. and Beatty, Sean and Kong, Esther and Radisavljevic, Nina and Cirstea, Mihai and Chafe, Shawn and McDonald, Paul C. and Aparicio, Sam and Finlay, B. Brett and Dedhar, Shoukat},
    number = {3},
    month = {3},
    volume = {16},
    publisher = {American Society for Microbiology},
    url = {/doi/pdf/10.1128/mbio.03595-24?download=true},
    doi = {10.1128/MBIO.03595-24;SUBPAGE:STRING:FULL},
    issn = {21507511},
    pmid = {39932300}
}

@article{Qualls2015ImmunometabolismRespiration,
    title = {{Immunometabolism within the tuberculosis granuloma: amino acids, hypoxia, and cellular respiration}},
    year = {2015},
    journal = {Seminars in Immunopathology 2015 38:2},
    author = {Qualls, Joseph E. and Murray, Peter J.},
    number = {2},
    month = {10},
    pages = {139--152},
    volume = {38},
    publisher = {Springer},
    url = {https://link.springer.com/article/10.1007/s00281-015-0534-0},
    doi = {10.1007/S00281-015-0534-0},
    issn = {1863-2300},
    pmid = {26490974}
}

@article{Takahashi2023ImpactTumours,
    title = {{Impact of tumoral structure and bacterial species on growth and biodistribution of live bacterial therapeutics in xenografted tumours}},
    year = {2023},
    journal = {Journal of Drug Targeting},
    author = {Takahashi, Maiko and Sukowati, Erike Widyasari and Nomura, Shoko and Kato, Akari and Mizuseki, Kenji and Watanabe, Yasuyoshi and Mukai, Hidefumi},
    number = {2},
    pages = {194--205},
    volume = {31},
    publisher = {Taylor and Francis Ltd.},
    url = {https://www.tandfonline.com/doi/pdf/10.1080/1061186X.2022.2122477},
    doi = {10.1080/1061186X.2022.2122477;WEBSITE:WEBSITE:TFOPB;PAGEGROUP:STRING:PUBLICATION},
    issn = {10292330},
    pmid = {36097977}
}

@article{Manifacier2024InDistances,
    title = {{In silico analysis shows that dynamic changes in curvature guide cell migration over long distances}},
    year = {2024},
    journal = {Biomechanics and Modeling in Mechanobiology},
    author = {Manifacier, Ian and Carlin, Gildas and Liu, Dongshu and Vassaux, Maxime and Pieuchot, Laurent and Luchnikov, Valeriy and Anselme, Karine and Milan, Jean Louis},
    number = {1},
    month = {2},
    pages = {315--333},
    volume = {23},
    publisher = {Springer Science and Business Media Deutschland GmbH},
    doi = {10.1007/s10237-023-01777-4},
    issn = {16177940},
    pmid = {37875692}
}

@article{Lin2024InImmunity,
    title = {{In situ immunomodulation of tumors with biosynthetic bacteria promote anti-tumor immunity}},
    year = {2024},
    journal = {Bioactive Materials},
    author = {Lin, Zhongda and Meng, Fanqiang and Ma, Yumeng and Zhang, Chi and Zhang, Zhirang and Yang, Zhaoxin and Li, Yuan and Hou, Linlin and Xu, Yuzhong and Liang, Xin and Zhang, Xudong},
    month = {2},
    pages = {12--27},
    volume = {32},
    publisher = {Elsevier},
    url = {https://www.sciencedirect.com/science/article/pii/S2452199X23002785},
    doi = {10.1016/J.BIOACTMAT.2023.09.007},
    issn = {2452-199X}
}

@article{Inui2025InModulation,
    title = {{In vitro impact of Streptococcus mitis on the inhibition of oral cancer cell proliferation via mitotic modulation}},
    year = {2025},
    journal = {Frontiers in Cellular and Infection Microbiology},
    author = {Inui, Inori and Mochizuki, Shinichi and Hirabayashi-Nishimuta, Fumika and Yoshioka, Yoshie and Takahashi, Osamu and Sasaguri, Masaaki and Habu, Manabu and Ariyoshi, Wataru and Yamasaki, Ryota},
    month = {5},
    pages = {1524820},
    volume = {15},
    publisher = {Frontiers Media SA},
    doi = {10.3389/FCIMB.2025.1524820/BIBTEX},
    issn = {22352988},
    pmid = {40415961}
}

@article{McCoy2024InInteractions,
    title = {{In Vitro Models for Investigating Intestinal Host–Pathogen Interactions}},
    year = {2024},
    journal = {Advanced Science},
    author = {McCoy, Reece and Oldroyd, Sophie and Yang, Woojin and Wang, Kaixin and Hoven, Darius and Bulmer, David and Zilbauer, Matthias and Owens, Róisín M.},
    number = {8},
    month = {2},
    pages = {2306727},
    volume = {11},
    publisher = {John Wiley and Sons Inc},
    url = {/doi/pdf/10.1002/advs.202306727 https://onlinelibrary.wiley.com/doi/abs/10.1002/advs.202306727 https://advanced.onlinelibrary.wiley.com/doi/10.1002/advs.202306727},
    doi = {10.1002/ADVS.202306727;REQUESTEDJOURNAL:JOURNAL:21983844;ISSUE:ISSUE:DOI},
    issn = {21983844},
    pmid = {38155358}
}

@article{Chung2019Interface-derivedCeramics,
    title = {{Interface-derived solid-state viscoelasticity exhibited by nanostructured and microstructured materials containing carbons or ceramics}},
    year = {2019},
    journal = {Carbon},
    author = {Chung, D. D.L.},
    month = {4},
    pages = {567--581},
    volume = {144},
    publisher = {Pergamon},
    url = {https://www.sciencedirect.com/science/article/pii/S0008622318312375},
    doi = {10.1016/J.CARBON.2018.12.097},
    issn = {0008-6223}
}

@article{Ma2024IntratumorMetastasis,
    title = {{Intratumor microbiome-derived butyrate promotes lung cancer metastasis}},
    year = {2024},
    journal = {Cell Reports Medicine},
    author = {Ma, Yi and Chen, Haiming and Li, Haoran and Zheng, Meiling and Zuo, Xianglin and Wang, Wenxiang and Wang, Shaodong and Lu, Yiming and Wang, Jun and Li, Yun and Wang, Jie and Qiu, Mantang},
    number = {4},
    month = {4},
    pages = {101488},
    volume = {5},
    publisher = {Cell Press},
    url = {https://www.cell.com/action/showFullText?pii=S2666379124001344 https://www.cell.com/action/showAbstract?pii=S2666379124001344 https://www.cell.com/cell-reports-medicine/abstract/S2666-3791(24)00134-4},
    doi = {10.1016/J.XCRM.2024.101488/ASSET/9ED26949-CDC2-4A7F-8F09-A7462917287C/MAIN.ASSETS/GR7.JPG},
    issn = {26663791},
    pmid = {38565146}
}

@article{Mahadevan2023KRASG12DCells,
    title = {{KRASG12D inhibition reprograms the microenvironment of early and advanced pancreatic cancer to promote FAS-mediated killing by CD8+ T cells}},
    year = {2023},
    journal = {Cancer Cell},
    author = {Mahadevan, Krishnan K. and McAndrews, Kathleen M. and LeBleu, Valerie S. and Yang, Sujuan and Lyu, Hengyu and Li, Bingrui and Sockwell, Amari M. and Kirtley, Michelle L. and Morse, Sami J. and Moreno Diaz, Barbara A. and Kim, Michael P. and Feng, Ningping and Lopez, Anastasia M. and Guerrero, Paola A. and Paradiso, Francesca and Sugimoto, Hikaru and Arian, Kent A. and Ying, Haoqiang and Barekatain, Yasaman and Sthanam, Lakshmi Kavitha and Kelly, Patience J. and Maitra, Anirban and Heffernan, Timothy P. and Kalluri, Raghu},
    number = {9},
    month = {9},
    pages = {1606--1620},
    volume = {41},
    publisher = {Cell Press},
    url = {https://www.cell.com/action/showFullText?pii=S1535610823002428 https://www.cell.com/action/showAbstract?pii=S1535610823002428 https://www.cell.com/cancer-cell/abstract/S1535-6108(23)00242-8},
    doi = {10.1016/j.ccell.2023.07.002},
    issn = {18783686},
    pmid = {37625401}
}

@article{Yu2025LactateAdenocarcinoma,
    title = {{Lactate production by tumor-resident Staphylococcus promotes metastatic colonization in lung adenocarcinoma}},
    year = {2025},
    journal = {Cell Host and Microbe},
    author = {Yu, Huan and Du, Yang and He, Yuling and Sun, Yifan and Li, Junfeng and Jia, Bo and Chen, Jiale and Peng, Xinya and An, Tongtong and Li, Jianjie and Chi, Yujia and Wang, Man and Cao, Lihua and Tai, Yidi and Zhai, Xiaoyu and Nuersulitan, Reyizha and Li, Sheng and Wu, Nan and Wang, Jia and Xiong, Hongchao and Wan, Shujie and Liu, Jiaming and Yang, Xuelian and Hu, Xingsheng and Lin, Dongmei and Sun, Wei and Wang, Yue and An, Guo and Ji, Xumeng and Kong, Lingdong and Ding, Lu and Ma, Yunan and Tian, Zhihua and Dong, Bin and Bi, Yujing and Wu, Jianmin and Wang, Ziping},
    number = {7},
    month = {7},
    pages = {1089--1105},
    volume = {33},
    publisher = {Cell Press},
    url = {https://www.cell.com/action/showFullText?pii=S1931312825002422 https://www.cell.com/action/showAbstract?pii=S1931312825002422 https://www.cell.com/cell-host-microbe/abstract/S1931-3128(25)00242-2},
    doi = {10.1016/j.chom.2025.06.013},
    issn = {19346069},
    pmid = {40639336}
}

@article{Feng2024M2-TypeMetastases,
    title = {{M2-Type Macrophages and Cancer-Associated Fibroblasts Combine to Promote Colorectal Cancer Liver Metastases}},
    year = {2024},
    journal = {OncoTargets and Therapy},
    author = {Feng, Yunpeng and Qiao, Shifeng and Chen, Jie and Wen, Xin and Chen, Yanlei and Xu, Jiaxin and Qiao, Xiucheng and Yang, Jing and Zhang, Shenshen and Feng, Yang and Song, Xiaoyu and Gao, Yu},
    pages = {243--260},
    volume = {17},
    publisher = {Dove Medical Press Ltd},
    url = {https://www.tandfonline.com/doi/pdf/10.2147/OTT.S447502},
    doi = {10.2147/OTT.S447502;REQUESTEDJOURNAL:JOURNAL:DOTT20;WGROUP:STRING:PUBLICATION},
    issn = {11786930}
}

@article{Xie2022MammaryFerroptosis,
    title = {{Mammary adipocytes protect triple-negative breast cancer cells from ferroptosis}},
    year = {2022},
    journal = {Journal of Hematology {\&} Oncology 2022 15:1},
    author = {Xie, Yizhao and Wang, Biyun and Zhao, Yannan and Tao, Zhonghua and Wang, Ye and Chen, Guangliang and Hu, Xichun},
    number = {1},
    month = {6},
    pages = {72-},
    volume = {15},
    publisher = {BioMed Central},
    url = {https://link.springer.com/articles/10.1186/s13045-022-01297-1 https://link.springer.com/article/10.1186/s13045-022-01297-1},
    doi = {10.1186/S13045-022-01297-1},
    issn = {1756-8722},
    pmid = {35659320}
}

@article{Galvez1997MathematicalLines,
    title = {{Mathematical modeling of adhesion of bacteria to host cell lines}},
    year = {1997},
    journal = {Bulletin of Mathematical Biology 1997 59:5},
    author = {G{\'{a}}lvez, Jesus and Lajarin, Francisco and Garda-Pelqarrubia, Pilar},
    number = {5},
    month = {9},
    pages = {833--856},
    volume = {59},
    publisher = {Springer},
    url = {https://link.springer.com/article/10.1007/BF02459995},
    doi = {10.1007/BF02459995},
    issn = {1522-9602}
}

@article{Suh2018MathematicalSpheroids,
    title = {{Mathematical Modeling of Bacteria-Enabled Drug Delivery System Penetration into Multicellular Tumor Spheroids}},
    year = {2018},
    journal = {Conference proceedings : ... Annual International Conference of the IEEE Engineering in Medicine and Biology Society. IEEE Engineering in Medicine and Biology Society. Annual Conference},
    author = {Suh, S. and Leaman, E. J. and Zhan, Ying and Behkam, B.},
    month = {7},
    pages = {6162--6165},
    volume = {2018},
    publisher = {NLM (Medline)},
    url = {https://ieeexplore.ieee.org/abstract/document/8513596},
    doi = {10.1109/EMBC.2018.8513596},
    issn = {1557170X},
    pmid = {30441741}
}

@article{Dominguez-Huttinger2017MathematicalTreatment,
    title = {{Mathematical modeling of Streptococcus pneumoniae colonization, invasive infection and treatment}},
    year = {2017},
    journal = {Frontiers in Physiology},
    author = {Dom{\'{i}}nguez-H{\"{u}}ttinger, Elisa and Boon, Neville J. and Clarke, Thomas B. and Tanaka, Reiko J.},
    number = {MAR},
    month = {3},
    pages = {245164},
    volume = {8},
    publisher = {Frontiers Research Foundation},
    url = {www.frontiersin.org},
    doi = {10.3389/FPHYS.2017.00115/BIBTEX},
    issn = {1664042X}
}

@article{Radzki2024MatrixPeriodontitis,
    title = {{Matrix Metalloproteinases in the Periodontium—Vital in Tissue Turnover and Unfortunate in Periodontitis}},
    year = {2024},
    journal = {International Journal of Molecular Sciences 2024, Vol. 25, Page 2763},
    author = {Radzki, Dominik and Negri, Alessandro and Kusiak, Aida and Obuchowski, Michał},
    number = {5},
    month = {2},
    pages = {2763},
    volume = {25},
    publisher = {Multidisciplinary Digital Publishing Institute},
    url = {https://www.mdpi.com/1422-0067/25/5/2763/htm https://www.mdpi.com/1422-0067/25/5/2763},
    doi = {10.3390/IJMS25052763},
    issn = {1422-0067},
    pmid = {38474009}
}

@article{Wu2023MatrixPathway,
    title = {{Matrix stiffening facilitates the collective invasion of breast cancer through the periostin-integrin mechanotransduction pathway}},
    year = {2023},
    journal = {Matrix Biology},
    author = {Wu, Tiantian and Xiong, Shanshan and Chen, Mimi and Tam, Bjorn T. and Chen, Wei and Dong, Ke and Ma, Zhenling and Wang, Zhe and Ouyang, Gaoliang},
    month = {8},
    pages = {22--40},
    volume = {121},
    publisher = {Elsevier},
    url = {https://www.sciencedirect.com/science/article/pii/S0945053X23000690},
    doi = {10.1016/J.MATBIO.2023.05.006},
    issn = {0945-053X},
    pmid = {37230256}
}

@article{Sun2025MatrixSignalling,
    title = {{Matrix stiffness induced gallbladder fibroblasts activation and paracrine SEMA7A promotes gallbladder cancer cell epithelial-mesenchymal transition and cancer stem cell-like properties by modulating AKT/p300 signalling}},
    year = {2025},
    journal = {Biology Direct 2025 20:1},
    author = {Sun, Liankang and Wang, Huanhuan and Li, Hanqi and Yang, Ruida and Dong, Xuyuan and Dong, Danfeng and Fan, Yangwei and Li, Enxiao and Wang, Hao and Wu, Yinying and Shi, Yu},
    number = {1},
    month = {8},
    pages = {1--20},
    volume = {20},
    publisher = {BioMed Central},
    url = {https://link.springer.com/articles/10.1186/s13062-025-00683-2 https://link.springer.com/article/10.1186/s13062-025-00683-2},
    doi = {10.1186/S13062-025-00683-2},
    issn = {1745-6150},
    pmid = {40830485}
}

@article{Rice2017MatrixCells,
    title = {{Matrix stiffness induces epithelial–mesenchymal transition and promotes chemoresistance in pancreatic cancer cells}},
    year = {2017},
    journal = {Oncogenesis 2017 6:7},
    author = {Rice, A. J. and Cortes, E. and Lachowski, D. and Cheung, B. C.H. and Karim, S. A. and Morton, J. P. and Del R{\'{i}}o Hern{\'{a}}ndez, A.},
    number = {7},
    month = {7},
    pages = {e352-e352},
    volume = {6},
    publisher = {Nature Publishing Group},
    url = {https://www.nature.com/articles/oncsis201754},
    doi = {10.1038/oncsis.2017.54},
    issn = {2157-9024}
}

@article{Zhou2025MechanicalProgression,
    title = {{Mechanical cues shaping cancer cell-microbiota interaction dynamics in tumor progression}},
    year = {2025},
    journal = {Cell Reports Physical Science},
    author = {Zhou, Yan and Zhou, Xuan and Zhu, Xiaoyu and Xu, Feng and Li, Fei},
    number = {6},
    month = {6},
    pages = {102612},
    volume = {6},
    publisher = {Cell Press},
    url = {https://www.cell.com/action/showFullText?pii=S2666386425002115 https://www.cell.com/action/showAbstract?pii=S2666386425002115 https://www.cell.com/cell-reports-physical-science/abstract/S2666-3864(25)00211-5},
    doi = {10.1016/j.xcrp.2025.102612},
    issn = {26663864}
}

@article{Bastounis2022MechanicalPathogens,
    title = {{Mechanical Forces Govern Interactions of Host Cells with Intracellular Bacterial Pathogens}},
    year = {2022},
    journal = {Microbiology and Molecular Biology Reviews},
    author = {Bastounis, Effie E. and Radhakrishnan, Prathima and Prinz, Christopher K. and Theriot, Julie A.},
    number = {2},
    month = {6},
    volume = {86},
    publisher = {American Society for Microbiology1752 N St., N.W., Washington, DC},
    url = {/doi/pdf/10.1128/mmbr.00094-20?download=true},
    doi = {10.1128/MMBR.00094-20},
    issn = {1092-2172},
    pmid = {35285720}
}

@article{Wang2021MechanicalMatrices,
    title = {{Mechanical regulation of cell volume in 3D extracellular matrices}},
    year = {2021},
    journal = {Extreme Mechanics Letters},
    author = {Wang, Pudi and Zhang, Qing and Fang, Xu and Lin, Feng and Huang, Jianyong},
    month = {11},
    pages = {101498},
    volume = {49},
    publisher = {Elsevier},
    url = {https://www.sciencedirect.com/science/article/pii/S2352431621001929},
    doi = {10.1016/J.EML.2021.101498},
    issn = {2352-4316}
}

@article{Hayek2020MechanismsConditions,
    title = {{Mechanisms controlling bacterial infection in myeloid cells under hypoxic conditions}},
    year = {2020},
    journal = {Cellular and Molecular Life Sciences 2020 78:5},
    author = {Hayek, Inaya and Schatz, Valentin and Bogdan, Christian and Jantsch, Jonathan and L{\"{u}}hrmann, Anja},
    number = {5},
    month = {10},
    pages = {1887--1907},
    volume = {78},
    publisher = {Springer},
    url = {https://link.springer.com/article/10.1007/s00018-020-03684-8},
    isbn = {0123456789},
    doi = {10.1007/S00018-020-03684-8},
    issn = {1420-9071},
    pmid = {33125509}
}

@article{Dufrene2020Mechanomicrobiology:Forces,
    title = {{Mechanomicrobiology: how bacteria sense and respond to forces}},
    year = {2020},
    journal = {Nature Reviews Microbiology 2020 18:4},
    author = {Dufr{\^{e}}ne, Yves F. and Persat, Alexandre},
    number = {4},
    month = {1},
    pages = {227--240},
    volume = {18},
    publisher = {Nature Publishing Group},
    url = {https://www.nature.com/articles/s41579-019-0314-2},
    doi = {10.1038/s41579-019-0314-2},
    issn = {1740-1534},
    pmid = {31959911}
}

@article{Hajar2021MedicinePresent,
    title = {{Medicine from Galen to the Present}},
    year = {2021},
    journal = {Heart Views},
    author = {Hajar, Rachel},
    number = {4},
    month = {10},
    pages = {307--308},
    volume = {22},
    url = {https://journals.lww.com/10.4103/heartviews.heartviews_125_21},
    isbn = {2021;22:3078},
    doi = {10.4103/heartviews.heartviews{\_}125{\_}21},
    issn = {1995-705X}
}

@article{Oudin2016MenaINVHaptotaxis,
    title = {{MenaINV mediates synergistic cross-talk between signaling pathways driving chemotaxis and haptotaxis}},
    year = {2016},
    journal = {https://doi.org/10.1091/mbc.e16-04-0212},
    author = {Oudin, Madeleine J. and Miller, Miles A. and Klazen, Joelle A.Z. and Kosciuk, Tatsiana and Lussiez, Alisha and Hughes, Shannon K. and Tadros, Jenny and Bear, James E. and Lauffenburger, Douglas A. and Gertler, Frank B.},
    number = {20},
    month = {8},
    pages = {3085--3094},
    volume = {27},
    publisher = { The American Society for Cell Biology },
    url = {https://www.molbiolcell.org/doi/pdf/10.1091/mbc.e16-04-0212},
    doi = {10.1091/MBC.E16-04-0212},
    issn = {19394586},
    pmid = {27559126}
}

@article{Palma2022MethodsInteractions,
    title = {{Methods to Evaluate Bacterial Motility and Its Role in Bacterial–Host Interactions}},
    year = {2022},
    journal = {Microorganisms 2022, Vol. 10, Page 563},
    author = {Palma, Victoria and Guti{\'{e}}rrez, María Soledad and Vargas, Orlando and Parthasarathy, Raghuveer and Navarrete, Paola},
    number = {3},
    month = {3},
    pages = {563},
    volume = {10},
    publisher = {Multidisciplinary Digital Publishing Institute},
    url = {https://www.mdpi.com/2076-2607/10/3/563/htm https://www.mdpi.com/2076-2607/10/3/563},
    doi = {10.3390/MICROORGANISMS10030563},
    issn = {2076-2607}
}

@article{Jin2024MicrobesFeedback,
    title = {{Microbes in porous environments: from active interactions to emergent feedback}},
    year = {2024},
    journal = {Biophysical Reviews 2024 16:2},
    author = {Jin, Chenyu and Sengupta, Anupam},
    number = {2},
    month = {4},
    pages = {173--188},
    volume = {16},
    publisher = {Springer},
    url = {https://link.springer.com/article/10.1007/s12551-024-01185-7},
    isbn = {0123456789},
    doi = {10.1007/S12551-024-01185-7},
    issn = {1867-2469},
    arxivId = {2311.11440}
}

@article{Sengupta2020MicrobialFramework,
    title = {{Microbial Active Matter: A Topological Framework}},
    year = {2020},
    journal = {Frontiers in Physics},
    author = {Sengupta, Anupam},
    month = {6},
    pages = {517227},
    volume = {8},
    publisher = {Frontiers Media SA},
    url = {www.frontiersin.org},
    doi = {10.3389/FPHY.2020.00184/TEXT},
    issn = {2296424X},
    arxivId = {2004.08904}
}

@article{Lehr2023MicrobialCancer,
    title = {{Microbial composition of tumorous and adjacent gastric tissue is associated with prognosis of gastric cancer}},
    year = {2023},
    journal = {Scientific Reports 2023 13:1},
    author = {Lehr, Konrad and Nikitina, Darja and Vilchez-Vargas, Ramiro and Steponaitiene, Ruta and Thon, Cosima and Skieceviciene, Jurgita and Schanze, Denny and Zenker, Martin and Malfertheiner, Peter and Kupcinskas, Juozas and Link, Alexander},
    number = {1},
    month = {3},
    pages = {1--11},
    volume = {13},
    publisher = {Nature Publishing Group},
    url = {https://www.nature.com/articles/s41598-023-31740-3},
    isbn = {0123456789},
    doi = {10.1038/s41598-023-31740-3},
    issn = {2045-2322},
    pmid = {36944721}
}

@article{Cullin2021MicrobiomeCancer,
    title = {{Microbiome and cancer}},
    year = {2021},
    journal = {Cancer Cell},
    author = {Cullin, Nyssa and Azevedo Antunes, Camila and Straussman, Ravid and Stein-Thoeringer, Christoph K. and Elinav, Eran},
    number = {10},
    month = {10},
    pages = {1317--1341},
    volume = {39},
    publisher = {Elsevier},
    url = {https://www.cell.com/action/showFullText?pii=S1535610821004463 https://www.cell.com/action/showAbstract?pii=S1535610821004463 https://www.cell.com/cancer-cell/abstract/S1535-6108(21)00446-3},
    doi = {10.1016/J.CCELL.2021.08.006},
    issn = {1535-6108},
    pmid = {34506740}
}

@article{Zhu2023MicrobiomeMice,
    title = {{Microbiome and spatially resolved metabolomics analysis reveal the anticancer role of gut Akkermansia muciniphila by crosstalk with intratumoral microbiota and reprogramming tumoral metabolism in mice}},
    year = {2023},
    journal = {Gut Microbes},
    author = {Zhu, Zhuxian and Cai, Jixu and Hou, Weiwei and Xu, Ke and Wu, Xuxiao and Song, Yuanlin and Bai, Chunxue and Mo, Yin Yuan and Zhang, Ziqiang},
    number = {1},
    month = {12},
    volume = {15},
    publisher = {Taylor and Francis Ltd.},
    url = {https://www.tandfonline.com/doi/pdf/10.1080/19490976.2023.2166700},
    doi = {10.1080/19490976.2023.2166700;PAGE:STRING:ARTICLE/CHAPTER},
    issn = {19490984},
    pmid = {36740846}
}

@article{Meyers2025MicrobiomeMethods,
    title = {{Microbiome in cancer metastasis: biological insights and emerging spatial omics methods}},
    year = {2025},
    journal = {Frontiers in Cellular and Infection Microbiology},
    author = {Meyers, Marianne and Stoffels, Charlotte B.A. and Frache, Gilles and Letellier, Elisabeth and Feucherolles, Maureen},
    month = {6},
    pages = {1559870},
    volume = {15},
    publisher = {Frontiers Media SA},
    url = {https://BioRender.com/q18n156},
    doi = {10.3389/FCIMB.2025.1559870/FULL},
    issn = {22352988},
    pmid = {40535543}
}

@misc{Mai2024ModulatingImmunotherapy,
    title = {{Modulating extracellular matrix stiffness: a strategic approach to boost cancer immunotherapy}},
    year = {2024},
    booktitle = {Cell Death and Disease},
    author = {Mai, Zizhao and Lin, Yunfan and Lin, Pei and Zhao, Xinyuan and Cui, Li},
    number = {5},
    month = {5},
    volume = {15},
    publisher = {Springer Nature},
    doi = {10.1038/s41419-024-06697-4},
    issn = {20414889},
    pmid = {38693104}
}

@article{You2021Mono-to-multilayerColonies,
    title = {{Mono-to-multilayer transition in growing bacterial colonies}},
    year = {2021},
    journal = {Physical Review Letters},
    author = {You, Zhihong and Pearce, Daniel J. G. and Sengupta, Anupam and Giomi, Luca},
    number = {17},
    month = {3},
    volume = {123},
    publisher = {American Physical Society},
    url = {http://arxiv.org/abs/1811.08875 http://dx.doi.org/10.1103/PhysRevLett.123.178001},
    doi = {10.1103/PhysRevLett.123.178001},
    arxivId = {1811.08875v2}
}

@article{Zhang2021MorphogenesisBiofilms,
    title = {{Morphogenesis and cell ordering in confined bacterial biofilms}},
    year = {2021},
    journal = {Proceedings of the National Academy of Sciences of the United States of America},
    author = {Zhang, Qiuting and Li, Jian and Nijjer, Japinder and Lu, Haoran and Kothari, Mrityunjay and Alert, Ricard and Cohen, Tal and Yan, Jing},
    number = {31},
    month = {8},
    pages = {e2107107118},
    volume = {118},
    publisher = {National Academy of Sciences},
    url = {/doi/pdf/10.1073/pnas.2107107118?download=true},
    doi = {10.1073/PNAS.2107107118;PAGE:STRING:ARTICLE/CHAPTER},
    issn = {10916490},
    pmid = {34330824}
}

@article{Toley2012MotilityTissue,
    title = {{Motility is critical for effective distribution and accumulation of bacteria in tumor tissue}},
    year = {2012},
    journal = {Integrative Biology},
    author = {Toley, Bhushan J. and Forbes, Neil S.},
    number = {2},
    month = {2},
    pages = {165--176},
    volume = {4},
    publisher = {Oxford Academic},
    url = {https://dx.doi.org/10.1039/c2ib00091a},
    doi = {10.1039/C2IB00091A},
    issn = {17579694},
    pmid = {22193245}
}

@article{Wu2023MUC16Cancer,
    title = {{MUC16 stimulates neutrophils to an inflammatory and immunosuppressive phenotype in ovarian cancer}},
    year = {2023},
    journal = {Journal of Ovarian Research},
    author = {Wu, Yuliang and Liu, Qi and Xie, Yan and Zhu, Jihui and Zhang, Sai and Ge, Yao and Guo, Jing and Luo, Ning and Huang, Wei and Xu, Runping and Liu, Shupeng and Cheng, Zhongping},
    number = {1},
    month = {12},
    pages = {181-},
    volume = {16},
    publisher = {BioMed Central Ltd},
    url = {https://link.springer.com/articles/10.1186/s13048-023-01207-0 https://link.springer.com/article/10.1186/s13048-023-01207-0},
    doi = {10.1186/S13048-023-01207-0/FIGURES/4},
    issn = {17572215},
    pmid = {37644468}
}

@article{Nguyen2023NitricBacteria,
    title = {{Nitric Oxide-Induced Morphological Changes to Bacteria}},
    year = {2023},
    journal = {ACS Infectious Diseases},
    author = {Nguyen, Huan K. and Picciotti, Samantha L. and Duke, Magdalena M. and Broberg, Christopher A. and Schoenfisch, Mark H.},
    number = {11},
    month = {11},
    pages = {2316--2324},
    volume = {9},
    publisher = {American Chemical Society},
    url = {https://pubs.acs.org/doi/abs/10.1021/acsinfecdis.3c00415},
    doi = {10.1021/ACSINFECDIS.3C00415},
    issn = {23738227},
    pmid = {37831756}
}

@article{Soleimani2023NumericalCo-aggregation,
    title = {{Numerical and experimental investigation of multi-species bacterial co-aggregation}},
    year = {2023},
    journal = {Scientific Reports 2023 13:1},
    author = {Soleimani, Meisam and Szafranski, Szymon P. and Qu, Taoran and Mukherjee, Rumjhum and Stiesch, Meike and Wriggers, Peter and Junker, Philipp},
    number = {1},
    month = {7},
    pages = {11839-},
    volume = {13},
    publisher = {Nature Publishing Group},
    url = {https://www.nature.com/articles/s41598-023-38806-2},
    isbn = {0123456789},
    doi = {10.1038/s41598-023-38806-2},
    issn = {2045-2322},
    pmid = {37481628}
}

@article{Ruffer1893OnTumours,
    title = {{On some parasitic protozoa found in cancerous tumours}},
    year = {1893},
    journal = {The Journal of Pathology and Bacteriology},
    author = {Ruffer, M. Armand and Walker, J. Herbert},
    pages = {198--215},
    volume = {1},
    address = {Edinburgh},
    language = {English}
}

@article{Chen2022OncogenicCancer,
    title = {{Oncogenic collagen I homotrimers from cancer cells bind to {$\alpha$}3{$\beta$}1 integrin and impact tumor microbiome and immunity to promote pancreatic cancer}},
    year = {2022},
    journal = {Cancer Cell},
    author = {Chen, Yang and Yang, Sujuan and Tavormina, Jena and Tampe, Desiree and Zeisberg, Michael and Wang, Huamin and Mahadevan, Krishnan K. and Wu, Chang Jiun and Sugimoto, Hikaru and Chang, Chia Chi and Jenq, Robert R. and McAndrews, Kathleen M. and Kalluri, Raghu},
    number = {8},
    month = {8},
    pages = {818--834},
    volume = {40},
    publisher = {Cell Press},
    url = {https://www.cell.com/action/showFullText?pii=S1535610822002756 https://www.cell.com/action/showAbstract?pii=S1535610822002756 https://www.cell.com/cancer-cell/abstract/S1535-6108(22)00275-6},
    doi = {10.1016/J.CCELL.2022.06.011/ASSET/77E93CD9-36CF-4578-BDE6-9757EDD9957D/MAIN.ASSETS/GR7{\_}LRG.JPG},
    issn = {18783686},
    pmid = {35868307}
}

@article{Yang2024Pan-cancerSubtypes,
    title = {{Pan-cancer single-cell dissection reveals phenotypically distinct B cell subtypes}},
    year = {2024},
    journal = {Cell},
    author = {Yang, Yu and Chen, Xueyan and Pan, Jieying and Ning, Huiheng and Zhang, Yaojun and Bo, Yufei and Ren, Xianwen and Li, Jiesheng and Qin, Shishang and Wang, Dongfang and Chen, Min Min and Zhang, Zemin},
    number = {17},
    month = {8},
    pages = {4790--4811},
    volume = {187},
    publisher = {Elsevier},
    url = {https://www.cell.com/action/showFullText?pii=S0092867424007128 https://www.cell.com/action/showAbstract?pii=S0092867424007128 https://www.cell.com/cell/abstract/S0092-8674(24)00712-8},
    doi = {10.1016/J.CELL.2024.06.038},
    issn = {0092-8674},
    pmid = {39047727}
}

@article{Guttenplan2013RegulationFormation,
    title = {{Regulation of flagellar motility during biofilm formation}},
    year = {2013},
    journal = {FEMS Microbiol. Rev.},
    author = {Guttenplan, Sarah B. and Kearns, Daniel B.},
    number = {6},
    month = {11},
    pages = {849--871},
    volume = {37},
    doi = {10.1111/1574-6976.12018},
    issn = {01686445},
    pmid = {23480406}
}
\printglossary

\end{document}